\documentclass[letter,11pt]{article}
\pdfoutput=1
\usepackage{jheppub}
\usepackage{graphicx,epstopdf,amsmath,amsfonts,amssymb,appendix,comment}
\usepackage{color,slashed,subfigure,braket,multirow}
\usepackage{epsfig,mathrsfs,latexsym,color,url,etoolbox}
\usepackage{bbm}
\usepackage{ulem}
\usepackage{mathtools}
\usepackage{footnote}
\usepackage[usenames,dvipsnames,table]{xcolor}

\definecolor{nicered}{rgb}{0.7,0.1,0.1}
\definecolor{nicegreen}{rgb}{0.1,0.5,0.1}
\DeclareMathAlphabet{\mathbbold}{U}{bbold}{m}{n}

\newcommand{\beq}{\begin{equation}}
\newcommand{\eeq}{\end{equation}}
\newcommand\bout{\bgroup\markoverwith{\textcolor{blue}{\rule[0.5ex]{4pt}{0.8pt}}}\ULon}

\newcommand*{\absq}[1]{\lvert #1 \rvert^2}
\newcommand*{\eone}{\left \lvert U_{e1}\right\rvert^2}
\newcommand*{\etwo}{\left \lvert U_{e2}\right\rvert^2}
\newcommand*{\ethr}{\left \lvert U_{e3}\right\rvert^2}
\newcommand*{\mone}{\left \lvert U_{\mu 1}\right\rvert^2}
\newcommand*{\mtwo}{\left \lvert U_{\mu 2}\right\rvert^2}
\newcommand*{\mthr}{\left \lvert U_{\mu 3}\right\rvert^2}

\newcommand*{\tthr}{\left \lvert U_{\tau 3}\right\rvert^2}

\usepackage{hyperref}
\hypersetup{colorlinks,citecolor= nicegreen,linkcolor= nicered}

\def\Fermilab{Theoretical Physics Department, Fermilab, P.O. Box 500, Batavia, IL 60510, USA}
\def\SLAC{SLAC National Accelerator Laboratory, 2575 Sand Hill Road, Menlo Park, CA, USA}

\begin{document}

\title{Current and Future Neutrino Oscillation Constraints on Leptonic Unitarity}

\author[1]{Sebastian A. R. Ellis}
\author[2]{Kevin J. Kelly}
\author[1]{Shirley Weishi Li}

\affiliation[1]{\SLAC}
\affiliation[2]{\Fermilab}

\emailAdd{sarellis@slac.stanford.edu}
\emailAdd{kkelly12@fnal.gov}
\emailAdd{shirleyl@slac.stanford.edu}

\date\today

\abstract{
The unitarity of the lepton mixing matrix is a critical assumption underlying the standard neutrino-mixing paradigm. However, many models seeking to explain the as-yet-unknown origin of neutrino masses predict deviations from unitarity in the mixing of the active neutrino states. Motivated by the prospect that future experiments may provide a precise measurement of the lepton mixing matrix, we revisit current constraints on unitarity violation from oscillation measurements and project how next-generation experiments will improve our current knowledge. With the next-generation data, the normalizations of all rows and columns of the lepton mixing matrix will be constrained to $\lesssim$10\% precision, with the $e$-row best measured at $\lesssim$1\% and the $\tau$-row worst measured at ${\sim}10\%$ precision. The measurements of the mixing matrix elements themselves will be improved on average by a factor of $3$. We highlight the complementarity of DUNE, T2HK, JUNO, and IceCube Upgrade for these improvements, as well as the importance of $\nu_\tau$ appearance measurements and sterile neutrino searches for tests of leptonic unitarity.
}

\preprint{FERMILAB-PUB-20-364-T, SLAC-PUB-17552}

\setcounter{tocdepth}{2}

\maketitle
\flushbottom


\section{Introduction}

With the discovery that neutrinos oscillate came a new understanding of the standard model (SM) of particle physics -- neutrinos have mass and leptons mix. Many experiments have since been performed, with more planned, to deepen our understanding of the nature and origin of neutrino masses and their mixing. A coherent picture is forming regarding leptonic mixing and the three-massive-neutrinos paradigm through the experimental data gathered to date. However, open questions regarding the dynamics of the neutrino sector remain, with substantial room for new physics to provide answers.

Unitarity, the requirement that the matrix governing the transformation between two eigenbases satisfies $U^\dagger U = U U^\dagger = \mathbb{I}$, forms the basis of our understanding of SM fermion mixing~\cite{Maki:1962mu,Cabibbo:1963yz, Pontecorvo:1967fh,Kobayashi:1973fv}. This theoretical paradigm has been thoroughly tested to great acclaim in the quark sector~\cite{Hocker:2001xe,Bona:2006ah,CKMfitter,UTfit,Tanabashi:2018oca,Wolfenstein:1983yz,Buras:1994ec,Charles:2004jd}. However, our understanding of the corresponding leptonic mixing matrix (LMM) remains limited~\cite{Qian:2013ora,Parke:2015goa,Ellis:2020ehi}. The phenomenon of neutrino mixing predicates nonzero neutrino masses, and yet the SM does not provide a mechanism for such masses to exist. As a result, a plethora of models has been postulated to explain the origin of neutrino masses, and hence oscillations, involving new physics beyond the standard model (BSM)~\cite{Minkowski:1977sc,Mohapatra:1979ia,Schechter:1980gr,Mohapatra:1980yp,Cheng:1980qt,Zee:1980ai,Gelmini:1980re,Ma:2006km}. A key feature of many such models is that they predict the existence of new neutrino eigenstates, leading to non-unitarity of the active neutrino LMM used to characterize neutrino oscillations~\cite{Minkowski:1977sc,Weinberg:1979sa,Mohapatra:1979ia,Wyler:1982dd,Langacker:1988up,Hewett:1988xc,Nardi:1993ag,Tommasini:1995ii,Gluza:2002vs,Abazajian:2012ys}.

Many studies have been undertaken to study the effect of LMM non-unitarity and to determine existing and projected constraints on non-unitarity~\cite{Bilenky:1992wv,Bergmann:1998rg,Czakon:2001em,Bekman:2002zk,Antusch:2006vwa,FernandezMartinez:2007ms,Goswami:2008mi,Antusch:2009pm,Antusch:2014woa,Li:2015oal,Fernandez-Martinez:2016lgt,Blennow:2016jkn,Fong:2016yyh,Fong:2017gke,Coutinho:2019aiy}. Such constraints can be derived from analyzing a multitude of processes, such as decays involving leptons, and, crucially, neutrino oscillations. The latter are among the most theoretically clean probes of LMM unitarity. With this in mind, and given that future neutrino oscillation experiments will be capable of precise measurements, we revisit current constraints and project future constraints on the unitarity of the LMM from oscillation experiments. A previous exploration of oscillation constraints on LMM unitarity was performed in 2015 in Ref.~\cite{Parke:2015goa}, utilizing contemporary data. Experimental precision has since improved, with better precision expected in near-future experiments, motivating our in-depth study.

In this work, expanding on the set up in Ref.~\cite{Parke:2015goa}, we offer a more comprehensive perspective on leptonic unitarity.  We explore a set of reasonable assumptions regarding the possible origin of unitarity violation and discuss how they can affect tests of unitarity.  We also break down how different subsets of experiments contribute to the constraints on specific rows and columns of the LMM, highlighting the importance of sterile neutrino searches and the uniqueness of $\tau$-appearance searches. While we are interested specifically in oscillation-based constraints on unitarity, we discuss other probes, and their model-dependence as well. We include all existing oscillation measurements that make major contributions to unitarity constraints, as well as projections for oscillation-based constraints through the next decade. These include the planned IceCube Upgrade, Jiangmen Underground Neutrino Observatory (JUNO), Deep Underground Neutrino Experiment (DUNE), and Tokai to Hyper-Kamiokande (T2HK) experiments. In our companion paper~\cite{Ellis:2020ehi}, we explored this combination of current and future data to address the unitarity constraints and CP violation present in the LMM through unitarity triangles, an approach familiarized by studies of the quark mixing matrix.

This manuscript is organized as follows. Section~\ref{sec:Oscillations} introduces the formalisms we adopt when computing neutrino oscillations, including the theoretical assumptions one can adopt when performing an analysis of non-unitarity, and how these assumptions impact results. In Sections~\ref{sec:CurrentExps} and~\ref{sec:FutureExps}, we explain the current and future datasets included in our analyses, respectively. In Section~\ref{sec:Results}, we present the primary results of our analyses in a number of ways, resulting in our constraints on the unitarity conditions $U U^\dagger = U^\dagger U = \mathbb{I}$ in Section~\ref{subsec:NormClos}. We consider some alternate assumptions that impact the results, and present the results in light of these alternate assumptions, in Section~\ref{sec:AltResults}. Finally, in Section~\ref{sec:Conclusions} we provide discussion on our results and conclude.

We also wish to highlight the results that are included in our appendices. In Appendix~\ref{app:RareDecays}, we discuss how non-oscillation probes, such as rare charged-lepton decays, can be used in certain scenarios to constrain the unitarity of the LMM. Appendix~\ref{app:Derivations} derives neutrino oscillation probabilities (both for appearance and disappearance/survival) in vacuum when unitarity is not assumed. In Appendix~\ref{app:Bayesian}, we discuss the Bayesian approach used in many of our analyses, and the priors that enter this type of analysis. Appendix~\ref{app:Phases} includes the measurement of the phases present in the LMM, a parameterization-dependent measurement. Lastly, Appendix~\ref{app:LSNDMB} offers some discussion regarding the LSND and MiniBooNE anomalies, whether they can be resolved in this framework, and how they may be tested in next-generation experiments.

 
\section{Neutrino Oscillations and the Leptonic Mixing Matrix}
\label{sec:Oscillations}

In this section, we summarize the phenomenon of neutrino oscillations, and how the structure of the LMM enters the calculations for oscillation probabilities. We introduce the formalism we use throughout our analyses, which allows for the possibility that the LMM is not unitary. Given that we allow this possibility, we discuss the possible origins of the unitarity violation and different theoretical assumptions that map on to these different origins. These different theoretical assumptions will affect our analyses, and so we will spend considerable time discussing their effects.


\subsection{Unitarity of the Leptonic Mixing Matrix}
\label{sec:LMM}

Neutrino oscillation studies are generally carried out assuming a unitary $3\times3$ mixing matrix for rotating between eigenstates of flavor and mass. However, this assumption only strictly holds in a rather limited number of models for neutrino masses, some of which suffer from fine-tuning issues.

In many models for neutrino masses, while there is non-unitarity of the lepton mixing matrix, it is expected to be small. For example, in a generic type-I see-saw scenario, the non-unitarity of the light-neutrino mixing comes from the mixing (angle squared) between the light-heavy states, which is proportional to the mass ratio between the light and heavy states (see Appendix~\ref{app:RareDecays} for details). We expect the deviation from unitarity to be at most $\mathcal{O}(10^{-13})$ even for an $\mathcal{O}(\text{TeV})$ seesaw scale.

There are nevertheless abundant examples of neutrino mass models that can lead to large non-unitarity.  It has been shown by various groups that for non-trivial neutrino Yukawa textures, the see-saw mechanism \textit{can} lead to substantial deviations from unitarity (see e.g.~\cite{Buchmuller:1991tu,Ingelman:1993ve,Chang:1994hz,Loinaz:2003gc,Branco:2019avf}). In addition, mass models invoking symmetry arguments may also produce large unitarity violation~\cite{Wyler:1982dd,Hewett:1988xc,Langacker:1988up,Nardi:1993ag,Tommasini:1995ii}. It is therefore important to test the unitarity of the $3\times3$ lepton mixing matrix with experimental data.

To test unitarity with oscillation data, one could adopt mathematical assumptions on the $3\times3$ matrix corresponding to different theoretical assumptions on the origin of the unitarity violation. To clearly describe our choices, let us first look at how a neutrino state is defined. A flavor eigenstate neutrino field $\nu_\alpha(x)$ can be written as a linear combination of mass eigenstate fields $\nu_k(x)$:
\begin{equation}
\nu_\alpha(x) = \sum_k \mathcal{U}_{\alpha k} \nu_k(x)\ .
\end{equation}
The flavor index $\alpha \in e,~ \mu,~ \tau,~\ldots ~m$ includes the usual SM flavor fields, along with $m-3$ possible additional right-handed (sterile) fields, while the mass index $k \in 1,~2,~3,~\ldots ~n$, allowing for $n-3$ additional mass eigenstate fields. $\mathcal{U}_{\alpha k}$ is thus an $m\times n$ matrix. In an oscillation experiment, neutrinos are produced and detected via weak interactions. The produced state, neglecting the effect of neutrino masses, can be defined as~\cite{Giunti:2004zf,Antusch:2006vwa}\footnote{Reference~\cite{Giunti:2003qt} provides a thorough explanation for why $\mathcal{U}_{\alpha k}^*$ is the object appearing in Eq.~\eqref{eq:state_def}, contrary to the definition of the flavor eigenstate field above. The assumption of zero neutrino masses is reasonable, as the neutrinos are produced ultra-relativistically. See Ref.~\cite{Cohen:2008qb} for a discussion of how this assumption can break down.}
\begin{equation}
\ket{\nu_\alpha} = \frac{1}{\sqrt{\left(\mathcal{U}\mathcal{U}^\dagger\right)_{\alpha\alpha}}}\sum_k \mathcal{U}_{\alpha k}^*\ket{\nu_k}\ ,
\label{eq:state_def}
\end{equation}
where the sum over $k$ in the normalization factor is implicit. Only active flavors participate in weak interactions, so the flavor index $\alpha \in e,~ \mu,~ \tau$. However, it is important to note that the mass eigenstate index $k$ does not run to $n$, the total number of mass eigenstates. Instead, the sum over $k$ is only performed over the total number of kinematically accessible mass eigenstates~\cite{Goswami:2008mi,Li:2015oal,Blennow:2016jkn}.  For pion-decay sources which are used for most experiments, a conservative cutoff is to include all mass states below 140~MeV.\footnote{Heavier sterile neutrinos may be produced in such sources either off-shell or from heavier meson decays. However, subtle non-standard oscillation effects from these neutrinos are not observable because of their reduced fluxes. We therefore ignore their contribution.}  For models where additional sterile neutrinos are heavier than the electroweak scale, this sum truncates at $k = 3$. The normalization factor in Eq.~\eqref{eq:state_def} guarantees that the flavor states are always properly normalized: $\langle\nu_\alpha|\nu_\alpha\rangle = 1$. Note that if $\mathcal{U}$ is not unitary, the flavor states are not necessarily orthogonal: $\langle\nu_\alpha|\nu_\beta\rangle \neq \delta_{\alpha\beta}$.  

In this work, we focus on understanding the structure of the $3\times 3$ (sub-)matrix of the full $m\times n$ mixing matrix $\mathcal{U}$. We will refer to the $3\times 3$ mixing matrix as the LMM or $U_\mathrm{LMM}$, or simply $U_{\alpha k}$, where now $\alpha = e,~ \mu,~ \tau$ and $k = 1,~2,~3$.  We will parameterize and derive oscillation formulae using only $U$. An implicit assumption we adopt at this step is that there exist no additional sterile states with masses below 140 MeV. In this case, oscillation measurements provide a direct test of unitarity (see Appendix~\ref{app:RareDecays} for discussions on the model-dependence of charged lepton decay searches). For sterile neutrinos with masses between 0.1--10~eV, dedicated searches for spectral distortions in oscillation experiments are very sensitive~\cite{Dentler:2018sju,Diaz:2019fwt,Boser:2019rta}. For heavier steriles that are still kinematically accessible, see Refs.~\cite{Fong:2016yyh,Fong:2017gke} for detailed discussions on their oscillation signatures. Sterile neutrinos in other mass ranges can be probed e.g. via beta decay, meson decay and neutrino-less double beta decay. See Refs.~\cite{deGouvea:2015euy,Bolton:2019pcu} for comprehensive discussions of these experimental searches.

We organize our discussion of oscillations around the following three cases:
\begin{enumerate}
	\item The ``standard'' case, where $m=n=3$, $U = \mathcal{U}$.
	\item The ``sub-matrix'' case, where $m=n>3$. $\mathcal{U}$ is unitary, and $U$ is not.
	\item The ``agnostic'' case, $m\geqslant 3$ and/or $n\geqslant 3$. $\mathcal{U}_{\alpha k}$ is not assumed to be unitary, and $U$ is not.
\end{enumerate}
The standard case is the most commonly adopted in oscillation studies.  It is worth pointing out that even when this is phenomenological applicable, it nevertheless involves fields beyond the SM, highlighting the need for BSM physics to fully understand the neutrino sector. The sub-matrix scenario is the most commonly adopted in unitarity-violation studies. It applies to all cases when the unitarity violation is induced by the existence of new particles. The agnostic case is a peculiar one, where the full $m\times n$ mixing matrix is not assumed to be unitary. This is of course a difficult case to realize, as unitarity is one of the fundamental principles upon which theories are typically built. However, it is useful to consider this possibility so as to verify whether experimental data support the theoretical bias that $\mathcal{U}_{\alpha k}$ should be unitary. Additionally, Refs.~\cite{Meloni:2009cg,Blennow:2016jkn} explored scenarios in which non-standard neutrino interactions during neutrino propagation through matter may be mapped on to the effects of a non-unitary mixing matrix. While this is a specific scenario, it is one in which the agnostic case applies, and provides motivation for adopting this case to allow for generality in the form of $\mathcal{U}_{\alpha k}$.

For the rest of this paper, we adopt the agnostic assumption as our default scenario, aimed to be the most conservative with our bounds on non-unitarity. We distinguish the sub-matrix and agnostic cases because the sub-matrix assumption imposes additional criteria on the structure of $U$ and hence leads to more stringent bounds. We discuss what these criteria are and how they improve certain bounds throughout our analysis.


\subsection{Mixing Matrix Parameterizations}
\label{sec:Parameterizations}

In the standard scenario, $U_\mathrm{LMM}$ is a $3\times 3$ unitary matrix. It is well-known that in order to parameterize such a matrix, three angles and three complex phases are required. Two of the (Majorana) phases are irrelevant for neutrino oscillations, and are unphysical if neutrinos are Dirac particles. The standard parameterization employs three mixing angles, $\theta_{12}$, $\theta_{13}$, and $\theta_{23}$, and one complex phase $\delta_{\text{CP}}$. Often referred to as the PMNS~\cite{Pontecorvo:1967fh,Maki:1962mu} or PDG~\cite{Tanabashi:2018oca} parameterization, this form of the LMM is
\begin{equation}
U_\mathrm{LMM} = U_\mathrm{PMNS} \equiv \left(\begin{array}{c c c} c_{12} c_{13} & s_{12} c_{13} & s_{13} e^{-i\delta_{\text{CP}}} \\ -s_{12} c_{23} - c_{12} s_{13} s_{23} e^{i\delta_{\text{CP}}} & c_{12} c_{23} - s_{12} s_{13} s_{23} e^{i\delta_{\text{CP}}} & s_{23} c_{13} \\ s_{12} s_{23} - c_{12} c_{23} s_{13} e^{i\delta_{\text{CP}}} & -c_{12} s_{23} - s_{12} c_{23} s_{13} e^{i\delta_{\text{CP}}} & c_{13} c_{23} \end{array}\right), 
\end{equation}
where $s_{ij}\equiv \sin\theta_{ij}$ and $c_{ij}\equiv \cos\theta_{ij}$. The mixing angles are often referred to by the regime of neutrino oscillations in which they have been studied in the most detail: solar ($\theta_{12}$), reactor ($\theta_{13}$), and atmospheric ($\theta_{23}$). A number of global fit efforts in the three-flavor hypothesis have been performed, leading to relatively precise understanding of the mixing angles under this hypothesis~\cite{Esteban:2018azc,deSalas:2017kay,Capozzi:2020qhw,Esteban:2020cvm}.

More generally, a complex $3\times 3$ matrix $U$ can be described by eighteen real parameters. There are \textit{\textbf{9}} conditions for relating a generic complex matrix for leptonic mixing to a unitary one. These conditions can be obtained from the requirement that a unitary matrix satisfies $U^\dagger U = \mathbb{I}$. This is equivalent to requiring that all columns of the matrix are normalized to one:
\begin{equation}
N_k \equiv \left \lvert U_{e k}\right\rvert^2 + \left \lvert U_{\mu k}\right\rvert^2 + \left \lvert U_{\tau k}\right\rvert^2 = 1\quad (k = 1,~2,~3), 
\label{eq:N_k}
\end{equation}
as well as requiring that the column unitarity triangles close:
\begin{equation}
t_{kl} \equiv U_{e k}^* U_{e l} + U_{\mu k}^* U_{\mu l} + U_{\tau k}^* U_{\tau l} = 0\quad (k \neq l; \quad k,~l = 1,~2,~3).
\label{eq:t_kl}
\end{equation}
Note that these are nine real constraints as $t_{kl}$ can be complex. Because $U^\dagger U$ is Hermitian, the unitarity condition can equivalently be written as $U U^\dagger = \mathbb{I}$, which can be translated to row normalization conditions:
\begin{equation}
N_\alpha \equiv \left \lvert U_{\alpha 1}\right\rvert^2 + \left \lvert U_{\alpha 2}\right\rvert^2 + \left \lvert U_{\alpha 3}\right\rvert^2 = 1\quad (\alpha = e,~\mu,~\tau),
\label{eq:N_a}
\end{equation}
and the closure of row unitarity triangles:
\begin{equation}
t_{\alpha\beta} \equiv U_{\alpha 1}^* U_{\beta 1} + U_{\alpha 2}^* U_{\beta 2} + U_{\alpha 3}^* U_{\beta 3} = 0\quad (\alpha \neq \beta; \quad\alpha,~\beta = e,~\mu,~\tau).
\label{eq:t_ab}
\end{equation}

For the general case where $U$ is a non-unitary $3\times 3$ matrix, the number of real parameters needed to describe the matrix for neutrino oscillation is $18-3-2=13$, where 3 phases can be absorbed by charged lepton fields and 2 Majorana phases do not participate in oscillations. Equivalently, one can see that 13 parameters are required, as a unitary LMM would have 4 parameters, and the extension to include potential non-unitarity involves relaxing 9 unitarity conditions. The magnitudes of the elements of the mixing matrix are parameterization-independent, therefore we choose to adopt the following parameterization:
\begin{equation}
U_\mathrm{LMM} \equiv \left(\begin{array}{l l l} \left\lvert U_{e1}\right\rvert & \left\lvert U_{e2}\right\rvert e^{i\phi_{e2}} & \left\lvert U_{e3}\right\rvert e^{i\phi_{e3}} \\
\left\lvert U_{\mu 1}\right\rvert & \left\lvert U_{\mu 2}\right\rvert & \left\lvert U_{\mu 3}\right\rvert \\
\left\lvert U_{\tau 1}\right\rvert & \left\lvert U_{\tau 2}\right\rvert e^{i\phi_{\tau 2}} & \left\lvert U_{\tau 3}\right\rvert e^{i\phi_{\tau 3}} \end{array}\right).
\label{eq:LMM}
\end{equation}
Here, we have nine magnitudes and four CP-violating phases.\footnote{The four phases can be assigned to any $2\times 2$ sub-matrix.} Going forward, we refer to the parameterization given in Eq.~\eqref{eq:LMM} as the Magnitudes \& Phases (MP) parameterization. Note that in this case, the row and column normalizations can be larger than 1, and the 13 parameters are completely independent of each other. This parameterization applies straightforwardly to the agnostic case described above.

When $U_\mathrm{LMM}$ is unitary, we can relate the MP parameterization in Eq.~\eqref{eq:LMM} to the PMNS parameterization, with a straightforwardly obtainable correspondence between parameters. The phases can be related to the PMNS parameterization by using Jarlskog factors $J_{\alpha i}$, which are defined as
\begin{equation}
\varepsilon_{\alpha\beta\gamma} \varepsilon_{ijk} J_{\alpha i} \equiv \mathrm{Im}\left( U_{\beta j} U_{\gamma k} U_{\beta k}^* U_{\gamma j}^*\right),
\end{equation}
where the $\varepsilon$ are Levi-Civita tensors. It is straightforward to see that 
\begin{eqnarray}
\sin\phi_{e2} &=& -\frac{J_{e2}}{\left\lvert U_{e1}\right\rvert \left\lvert U_{e2}\right\rvert \left\lvert U_{\mu 1}\right\rvert \left\lvert U_{\mu 2}\right\rvert}, \quad
\sin\phi_{e3} = \frac{J_{e3}}{\left\lvert U_{e1}\right\rvert \left\lvert U_{e3}\right\rvert \left\lvert U_{\mu 1}\right\rvert \left\lvert U_{\mu 3}\right\rvert}, \\
\sin\phi_{\tau 2} &=& \frac{J_{\tau 2}}{\left\lvert U_{\tau 1}\right\rvert \left\lvert U_{\tau 2}\right\rvert \left\lvert U_{\mu 1}\right\rvert \left\lvert U_{\mu 2}\right\rvert}, \quad
\sin\phi_{\tau 3} = -\frac{J_{\tau 3}}{\left\lvert U_{\tau 1}\right\rvert \left\lvert U_{\tau3}\right\rvert \left\lvert U_{\mu 1}\right\rvert \left\lvert U_{\mu 3}\right\rvert}.
\label{eq:phases}
\end{eqnarray}
If $U_\mathrm{LMM}$ is unitary, all constructible Jarlskog factors must be equal to each other and equal to the PMNS matrix Jarlskog invariant~\cite{Jarlskog:1985ht}:~\footnote{This can be used to test the unitarity of the LMM.  For details, see Ref.~\cite{Ellis:2020ehi}.} 
\begin{equation}
J_{\rm PMNS} = c_{12} s_{12} c_{23} s_{23} c_{13}^2 s_{13} \sin\delta_{\rm CP}.
\end{equation}
Enforcing $J_{\tau 2} = J_{\tau 3} = J_{e2} = J_{e3} = J_{\rm PMNS}$ in Eq.~\eqref{eq:phases} allows for the simple derivation of a relation between the phases  $\phi_{e2}$, $\phi_{e3}$, $\phi_{\tau2}$, and $\phi_{\tau3}$ and PMNS parameters when  $U_{\rm LMM} = U_{\rm PMNS}$.

Finally, we briefly discuss the sub-matrix case, where $U$ is a $3\times 3$ sub-matrix of a larger unitary matrix $\mathcal{U}$. This introduces two additional constraints on the structure of $U$: the row and column normalizations of LMM must not exceed unity:
\begin{equation}
N_\alpha\leqslant 1 , \quad N_k \leqslant 1 .
\label{eq:norm_one}
\end{equation}
Further, by applying the Cauchy-Schwarz inequality on the vectors $\big\{\mathcal{U}_{\alpha k}\big\}$, where $\alpha$ or $k$ runs from 4, 5... $n$, we obtain the following inequalities~\cite{Parke:2015goa}~\footnote{One can directly apply Cauchy-Schwarz inequality on the matrix $U$, which leads to the weaker conditions
\begin{equation}
\left|t_{\alpha\beta}\right|^2  \leqslant N_\alpha N_\beta, \quad 
\left|t_{kl}\right|^2  \leqslant N_k N_j.
\end{equation}
These inequalities hold for both the sterile and agnostic cases. However, we know that LMM is at least very close to unitary, so these conditions are met for all viable parameter space.}:
\begin{eqnarray}
\left|t_{\alpha\beta}\right|^2  &\leqslant& \left(1-N_\alpha\right)\left(1-N_\beta\right) \nonumber \\ 
\left|t_{kl}\right|^2  &\leqslant& \left(1-N_k\right)\left(1-N_l\right) .
\label{eq:Cauchy-Schwartz}
\end{eqnarray}
In this work, the bulk of our results will be presented under the minimal set of theoretical assumptions, corresponding to the agnostic case. Where we discuss sub-matrix case results, the conditions of Eqs.~\eqref{eq:norm_one} and~\eqref{eq:Cauchy-Schwartz} are imposed on $U_{\rm LMM}$, and the comparison with the agnostic case will be analyzed. Such comparisons will appear throughout our analysis, as well as in Section~\ref{sec:AltResults}.

The most commonly adopted parameterization in the sub-matrix case~\cite{Xing:2007zj, Escrihuela:2015wra,Blennow:2016jkn,C:2017scx} is the following: 
\begin{equation}
U_\mathrm{LMM} \equiv N U_\mathrm{PMNS} = \left(\begin{array}{c c c} \alpha_{11} & 0 & 0 \\ \alpha_{21} & \alpha_{22} & 0 \\ \alpha_{31} & \alpha_{32} & \alpha_{33}\end{array}\right) U_\mathrm{PMNS}.
\end{equation}
When there are three active neutrinos and any number of sterile neutrinos, one can express $\alpha_{kl}$ in terms of mixing angles and phases between active and sterile neutrino mixing.  For the full expressions, see Ref~\cite{Escrihuela:2015wra}. Here, unitarity is achieved in the limit that $\alpha_{kl} \to \delta_{kl}$. The off-diagonal $\alpha_{kl}$ may be complex, so there are nine free parameters corresponding to the nine constraints discussed above. We note here that this parameterization is useful in that unitarity is obtained in a relatively simple limit, i.e., $\alpha_{kk} \to 1$ and $\alpha_{k\neq l} \to 0$, compared to the MP parameterization. However, it is not straightforward to map between the ``$\alpha$'' parameterization and the individual, specific constraints of unitarity -- the normalizations and closures of columns and rows of $U_{\rm LMM}$. As constructed, the map between the $\alpha_{kl}$ and the normalization of rows and closures between two different rows is relatively simple:
\begin{eqnarray}
N_{e} &=& \alpha_{11}^2, \\
N_{\mu} &=& \alpha_{22}^2 + \left\lvert \alpha_{21}\right\rvert^2, \\
N_\tau &=& \alpha_{33}^2 + \left\lvert \alpha_{32}\right\rvert^2 + \left \lvert \alpha_{31}\right\rvert^2, \\
t_{e\mu} &=& \alpha_{11}\alpha_{21}\ , \\
t_{e\tau} &=& \alpha_{11}\alpha_{32}\ , \\
t_{\mu\tau} &=& \alpha_{21}^*\alpha_{31}+\alpha_{22}\alpha_{32}\ .
\end{eqnarray}
On the other hand, for the normalizations of the columns $N_{k}$ and the closures of the triangles between different columns $t_{kl}$, such a mapping depends on both the $\alpha_{kl}$ as well as the mixing angles and $\delta_{\rm CP}$ from the PMNS parameterization. 

Since our goal in this work is to determine the current and future constraints on $N_{\alpha}$, $N_{k}$, $t_{\alpha\beta}$, and $t_{kl}$, we use the MP parameterization which has a straightforward map between the input parameters and these quantities. Using the MP paramaterization has a second advantage, in that the majority of the underlying inputs are parameterization-independent. Specifically, all of the magnitudes-squared $\left\lvert U_{\alpha k}\right\rvert^2$ are independent of the adopted parameterization, which makes translation between different experimental results in this context simpler.


\subsection{Oscillation Probabilities}
\label{subsec:OscProbs}

We now review the oscillation formulae for different parameterizations. See Refs.~\cite{Antusch:2006vwa,Li:2015oal} for further discussions. In vacuum, the mass states $|\nu_k\rangle$ are eigenstates of the Hamiltonian, such that they form an orthogonal basis, $\langle\nu_k|\nu_l\rangle = \delta_{kl}$, and evolve in time as 
\begin{equation}
|\nu_k(t)\rangle = e^{-iE_kt}|\nu_k\rangle.
\end{equation}
A flavor state is created as Eq.~\eqref{eq:state_def}. After traveling for time $t$, the flavor state is evolved to
\begin{equation}\label{eq:state_evolve}
|\nu_\alpha(t)\rangle = \frac{1}{\sqrt{N_\alpha}}\sum_{k=1}^3 U_{\alpha k}^* e^{-iE_kt}|\nu_k\rangle.
\end{equation}
The oscillation probability for a neutrino of energy $E$ produced as $\nu_\alpha$ and detected as $\nu_\beta$ after propagating $L = c t$ is therefore
\begin{eqnarray}
P(\nu_\alpha\rightarrow\nu_\beta) &=& |\langle\nu_\beta|\nu_\alpha(t)\rangle|^2 = \frac{1}{N_\alpha N_\beta} \left|\displaystyle\sum_{k=1}^3U_{\alpha k}^*U_{\beta k}e^{-iE_kt}\right|^2.
\label{eq:ProbAB}
\end{eqnarray}
Here, $E_k = m_k^2/2E_\nu$. In order to better understand the behavior of the oscillation probability, we separate the discussion here into two cases, $\alpha = \beta$ (disappearance/survival probability) and $\alpha \neq \beta$ (appearance probability). Defining $\Delta_{ij} \equiv \Delta m_{ij}^2 L/2E_\nu$, the disappearance/survival probability may be written as
\begin{eqnarray}
P(\nu_\alpha\rightarrow\nu_\alpha) &=& 1 - \frac{4\left\lvert U_{\alpha 2}\right\rvert^2 \left( \left\lvert U_{\alpha 1}\right\rvert^2 + \left\lvert U_{\alpha 3}\right\rvert^2 \right)}{N_\alpha^2} \sin^2\left(\frac{\Delta_{21}}{2}\right) - \frac{4\left\lvert U_{\alpha 3}\right\rvert^2 \left( \left\lvert U_{\alpha 1}\right\rvert^2 + \left\lvert U_{\alpha 2}\right\rvert^2 \right)}{N_\alpha^2}\sin^2\left(\frac{\Delta_{31}}{2}\right) \nonumber \\
&+& \frac{8 \left\lvert U_{\alpha 2}\right\rvert^2 \left\lvert U_{\alpha 3}\right\rvert^2}{N_\alpha^2} \sin\left(\frac{\Delta_{21}}{2}\right) \sin\left(\frac{\Delta_{31}}{2}\right) \cos\left(\frac{\Delta_{32}}{2}\right ). 
\label{eq:Paa}
\end{eqnarray}
For the appearance probability, we define $\varphi_{k}^{\alpha\beta} \equiv \mathrm{arg}\left( U_{\alpha k}^* U_{\beta k}\right)$ and $\varphi^{\alpha\beta} \equiv \mathrm{arg}\left(t_{\alpha \beta}\right)$. The appearance probability may be written as
\begin{eqnarray}
P(\nu_\alpha \rightarrow \nu_\beta) &=& \frac{\left\vert t_{\alpha\beta}\right\rvert^2}{N_\alpha N_\beta} + \frac{4 \left\lvert U_{\alpha 2}\right\rvert^2 \left\lvert U_{\beta 2}\right\rvert^2}{N_\alpha N_\beta} \sin^2\left(\frac{\Delta_{21}}{2}\right) + \frac{4 \left\lvert U_{\alpha 3}\right\rvert^2 \left\lvert U_{\beta 3}\right\rvert^2}{N_\alpha N_\beta} \sin^2\left(\frac{\Delta_{31}}{2}\right) \nonumber \\
&+& \frac{8\left\lvert U_{\alpha 2}\right\rvert \left\lvert U_{\beta 2}\right\rvert \left \lvert U_{\alpha 3}\right\rvert \left \lvert U_{\beta 3}\right \rvert}{N_\alpha N_\beta} \sin\left(\frac{\Delta_{21}}{2}\right) \sin\left(\frac{\Delta_{31}}{2}\right) \cos\left(\frac{\Delta_{32}}{2} + \varphi_{2}^{\alpha\beta} - \varphi_{3}^{\alpha\beta} \right) \nonumber \\
&+& \frac{4\left\lvert t_{\alpha \beta}\right\rvert}{N_\alpha N_\beta} \left[ \left\lvert U_{\alpha 2}\right\rvert \left\lvert U_{\beta2}\right\rvert \sin\left( \frac{\Delta_{21}}{2}\right) \sin\left(\frac{\Delta_{21}}{2} + \varphi^{\alpha\beta} - \varphi^{\alpha\beta}_{2}\right) \right. \nonumber \\
&+& \left. \left\lvert U_{\alpha 3}\right\rvert \left\lvert U_{\beta3}\right\rvert \sin\left( \frac{\Delta_{31}}{2}\right) \sin\left(\frac{\Delta_{31}}{2} + \varphi^{\alpha\beta} - \varphi^{\alpha\beta}_{3}\right) \right]  .
\label{eq:Pab}
\end{eqnarray}
When $U$ is unitary, Eqs.~\eqref{eq:Paa} and~\eqref{eq:Pab} reduce to the standard oscillation formulae. In most neutrino experiments, one or two mass terms dominate the oscillation behavior due to the large hierarchy between $\Delta m_{21}^2$ and $\Delta m_{32}^2$, such that Eqs.~\eqref{eq:Paa} and~\eqref{eq:Pab} can be simplified. These oscillation formulae are derived in Appendix~\ref{app:Derivations}. We also provide the simplified expressions when either $\Delta m_{21}^2$ or $\Delta m_{31}^2$ is the dominant term of interest in Appendix~\ref{app:Derivations} and comment on which experiments belong in each of these regimes.

Equation~\eqref{eq:Pab} reveals an interesting consequence of a non-unitary mixing matrix, namely the zero-distance effect~\cite{Langacker:1988up,Antusch:2006vwa}. When $L \to 0$ (or in an experiment where $\Delta_{21}$ and $\Delta_{31} \ll 1$), the disappearance probability $P(\nu_\alpha \to \nu_\alpha) \to 1$, but the appearance probability becomes
\begin{equation}
P(\nu_\alpha\rightarrow\nu_\beta, L\sim 0) \simeq \frac{1}{N_\alpha N_\beta}\left|\sum_{k=1}^3U_{\alpha k}^* U_{\beta k}\right|^2 = \frac{\left\lvert t_{\alpha\beta}\right\rvert^2}{N_\alpha N_\beta}
\end{equation}
This implies that searches for short-baseline anomalous appearance from one flavor eigenstate $\alpha$ to another $\beta$ provides a direct constraint on the closure between the $\alpha$ and $\beta$ rows. We discuss how these searches, typically interpreted in the context of searches for light, coherently-oscillating sterile neutrinos, may be applied to our scenario in Section~\ref{subsec:SterileSearch}.

\textbf{Matter effects in the context of non-unitarity:} Even though it is a good approximation to use the vacuum oscillation probabilities in Eqs.~\eqref{eq:Paa} and~\eqref{eq:Pab} for many oscillation experiments of interest (as well as providing useful analytic interpretation of results), matter effects are important for several existing experiments, and crucial for the future DUNE experiment. Interactions of neutrinos with matter as they traverse can be included by adding a potential to the Hamiltonian that governs the time-evolution of the neutrino states, which is diagonal in the flavor basis:
\begin{equation}
V_{\alpha\beta} = \sqrt{2}G_\mathrm{F}\left(\begin{array}{c c c} n_e-n_n/2 & 0 & 0 \\ 0 & -n_n/2 & 0 \\ 0 & 0 & -n_n/2\end{array}\right), 
\end{equation}
where $n_e$ and $n_n$ are the electron and neutron density in the medium, respectively. This potential is rotated by $U_{\rm LMM}$ into the mass basis, and combined with the (mass-basis-diagonal) energy values $\Delta_{k1}/(2E_\nu)$. Typically, the neutron density is removed as its contribution to the total Hamiltonian is proportional to the identity matrix, and represents a phase common to the propagation of all three neutrino states. However, since $U^\dagger U = \mathbb{I}$ is not assumed in our analysis, $n_n$ must be included in our calculation, as the phase is no longer common to all three propagating states. We note here that the inclusion of matter effects in light of non-unitary mixing depends strongly on the assumptions regarding any new neutrino states (for instance, whether they interact with matter via the standard weak interactions or any other new interactions, and what their masses are). The form of $V_{\alpha\beta}$ above is that obtained in the minimal-unitarity-violation context, in which the new physics scale is assumed to be much higher than the electroweak scale~\cite{Antusch:2006vwa}, and we adopt it for the remainder of our work.


\subsection{Normalization Effects}
\label{subsec:normalization}

In this subsection we discuss further non-trivial consequences of a non-unitary LMM. Much of this discussion is adapted from Ref.~\cite{Antusch:2006vwa}. Without the assumption of the LMM being unitary, the expected flux of a given neutrino flavor (e.g. produced by pion decay-in-flight) will be modified from the unitary expectation. The same will be true of charged-current (CC) scattering cross sections, where non-unitarity of the LMM leads to deviations in the rate of charged leptons produced from interactions involving neutrinos of the corresponding flavor. The flux of neutrinos of flavor $\alpha$, and the corresponding CC cross section, may be expressed (relative to their unitary, SM expectations) as
\begin{equation}
\Phi_{\alpha}(E) = N_\alpha \Phi^{\rm SM}_{\alpha}(E),\quad\quad \sigma_{\beta}(E) =  N_\beta \sigma_{\beta}^{\rm SM}(E).
\end{equation}
The neutral-current cross section is also modified,
\begin{align}
\sigma_k^{\rm NC}(E) &= \sigma_{\rm SM}^{\rm NC}(E) \sum_l \Big\vert (U^\dagger U)_{kl}\Big\vert^2 \ .
\end{align}

Neutrino oscillation experiments infer oscillation probabilities by measuring event rates and spectra, which are a convolution of fluxes, cross sections, and efficiencies of detection. The number of detected neutrinos of flavor $\beta$, $n_\beta$, is given by 
\begin{equation}
n_\beta \propto \Phi_\alpha P(\nu_\alpha\rightarrow\nu_\beta) \sigma_\beta = N_\alpha \Phi^{\rm SM}_{\alpha} P(\nu_\alpha\rightarrow\nu_\beta) N_\beta \sigma_{\beta}^{\rm SM} = \Phi^{\rm SM}_{\alpha} \hat{P}(\nu_\alpha\rightarrow\nu_\beta) \sigma_{\beta}^{\rm SM} \ .
\label{eq:norm_ab}
\end{equation}
Thus, if an experimental analysis is performed assuming SM predictions of fluxes and cross-sections as truth, the measurement of $n_\beta$ corresponds to an inferred measurement of $\hat{P}(\nu_\alpha\rightarrow\nu_\beta)$. Recalling the vacuum oscillation formula of Eq.~\eqref{eq:Pab}, this leads to the conclusion that experiments making such assumptions are inferring the oscillation probability given by
\begin{equation}
\hat{P}(\nu_\alpha\rightarrow\nu_\beta) = \left|\sum_{k=1}^3 U_{\alpha k}^* U_{\beta k} e^{-iE_k t}\right|^2 \ ,
\end{equation}
such that these measurements are not sensitive to the normalization factors $N_\alpha$ and $N_\beta$.

Experiments often use a near detector to measure the neutrino flux. This is the case in, e.g., DUNE. Normalization effects will therefore manifest themselves differently for appearance and disappearance probabilities, as well as for near-detector-only measurements compared with near-to-far ratio measurements. These effects are both taken into account in our analysis. 

Let us first consider disappearance measurements at near detectors, i.e., sterile neutrino searches. These experiments measure an energy spectrum of events $n_\alpha^{\rm ND}(E)$, where $\alpha$ is the flavor label:
\begin{equation}
n_\alpha^{\rm ND}(E) \propto \Phi_{\alpha}(E) \sigma_\alpha (E) P_{\alpha\alpha}(E;L=0) \ ,
\end{equation}
At $L = 0$, the oscillation probability is $P_{\alpha \alpha} (E; L= 0) = 1$. Experiments that measure disappearance spectra at near detectors to constrain an oscillation probability rely on understanding of the (SM) predictions of $\Phi_{\alpha}$ and $\sigma_{\alpha}$, so the measured spectrum can be expressed as
\begin{equation}
\label{eq:ND}
n_{\alpha}^{\rm ND}(E) \propto \Phi_{\alpha}^{\rm SM}(E) \sigma_\alpha^{\rm SM}(E) N_\alpha^2.
\end{equation}
Therefore, with a sufficiently precise measurement of the spectrum, as well as understanding of the SM-expected flux and cross section, a constraint on $N_{\alpha}^2$ can be placed. These results are usually reported in the context of a limit on disappearance probabilities, i.e. $P_{\alpha\alpha}(E; L=0)$ is close to $1$ with some degree of confidence. In Section~\ref{subsec:SterileSearch} we discuss how these reported limits map onto constraints on $N_\alpha$. When considering searches for anomalous short-baseline appearance of a flavor $\alpha$ to a flavor $\beta$, Eq.~\eqref{eq:ND} must be modified accordingly, and we find that these searches are sensitive to 
\begin{equation}
n_{\beta}^{\rm ND}(E) \propto \left\lvert t_{\alpha\beta}\right\rvert^2 \ ,
\end{equation}
thereby providing constraints on the closure of row unitarity triangles.

Let us now consider near-to-far-ratio measurements. These experiments infer a far-detector oscillation probability $P_{\alpha\beta} \equiv P(\nu_\alpha \to \nu_\beta; E, L)$ by measuring a near detector spectrum $n_{\alpha}^{\rm ND}$ as above, as well as a far detector spectrum $n_\beta^{\rm FD}$,
\begin{equation}
n_{\beta}^{\rm FD}(E) \propto \Phi_{\alpha}(E) \sigma_\beta (E) P_{\alpha\beta}.
\end{equation}
We may express the measured ratio of oscillation probabilities as
\begin{equation}
P_{\alpha \beta} = \frac{n_{\beta}^{\rm FD}(E)}{n_{\alpha}^{\rm ND}(E)} \frac{\sigma_\alpha(E)}{\sigma_\beta(E)} = \frac{n_{\beta}^{\rm FD}(E)}{n_{\alpha}^{\rm ND}(E)} \frac{\sigma^{\rm SM}_\alpha(E)}{\sigma^{\rm SM}_\beta(E)} \frac{N_\alpha}{N_\beta} = \hat{P}_{\alpha\beta} \frac{N_\alpha}{N_\beta}.
\end{equation}
For disappearance measurements where $\alpha=\beta$, the measured probabilities are the true probabilities, e.g., Eq.~\eqref{eq:Pab} for the vacuum case. For appearance measurements where $\alpha\neq\beta$, the measured probabilities $\hat{P}_{\alpha\beta}$ are the true probabilities $P_{\alpha\beta}$ multiplied by a factor $N_\beta/N_\alpha$. 


\section{Current Experiments and Statistical Treatment}
\label{sec:CurrentExps}

Before turning to upcoming experiments, we first consider the most powerful set of existing experimental results that are sensitive to quantities of interest in the LMM. Our analysis is performed only over the experimental data which, when combined, provides the dominant sensitivity to the corresponding parameters in the mixing matrix. As such, it does not constitute a complete set of experimental results, but nevertheless captures our current knowledge of neutrino oscillation parameters. This section serves to describe the inputs to our fit, as well as the translation from experimental measurements in the PMNS parameterization to the MP parameterization we use for our analysis.

For most experiments, we take the reported event spectra and fit for LMM elements, i.e., mixing angles in the PMNS parameterization. Because our focus is on the LMM matrix elements and not the mass-squared splittings, we include the best measurement of the latter for each experiment as a Gaussian prior. We will specify throughout the values we take from each experiment. Included in this, we marginalize over the mass ordering (i.e., the sign of $\Delta m_{31}^2$). While there is a long-standing tension between solar and reactor experiment measurements of $\Delta m_{21}^2$, we find that allowing the mass-squared splittings to vary does not impact the results of measuring the elements of the mixing matrix.

Due to the complexity of its simulation (both the computational expense of simulating the oscillation probabilities and the many relevant systematic uncertainty parameters), we do not include the atmospheric neutrino results from Super-Kamiokande in our analysis~\cite{Abe:2017aap,Jiang:2019xwn}. While $\chi^2$ tables are provided by the collaboration for the results presented in Ref.~\cite{Abe:2017aap}, these assume three-neutrino mixing and a unitary leptonic mixing matrix. Since our goal is to keep the data analyzed consistent whether we are assuming unitarity or not, we do not include this in any of the following fits.

Table~\ref{tab:QuantitiesPMNSLMM} summarizes the experiments that enter our analysis. We show the dominant quantity to which each experiment is sensitive both when unitarity is assumed (the middle column, labeled ``PMNS Quantity''), and when unitarity is not assumed (the right column, labeled ``LMM Quantity'').

\begin{table}
\caption{Quantities to which each experiment is sensitive: using the PMNS parameterization when unitarity is assumed (center column), using the MP parametrization when unitarity is not assumed (right column). \label{tab:QuantitiesPMNSLMM}}
\centering
\begin{tabular}{||c||c|c||}\hline
\textbf{Experiment} & \textbf{PMNS Quantity} & \textbf{LMM Quantity} \\ \hline\hline
Solar Neutral Current & $1$ & $(\eone + \etwo)N_2^2 + \ethr N_3^2$ \\ \hline
Solar Charged Current & $\sin^2\theta_{12}\cos^4\theta_{13} + \sin^4\theta_{13}$ & $\etwo (\eone+\etwo) + \left\lvert U_{e3}\right\rvert^4$ \\ \hline
KamLAND & $\cos^4\theta_{13}\sin^2\left(2\theta_{12}\right)$ & $4\eone\etwo$ \\ \hline
Daya Bay & $\sin^2\left(2\theta_{13}\right)$ & $4\ethr(\eone+\etwo)/N_e^2$ \\ \hline
Sterile Neutrino $P_{\alpha\beta}$ ($\alpha\neq\beta$) & $0$ & $\left\lvert t_{\alpha\beta}\right\rvert^2$ \\ \hline
OPERA & $\cos^4\theta_{13} \sin^2\left(2\theta_{23}\right)$ & $4\mthr \tthr/N_\mu^2$ \\ \hline
Long-baseline $P_{\mu e}$ & \multirow{2}{*}{$\sin^2\theta_{23} \sin^2\left(2\theta_{13}\right)$} & \multirow{2}{*}{$4\ethr\mthr/N_\mu^2$} \\ (T2K, NOvA, DUNE, T2HK) & & \\ \hline
Long-baseline $P_{\mu\mu}$ & \multirow{2}{*}{$4\cos^2\theta_{13}\sin^2\theta_{23}\left(1 - \cos^2\theta_{13}\sin^2\theta_{23}\right)$} & \multirow{2}{*}{$4\mthr(\mone+\mtwo)/N_\mu^2$} \\ (T2K, NOvA, DUNE, T2HK)  & & \\ \hline
\end{tabular}
\end{table}


\subsection{Solar Neutrino Measurements}

The Sudbury Neutrino Experiment (SNO)~\cite{Aharmim:2011vm} measures solar neutrinos via neutral-current interactions. The oscillation probabilities for solar neutrinos can, to very good approximation, be calculated by considering an incoherent sum over neutrino mass eigenstates of their production in the sun and their scattering cross sections in a detector. Critically, this includes their journey from production to exiting the sun. The effective probability that a neutrino begins as an electron flavor eigenstate and exits as a mass eigenstate $\ket{\nu_k}$ is related to the effective matrix-element-squared in propagation in matter, which we call $\absq{\widetilde{U}_{ek}}$. Since all of the results we consider are in the regime where matter effects dominate, we focus on this region, where we can express these mixing angles as $\absq{\widetilde{U}_{ek}} = \left\lbrace 0, (\absq{U_{e1}} + \absq{U_{e2}})/N_e, \absq{U_{e3}}/N_e\right\rbrace$.

Following discussion similar to that of Sec.~\ref{subsec:normalization}, we can write the detected number of neutral current events as
\begin{align}
n_{\rm N} \propto \sum_k \Phi_k^{\rm exiting\ sun} \sigma_k^{\rm NC} &= \sum_k \left(\Phi_{\rm SM} N_e \absq{\widetilde{U}_{ek}}\right) \left(\sigma_{\rm SM}^{\rm NC} \left( \sum_l \Big\vert (U^\dagger U)_{kl}\Big\vert^2\right)\right), \\
&= \Phi_{\rm SM} \sigma_{\rm SM}^{\rm NC} \sum_k N_e \absq{\widetilde{U}_{ek}} \left(\sum_l \Big\vert (U^\dagger U)_{kl}\Big\vert^2\right), \\
&= \Phi_{\rm SM} \sigma_{\rm SM}^{\rm NC} \left[\left( \absq{U_{e1}} + \absq{U_{e2}}\right) \left(N_2^2 + \absq{t_{12}} + \absq{t_{23}}\right) + \absq{U_{e3}} \left(N_3^2 + \absq{t_{13}} + \absq{t_{23}}\right)\right], \\
&= \Phi_0 \sigma_{\rm SM}^{\rm NC} \hat{P}_{\rm NC},
\end{align}
where we define
\begin{align}
\hat{P}_\mathrm{NC} &\equiv \left( \absq{U_{e1}} + \absq{U_{e2}}\right) \left(N_2^2 + \absq{t_{12}} + \absq{t_{23}}\right) + \absq{U_{e3}} \left(N_3^2 + \absq{t_{13}} + \absq{t_{23}}\right), \\
&= \left(\absq{U_{e1}} + \absq{U_{e2}}\right)N_2^2 + \absq{U_{e3}} N_3^2 \nonumber \\
&+ \left(\absq{U_{e1}} + \absq{U_{e2}}\right)\left( \absq{t_{12}}+\absq{t_{23}}\right) + \absq{U_{e3}}\left(\absq{t_{13}}+\absq{t_{13}}\right).\label{eq:PNC}
\end{align}
The terms of the last line in Eq.~\eqref{eq:PNC} are small relative to those in the line above it, so we can express the measured oscillation probability at leading order as $\hat{P}_{\rm NC} \simeq (\absq{U_{e1}} + \absq{U_{e2}})N_2^2 + \absq{U_{e3}}N_3^2$. This measurement is limited by uncertainties associated with the $^8$B neutrino flux prediction from the standard solar model~\cite{Vinyoles:2016djt}. We conservatively assume that this probability is measured at the 25\% level, $\hat{P}_{\rm NC} = 1 \pm 0.25$.

In addition, SNO, and Super-Kamiokande (Super-K)~\cite{Abe:2016nxk} measure solar neutrinos via CC interactions. The oscillation probability is entirely dominated by matter effects. When assuming unitarity, the measured survival probability can be expressed as $P_{ee} = \sin^2\theta_{12}\cos^4\theta_{13}+\sin^4\theta_{13}$. We take the results from a preliminary joint SNO and Super-K analysis~\cite{SKNu2020}, which reports $\sin^2\theta_{12} = 0.306\pm 0.014$ when using a prior of $\sin^2\theta_{13}=0.0219\pm 0.0014$ in their analysis, and interpret it as a measurement of the survival probability $P_{ee} = 0.2932 \pm 0.0134$.\footnote{The results in Ref.~\cite{SKNu2020} result in slightly stronger constraints on $P_{ee}$ than those in the previously reported Ref.~\cite{Abe:2016nxk}. More interestingly, the results from Ref.~\cite{SKNu2020} prefer a larger value of $\Delta m_{21}^2$ ($6.11 \times 10^{-5}$ eV$^2$) than those of Ref.~\cite{Abe:2016nxk} ($4.8 \times 10^{-5}$ eV$^2$), more consistent with other observations from KamLAND.} If unitarity is not assumed, the survival probability is
\begin{equation}
P_{ee} = \etwo (\eone + \etwo) + \left\vert U_{e3}\right\rvert^4 .
\end{equation}
We also include the measured value of $\Delta m_{21}^2 = \left(6.11 \pm 1.21\right) \times 10^{-5}$ eV$^2$ from the joint analysis~\cite{SKNu2020}.


\subsection{KamLAND} 

The Kamioka Liquid Scintillator Antineutrino Detector (KamLAND) was an experiment that observed the oscillation of reactor electron antineutrinos $\overline{\nu}_e$ (with energies between $\simeq$2--10~MeV) at distances of roughly 180~km. This allows KamLAND to be sensitive to the solar mass-squared splitting, measuring $\Delta m_{21}^2 = \left( 7.50 \pm 0.20\right) \times 10^{-5}$ eV$^2$. We use a slightly older measurement from Ref.~\cite{Gando:2010aa} to be consistent with the corresponding measurement of the oscillation probability. A more recent analysis measures this mass-squared splitting slightly more precisely~\cite{Gando:2013nba}, but this does not affect our results\footnote{The more recent Ref.~\cite{Gando:2013nba} reports a slightly smaller preferred value of $\tan^2\theta_{12} = 0.436$ than Ref.~\cite{Gando:2010aa}'s $\tan^2\theta_{12} = 0.452$. While this shift is at the ${\sim}0.5\sigma$ level, the more powerful solar neutrino measurements are more important for the resulting fits than KamLAND.}.

Appendix B of Ref.~\cite{Gando:2010aa} gives weighted measurements and uncertainties on the oscillation probability $P(\overline{\nu}_e \to \overline{\nu}_e)$ for different values of $x \equiv \langle\sin^2{\left(\Delta_{21}/2\right)}\rangle$, where the averaging is performed over effective mixing angles in matter to incorporate matter effects in their calculations. If unitarity is assumed, this oscillation probability (since oscillations associated with $\Delta m_{31}^2$ have averaged out in KamLAND) is written as~\cite{Gando:2010aa}
\begin{equation}
P_{ee} = (\cos^4\theta_{13} + \sin^4\theta_{13}) - \cos^4\theta_{13} \sin^2\left(2\theta_{12}\right) x.
\end{equation}
When not assuming unitarity, this becomes
\begin{equation}
P_{ee} = 1 - \frac{2\ethr(\eone + \etwo)}{N_e^2} - \frac{4\eone\etwo}{N_e^2} x.
\end{equation}
KamLAND does not use a near detector in its analysis, so we need to re-scale this oscillation probability by $N_e^2$. The probability measured by KamLAND is therefore
\begin{equation}
\hat{P}_{ee} = |U_{e3}|^4 +  \left(\eone + \etwo\right)^2 - 4\eone\etwo x \ .
\end{equation}
The averaging over $x$ performed in Ref.~\cite{Gando:2010aa} depends on the neutrino matter effects during propagation, which we discussed in Section~\ref{subsec:OscProbs}. For long-baseline $\overline{\nu}_e$ oscillations such as those being measured at KamLAND, the deviations from the (small) unitarity-assumed matter effects are, to leading order, dependent on quantities such as $N_2 - N_1$ (the differences between two different column normalizations). Where our global fit is concerned, these differences are constrained to be greatly sub-dominant relative to other contributions to the matter-induced mixing angle and mass-squared splitting. Given this sub-dominance, and the fact that other experiments (solar neutrino experiments and Daya Bay, specifically) provide more powerful measurements of the electron row elements of the LMM, we find this method to be the most complete way of including KamLAND's results in such an analysis.


\subsection{Daya Bay}\label{subsec:DayaBay}

The Daya Bay experiment observes the disappearance of electron antineutrinos $P(\overline{\nu}_e \to \overline{\nu}_e)$. Daya Bay operates in the regime of $\Delta_{21} \ll 1$ and the coefficient of the dominant oscillation term is given by $\sin^2\left(2\theta_{13}\right)$ in the PMNS parameterization. Given the derivations in Appendix~\ref{app:Derivations}, we apply Eq.~\eqref{eq:Dis:SBL}, which shows that the disappearance probability (without assuming unitarity, and including near detector normalization effects) is
\begin{equation} 
\label{eq:DBSens}
P_{ee} \simeq 1 - \frac{4 \ethr (\eone + \etwo)}{N_e^2}\sin^2\left(\frac{\Delta_{31}}{2} \right)\ .
\end{equation}
The most recent measurement from Daya Bay is $\sin^2(2\theta_{13}) = 0.0856 \pm 0.0029$~\cite{Adey:2018zwh}, which we map on to a measurement of the coefficient in Eq.~\eqref{eq:DBSens} in our fit. Daya Bay's measurement of the mass-squared-splitting $\left\lvert \Delta m_{31}^2\right\rvert = \left(2.471 \pm 0.070\right) \times 10^{-3}$ eV$^2$ is also included in our analysis~\cite{Adey:2018zwh}.


\subsection{OPERA} 

The Oscillation Project with Emulsion-tRacking Apparatus (OPERA) collaboration has completed data collection and has provided results of searches for $\nu_\mu \to \nu_\tau$ oscillations, where the neutrinos have a mean energy of $17$ GeV and a baseline of $730$ km~\cite{Agafonova:2018auq}. Across four channels, the experiment observed 10 $\nu_\tau$ signal events with an expectation of $2.0 \pm 0.4$ background events, and $6.8 \pm 1.4$ signal events (assuming $\sin^2\left(2\theta_{23}\right) = 1$ and $\Delta m_{32}^2 = 2.5 \times 10^{-3}$ eV$^2$). This measurement predominantly constrains the quantity $|t_{\mu \tau}|^2$ and the product $|U_{\mu 3}|^2 |U_{\tau 3}|^2$. For our analysis, we assumed that OPERA measures the oscillation probability $P(\nu_\mu \to \nu_\tau)$ for a fixed energy of $17$ GeV and baseline of $730$ km. We compute the oscillation probability numerically (including matter effects) and multiply it by $N_\mu N_\tau$ to account for the normalization effects discussed in Section~\ref{subsec:normalization}. We find that this approximation reproduces the results reasonably well. Finally, we include OPERA's measurement of the mass-squared-splitting $\left\lvert \Delta m_{32}^2\right\rvert = \left( 2.7 \pm 0.7 \right)\times 10^{-3}$ eV$^2$~\cite{Agafonova:2018auq}.


\subsection{T2K} 

\begin{figure}
\centering
\includegraphics[width=\linewidth]{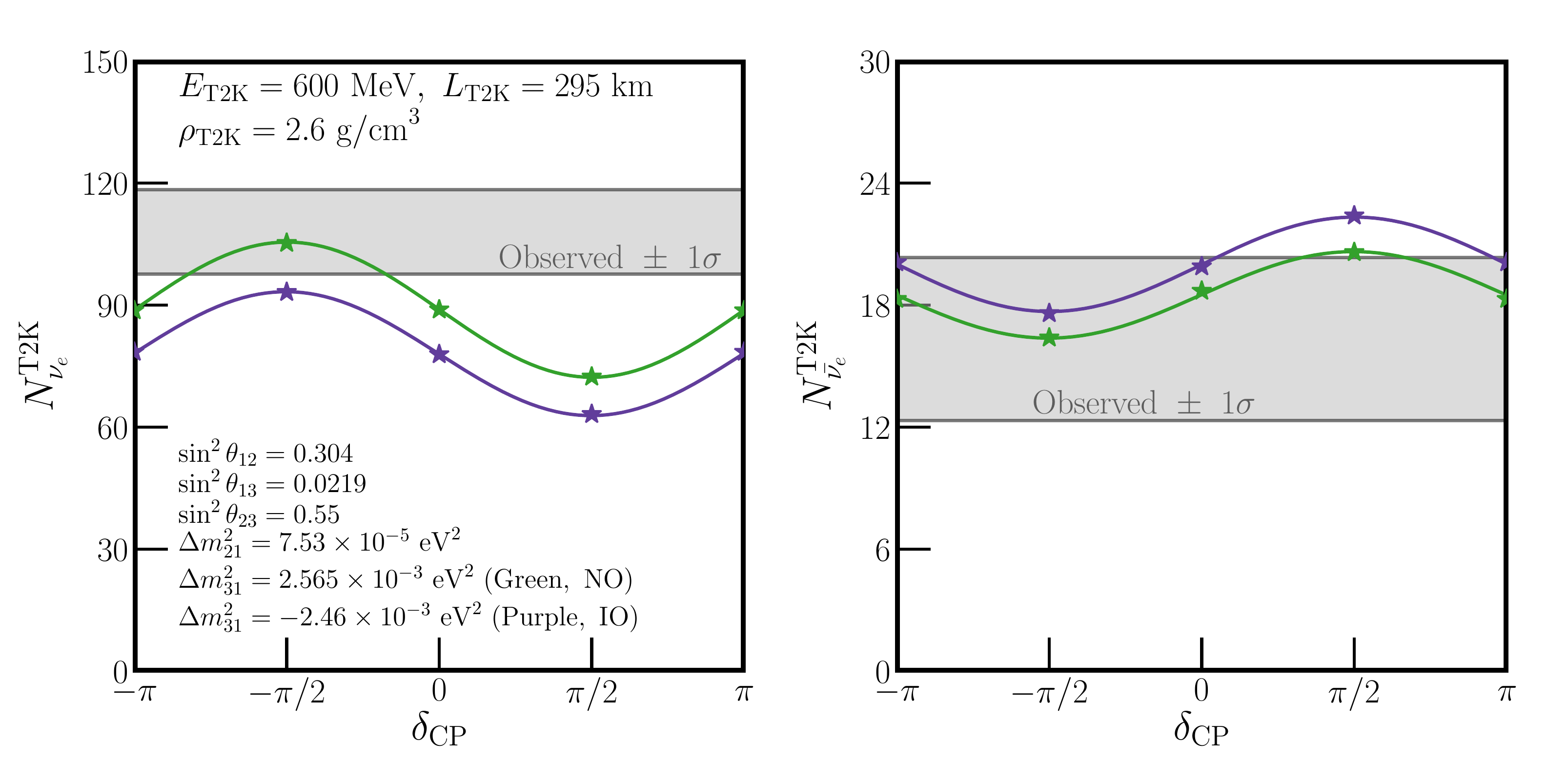}
\caption{Validation of our T2K analysis: each panel displays the number of expected and observed events for a different data sample at T2K: neutrino mode $\nu_e$ events (left) and antineutrino mode $\overline{\nu}_e$ events (right). All oscillation parameters except $\delta_{\rm CP}$ and $\Delta m_{31}^2$ are fixed to the values listed in the left panel. The green (purple) lines and stars assume the normal (inverted) mass ordering, with $\Delta m_{31}^2 = 2.565 (-2.46) \times 10^{-3}$ eV$^2$. Stars correspond to the values given for expected signal and background rates extracted from Ref.~\cite{T2KNu2020}, and the lines correspond to our estimates from Eqs.~(\ref{eq:T2Knu})-(\ref{eq:T2Knubar}). Grey regions indicate the observed number of events in each channel with a $\pm 1\sigma$ range from Poissonian statistics.
\label{fig:T2KValid}}
\end{figure}

The Tokai to Kamioka (T2K) experiment has performed searches for both electron (anti-)neutrino appearance and muon (anti-)neutrino disappearance in a mostly muon (anti-)neutrino beam. We include all searches possible, making some simplifications. We use the preliminary results reported in Ref.~\cite{T2KNu2020}, using the figures therein to extract the expected signal and background rates for $\nu_e$ and $\overline{\nu}_e$ events as a function of oscillation parameters.\footnote{In Ref.~\cite{Ellis:2020ehi}, we had used the detailed results of Ref.~\cite{Abe:2018wpn} to perform our simulations. We find that our updated simulation with the preliminary results of Ref.~\cite{T2KNu2020} are mostly consistent with the previous published result, up to the fact that newer data are now included. Reference~\cite{Abe:2019vii} also includes newer data than Ref.~\cite{Abe:2018wpn}, and its results are more-or-less consistent with ours.} In Ref.~\cite{T2KNu2020}, expected signal plus background rates are shown for the $\nu_\mu \to \nu_e$ and $\overline{\nu}_\mu \to \overline{\nu}_e$ channels for values of $\delta_{\rm CP}$ of $-\pi/2$, $0$, $\pi/2$, and $\pi$. These are given for both the normal and inverted mass orderings, assuming the other oscillation parameters are fixed. We make the simple assumption that T2K measures these event rates for a fixed energy $E_{\rm T2K} = 600$ MeV at a fixed baseline length $L_{\rm T2K} = 295$ km, with a constant matter density of $\rho_{\rm T2K} = 2.6$ g/cm$^3$~\cite{Abe:2018wpn}. Using the figures provided in Ref.~\cite{T2KNu2020}, we arrive at the following expected event rates:
\begin{align}
N_{\nu_e}^{\rm T2K} &= 21.90 + 1282.84\times P(\nu_\mu \to \nu_e, L_{\rm T2K}, E_{\rm T2K}), \label{eq:T2Knu}\\
N_{\overline{\nu}_e}^{\rm T2K} &= 10.66 + 179.59\times P(\bar{\nu}_\mu \to \bar{\nu}_e, L_{\rm T2K}, E_{\rm T2K}). \label{eq:T2Knubar}
\end{align}
Here, the values $21.90$ and $10.66$ are our extracted background rates for each of the two channels. In order to extract these predictions, we make the assumption that the background rates are independent of the other oscillation parameters. The pre-factors $1282.84$, $179.59$, can be interpreted as a weighted flux times cross-section for this particular energy, translating an oscillation probability into an expected number of signal events. Figure~\ref{fig:T2KValid} presents the expected number of signal plus background events for these two different channels, given by the formulae in Eqs.~(\ref{eq:T2Knu})-(\ref{eq:T2Knubar}), along with stars that indicate the values given by the figure in Ref.~\cite{T2KNu2020}. Note that, despite assuming a mono-energetic measurement, our curves intersect the stars nearly perfectly. Also shown in each panel of Fig.~\ref{fig:T2KValid} is the observed number of events in each channel (again, from Ref.~\cite{T2KNu2020}), and its $\pm 1\sigma$ statistical range. Since the statistical uncertainties are large, we only include them (and no other systematic uncertainties) in this T2K electron-neutrino appearance measurement. We compute the oscillation probabilities including matter effects numerically and multiply it by $N_e/N_\mu$ to account for the use of a near detector for T2K.

For T2K's measurement of muon-neutrino and muon-antineutrino disappearance, we find that instead of assuming a mono-energetic measurement, that we obtain results more compatible with those of the collaboration if we assume a fixed measurement of the relevant coefficient of the disappearance probability. The disappearance probability is 
\begin{align}
\label{eq:PMuMuT2K}
P_{\mu \mu} \simeq 1 - \frac{4 \mthr (\mone + \mtwo)}{N_\mu^2}\sin^2\left(\frac{\Delta_{31}}{2} \right)\ ,
\end{align}
such that the relevant coefficient is
\begin{equation}
\label{eq:AmpMuMu}
C_{\mu\mu}^{\mathrm{Dis.}} = \frac{4\mthr \left(\mone + \mtwo\right)}{N_\mu^2}.
\end{equation}
We include T2K's combined $\nu_\mu$ and $\overline{\nu}_\mu$ disappearance searches by assuming a measurement of $C_{\mu\mu} = 1.0\  \pm\ 0.03$, and find that it gives us results consistent with Refs.~\cite{Abe:2018wpn,Abe:2019vii,T2KNu2020}. Finally, when analyzing results from T2K (when studying its appearance channels, disappearance channels, or both), we include a Gaussian prior on the mass squared splitting $\left\lvert \Delta m_{32}^2\right\rvert = \left(2.49 \pm 0.082\right) \times 10^{-3}$ eV$^2$~\cite{T2KNu2020}.


\subsection{NOvA} 

The last long-baseline experiment we include is the NuMI Off-Axis $\nu_e$ Appearance (NOvA) experiment. It operates at similar values of $L/E_\nu$ as T2K, but at longer distance/higher energy. Like our T2K analysis, we assume that NOvA measures event rates at a fixed energy and baseline -- we assume this to be $E_{\rm NOvA} = 1.9$~GeV, $L_{\rm NOvA} = 810$~km, with a constant matter density along the path of propagation of $\rho_{\rm NOvA} = 2.84$~g/cm$^3$~\cite{NOvA:2018gge}. Using the preliminary results of Ref.~\cite{NOvANu2020}, we extract expected signal and background rates for different sets of oscillation parameters, using our mono-energetic assumption and assuming that the backgrounds are independent of neutrino oscillations.\footnote{As with T2K, we have also used the more detailed Ref.~\cite{Acero:2019ksn} to perform a similar analysis with less overall data as a cross-check. We find that our simulations match both Ref.~\cite{Acero:2019ksn} and Ref.~\cite{NOvANu2020} well.
}

Reference~\cite{NOvANu2020} reports observed event rates for both neutrino and antineutrino appearance, as well as expected background and signal rates for a set of oscillation parameters. We take these expected rates and our mono-energetic assumption to infer our expected event rates,
\begin{align}
N_{\nu,\ {\rm CCQE}}^{\rm NOvA} &= 29.09 + 1202.65 \times P\left(\nu_\mu \to \nu_e, L_{\rm NOvA}, E_{\rm NOvA}\right), \label{eq:NOvANu} \\
N_{\overline{\nu},\ {\rm CCQE}}^{\rm NOvA} &= 16.59 + 438.426 \times P\left(\overline{\nu}_\mu \to \overline{\nu}_e, L_{\rm NOvA}, E_{\rm NOvA}\right). \label{eq:NOvANuBar}
\end{align}
%

\begin{figure}
\centering
\includegraphics[width=0.5\linewidth]{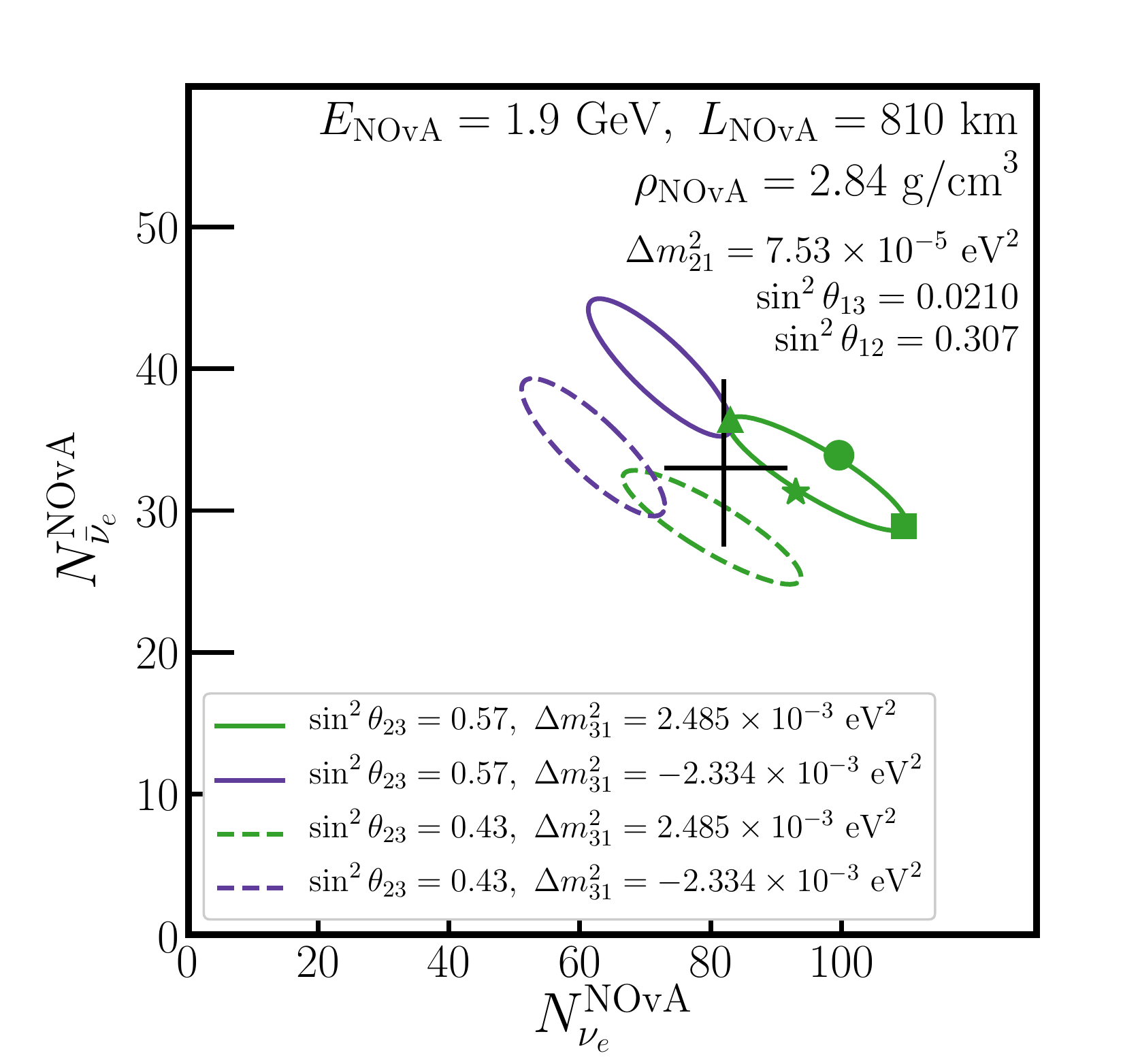}
\caption{Validation of our NOvA analysis: we present the number of expected signal events in neutrino mode ($x$-axis) and antineutrino mode ($y$-axis), for $\Delta m_{21}^2$, $\sin^2\theta_{13}$, and $\sin^2\theta_{12}$ values given in the upper-right corner of the figure. The four different ellipses correspond to different combinations of ($\sin^2\theta_{23}$, $\Delta m_{31}^2$) given in the legend where we vary $\delta_{\rm CP}$ between $-\pi$ and $\pi$. The black cross indicates the number of observed signal neutrino and antineutrino events with statistical uncertainties only. Colored markers on the solid green ellipse correspond to the expected values of event rates presented in Ref.~\cite{NOvANu2020} for $\delta_{\rm CP} = \pm\pi$ (star), $-\pi/2$ (square), $0$ (circle), and $\pi/2$ (triangle).
\label{fig:NOvAValid}}
\end{figure}

Instead of presenting the expected event rates for NOvA as a function of $\delta_{\rm CP}$ in separate panels, we choose to show a ``bi-event'' plot, which shows the two expected rates simultaneously, in Fig.~\ref{fig:NOvAValid}. We have fixed $\Delta m_{21}^2$, $\sin^2\theta_{12}$, and $\sin^2\theta_{13}$ as given by the values in the top-right of the panel. Each ellipse is generated by fixing $(\sin^2\theta_{23}, \Delta m_{31}^2)$ (given by the legend), and varying $\delta_{\rm CP}$, while using Eqs.~\eqref{eq:NOvANu} and~\eqref{eq:NOvANuBar}. The values of $\sin^2\theta_{23}$ and $\Delta m_{31}^2$ for the combination correspond to the upper octant of $\theta_{23} > 1/2$, and the normal mass ordering $\Delta m_{31}^2 > 0$ corresponds to the best-fit according to Ref.~\cite{NOvANu2020}. Since corresponding values for NOvA's preferred values in the inverted mass ordering and/or lower octant of $\theta_{23}$ are not provided, we simply choose values such that $\sin^2\left(2\theta_{23}\right)$ and $\left\lvert \Delta m_{32}\right\rvert^2$ are constant for these choices. For the normal ordering, upper octant (green solid line) choice, we display as markers the four expected event rates corresponding to the figure shown in Ref.~\cite{NOvANu2020}, displaying how well our results agree with the official collaboration ones. Finally, the black cross indicates the observed event rate of $N_{\nu_e}^{\rm NOvA} = 82$, $N_{\overline{\nu}_e}^{\rm NOvA} = 33$ with statistical uncertainties. As with T2K, we do not include any systematic uncertainties in this portion of the analysis.

For NOvA's measurement of $\nu_\mu$ and $\overline{\nu}_\mu$ disappearance, we find, similar to our analysis of T2K, that the reported results are more realistically matched if we simply assume a fixed measurement of the disappearance coefficient given in Eq.~\eqref{eq:AmpMuMu}. Since NOvA measures this to be slightly away from maximal, we include this as a measurement $C_{\mu\mu}^{\rm Dis.} = 0.99 \pm 0.02$, which replicates the results from Ref.~\cite{Acero:2019ksn,NOvANu2020} fairly well. When we analyze NOvA results, either appearance data alone, disappearance data alone, or combined, we include its measurement $\left\lvert \Delta m_{32}^2\right\rvert = \left(2.41 \pm 0.07\right) \times 10^{-3}$ eV$^2$~\cite{NOvANu2020}.


\subsection{Sterile Neutrino Searches}
\label{subsec:SterileSearch} 

When unitarity is not assumed, there could be additional zero-distance effects. Given our current knowledge of the mass-squared-splittings $\Delta m_{21}^2$ and $\Delta m_{31}^2$ and assuming that there are no additional neutrinos, any experiment, including near detectors, that operates at $L/E_\nu$ such that
\begin{equation}
\frac{\Delta m_{21}^2 L}{4E_\nu} \ll 1,\quad \quad \mathrm{and}\quad\quad \frac{\Delta m_{31}^2 L}{4 E_\nu} \ll 1 
\label{eq:ShortDistanceReqs},
\end{equation}
should see no neutrino oscillations, and therefore is sensitive to these zero-distance effects. In this regime of $L/E_\nu$, oscillation probabilities without assuming unitarity are
\begin{align}
P(\nu_\alpha \to \nu_\alpha) &= 1,\quad \mathrm{(Disappearance)}, \\
P(\nu_\alpha \to \nu_\beta) &= \frac{\left\lvert t_{\alpha\beta}\right\rvert^2}{N_\alpha N_\beta},\quad (\alpha\neq\beta\ \mathrm{Appearance}). 
\label{eq:NuZeroDistApp_true}
\end{align}
Due to the nature of such experiments, none of the sterile neutrino searches employ a supplementary ``near" detector. Based on our discussions in Section~\ref{subsec:normalization}, the measured oscillation probabilities must be rescaled due to Monte Carlo predictions as
\begin{align}
\hat{P}(\nu_\alpha \to \nu_\alpha) &= N_\alpha^2,\quad \mathrm{(Disappearance)}, \label{eq:NuZeroDistDis}\\
\hat{P}(\nu_\alpha \to \nu_\beta) &= \left\lvert t_{\alpha\beta}\right\rvert^2,\quad (\alpha\neq\beta\ \mathrm{Appearance}). 
\label{eq:NuZeroDistApp}
\end{align}

Sterile neutrino searches are typically carried out in the following way. If a fourth neutrino exists with a mass-squared splitting $\Delta m_{41}^2 \gg \left \lvert \Delta m_{31}^2\right\rvert$, and an experiment operates in the $L/E_\nu$ regime given by Eq.~\eqref{eq:ShortDistanceReqs}, then it can search for oscillations given by the probabilities
\begin{align}
P(\nu_\alpha \to \nu_\alpha) &= 1 - \sin^2\left(2\theta_{\alpha\alpha}\right) \sin^2\left(\frac{\Delta m_{41}^2 L}{4E_\nu}\right) = 1 - 4\left\lvert U_{\mu 4}\right\rvert^2\left(1 - \left\lvert U_{\mu 4}\right\rvert^2\right) \sin^2\left(\frac{\Delta m_{41}^2 L}{4E_\nu}\right),\ \mathrm{(Disappearance)} \\
P(\nu_\alpha \to \nu_\beta) &= \sin^2\left(2\theta_{\alpha\beta}\right) \sin^2\left(\frac{\Delta m_{41}^2 L}{4E_\nu}\right) = 4\left\lvert U_{\alpha 4}\right\rvert^2 \left\lvert U_{\beta 4}\right\rvert^2 \sin^2\left(\frac{\Delta m_{41}^2 L}{4E_\nu}\right),\ (\alpha\neq\beta\ \mathrm{Appearance}).
\end{align}
Limits or potential observations from these searches are presented in terms of $\sin^2\left(2\theta_{\alpha\beta}\right)$ and $\Delta m_{41}^2$ (see, for example, Refs.~\cite{Dentler:2018sju,Diaz:2019fwt,Boser:2019rta}) For any particular experiment situated at a baseline $L$ and measuring oscillations for a specific energy range, the oscillations associated with $\Delta m_{41}^2$ become very rapid (as a function of $E_\nu$) as $\Delta m_{41}^2$ becomes larger and larger. Eventually, these oscillations average out, and the term $\sin^2\left(\Delta m_{41}^2 L/4E_\nu\right) \to 1/2$. In this averaged-out regime, the oscillation probabilities become
\begin{align}
P(\nu_\alpha \to \nu_\alpha) &\to 1 - \frac{1}{2}\sin^2\left(2\theta_{\alpha\alpha}\right) = 1 - 2\left\lvert U_{\mu 4}\right\rvert^2\left(1 - \left\lvert U_{\mu 4}\right\rvert^2\right),\quad \mathrm{(Disappearance)}, \label{eq:SterileAvgDis}\\
P(\nu_\alpha \to \nu_\beta) &\to \frac{1}{2} \sin^2\left(2\theta_{\alpha\beta}\right) = 2\left\lvert U_{\alpha 4}\right\rvert^2 \left\lvert U_{\beta 4}\right\rvert^2,\quad (\alpha\neq\beta\ \mathrm{Appearance}). 
\label{eq:SterileAvgApp}
\end{align}
These have the same lack of energy-dependence as searches for unitarity violation. Equating Eq.~\eqref{eq:SterileAvgDis} with Eq.~\eqref{eq:NuZeroDistDis} therefore allows us to map constraints on $\sin^2\left(2\theta_{\alpha\alpha}\right)$ from sterile neutrino searches in the averaged-out regime onto constraints of $N_\alpha^2$, while comparing Eq.~\eqref{eq:SterileAvgApp} with Eq.~\eqref{eq:NuZeroDistApp} allows us to map $\sin^2\left(2\theta_{\alpha\beta}\right)$ onto $\left\lvert t_{\alpha\beta}\right\rvert^2$ for $\alpha\neq\beta$. Table~\ref{tab:Steriles} summarizes the null sterile neutrino searches included in our analysis: KARMEN, NOMAD, CHORUS, and MINOS/MINOS+.

\begin{table}
\caption{Sterile neutrino searches included in our analysis, and the associated 90\% CL limit on the effective mixing angle from the given experimental search. \label{tab:Steriles}}
\begin{tabular}{|c||c|c|c|}\hline
Search & 90\% CL Limit & Angle Constrained & Unitarity Constraint \\ \hline \hline
KARMEN $\overline{\nu}_\mu \to \overline{\nu}_e$~\cite{Armbruster:2002mp} & $1.8 \times 10^{-3}$ & $\sin^2\left(2\theta_{\mu e}\right)$ & $\left\lvert t_{\mu e}\right\rvert^2$ \\ \hline
NOMAD $\nu_\mu \to \nu_e$~\cite{Astier:2003gs} & $1.4 \times 10^{-3}$ & $\sin^2\left(2\theta_{\mu e}\right)$ & $\left\lvert t_{\mu e}\right\rvert^2$ \\ \hline
NOMAD $\nu_e \to \nu_\tau$~\cite{Astier:2001yj} & $1.5 \times 10^{-2}$ & $\sin^2\left(2\theta_{e\tau}\right)$ & $\left\lvert t_{e\tau}\right\rvert^2$ \\ \hline
NOMAD $\nu_\mu \to \nu_\tau$~\cite{Astier:2001yj} & $3.3 \times 10^{-4}$ & $\sin^2\left(2\theta_{\mu\tau}\right)$ & $\left\lvert t_{\mu\tau}\right\rvert^2$ \\ \hline
CHORUS $\nu_\mu \to \nu_\tau$~\cite{Eskut:2007rn} & $4.4 \times 10^{-4}$ & $\sin^2\left(2\theta_{\mu\tau}\right)$ & $\left\lvert t_{\mu\tau}\right\rvert^2$ \\ \hline
MINOS/MINOS+~\cite{Adamson:2017uda} $\nu_\mu \to \nu_\mu$ & $2.5 \times 10^{-2}$ & $\sin^2\left(2\theta_{\mu\mu}\right)$ & $ N_{\mu}^2$ \\ \hline
\end{tabular}
\end{table}

We do not consider any sterile neutrino searches for $P(\overline{\nu}_e \to \overline{\nu}_e)$ from reactor antineutrino experiments. The averaged-out regime of these searches, which is required to perform this mapping, depends on flux and cross section uncertainties to be well understood in a disappearance search. The overall flux of these searches is notoriously difficult to constrain~\cite{Huber:2011wv,Mueller:2011nm,Berryman:2019hme,Berryman:2020agd}, so these experiments do not place robust, strong limits in the high-$\Delta m_{41}^2$ regime.

Both the Liquid Scintillator Neutrino Detector (LSND)~\cite{Athanassopoulos:1996wc,Athanassopoulos:1997pv} and MiniBooNE~\cite{Aguilar-Arevalo:2018gpe} experiments have observed an excess of electron-like events in the presence of a beam that is mostly $\nu_\mu$ (or $\overline{\nu}_\mu$), which can be interpreted as a short-baseline oscillation with $P(\nu_\mu \to \nu_e) \approx 2.6 \times 10^{-3}$. A combined study of these two experiments favors $P(\nu_\mu \to \nu_e) \neq 0$ at roughly $6\sigma$. When analyzed in the context of a light sterile neutrino, the preferred parameter space is compatible with the averaged-out regime, however that is not where their best-fit point lies. We inspect the effect of including the favored $\left\lvert t_{\mu e}\right\rvert^2 \neq 0$ preference from LSND/MiniBooNE in Appendix~\ref{app:LSNDMB}.


\section{Future Experiments and Simulations}
\label{sec:FutureExps}

In this section, we describe the future experiments that we consider in our analysis. Specifically, we focus on the IceCube Upgrade~\cite{Ishihara:2019aao}, JUNO~\cite{An:2015jdp,Abusleme:2020bzt}, DUNE~\cite{Abi:2020evt,Abi:2020qib}, and T2HK~\cite{Abe:2018uyc} experiments.

When simulating future data, we assume that the LMM is unitary, and consistent with the best-fit-point of an analysis of current data with unitarity assumed. When we analyze all current data assuming unitarity (using the PMNS parameterization), we obtain the following best-fit point: $\sin^2\theta_{12} = 0.308$, $\sin^2\theta_{13} = 0.02190$, $\sin^2\theta_{23} = 0.551$, $\delta_{\rm CP} = -2.78 = 200.4^\circ$, $\Delta m_{21}^2 = 7.50 \times 10^{-5}$ eV$^2$, and $\Delta m_{31}^2 = 2.53 \times 10^{-3}$ eV$^2$.\footnote{With the recent update of oscillation data from T2K~\cite{T2KNu2020} and NOvA~\cite{NOvANu2020}, the preference for the normal mass ordering ($\Delta m_{31}^2 > 0$) over the inverted mass ordering ($\Delta m_{31}^2 < 0$) has diminished~\cite{Kelly:2020fkv,Esteban:2020cvm}. We choose the best-fit point according to the normal ordering, given data not included in our fit, specifically Super-Kamiokande's atmospheric neutrino sample.} We translate these values in to the values of $\absq{U_{\alpha k}}$ and $\phi_{\alpha k}$ and obtain

\begin{equation}
\left\lvert U_{\rm LMM}\right\rvert^2 = \left(\begin{array}{c c c} 0.677 & 0.302 & 0.022 \\ 0.083 & 0.378 & 0.534 \\ 0.240 & 0.320 & 0.439 \end{array}\right);\ \phi_{e2} = 3.00,\ \phi_{e3} = -0.47,\ \phi_{\tau2} = -0.24,\ \phi_{\tau3} = 2.97.
\end{equation}


\subsection{IceCube Upgrade}
\label{subsec:IceCube}

The IceCube experiment is capable of detecting atmospheric neutrinos over a broad range of energies, $1\ \mathrm{GeV}\lesssim E_\nu \lesssim 100\ \mathrm{GeV}$. By measuring track-like ($\nu_\mu$) and cascade-like ($\nu_e$ and $\nu_\tau$) events, IceCube is sensitive to effects of the mass-squared splitting $\Delta m_{31}^2$ and oscillations associated with that splitting. Of importance for constraining leptonic unitarity is IceCube's ability to constrain the normalization of $\nu_\mu \to \nu_\tau$ appearance in its data sample. Currently, this measurement has a precision of $\sim$40\%~\cite{Aartsen:2019tjl}.  However, in the coming years, $\mathcal{O}(10\%)$ precision will be attainable by considering either eight years of IceCube DeepCore data or one year of IceCube Upgrade data~\cite{summer_blot_2020_3959546}. Given the regime of oscillations that IceCube measures, for $\nu_\mu \to \nu_\tau$ appearance it is dominantly sensitive to $4\absq{U_{\mu3}}\absq{U_{\tau3}}/N_\mu^2$. We do not include any information from IceCube on this quantity in our current fits, but include a $10\%$ measurement of that quantity in our future projections.


\subsection{JUNO}
\label{subsec:JUNO}

JUNO is an upcoming reactor-based neutrino experiment (scheduled to start operation in 2022~\cite{JUNONu2020}), where neutrinos from 10 different nuclear reactors travel for baselines of $\simeq$53 km before reaching a 20 kiloton detector (with 2 additional reactors at a baseline of $\simeq$200 km), comprised of liquid scintillator. JUNO is primarily sensitive to the atmospheric mass-squared splitting $\Delta m_{31}^2$, as well as being sensitive to the solar mass-squared splitting $\Delta m_{21}^2$. Its design goal is to measure $\Delta m_{31}^2$ precisely enough to determine the mass ordering of the neutrinos, i.e. whether $m_1 < m_2 < m_3$ or $m_3 < m_1 < m_2$. Because it operates in neither the regime $\Delta_{21} \ll 1$ nor $\Delta_{31} \gg 1$, when considering JUNO we must use the full oscillation probability in Eq.~\eqref{eq:Pdis}. The effects of the matter potential on the oscillation probability $P(\overline{\nu}_e \to \overline{\nu}_e)$ are negligible for sensitivity studies on the solar sector mixing~\cite{An:2015jdp}, which is the main contribution from JUNO dataset to the global unitarity constraints, so we employ the form for said probability given in Eq.~\eqref{eq:Pdis}.

We simulate the expected event rate assuming six years of data collection, corresponding to a total of $1.2\times 10^5$ signal events from $\overline{\nu}_e$ inverse beta-decay scattering in the JUNO detector~\cite{An:2015jdp,Abrahao:2015rba}. Following Ref.~\cite{An:2015jdp}, we include the following sources of systematic uncertainties in our simulation: correlated (among different reactors) flux uncertainty of 2\%, uncorrelated flux uncertainty of 0.8\% for each reactor, spectrum shape uncertainty of 1\%. We do not include backgrounds in our simulation. Our projected sensitivity to $\sin^2\theta_{12}$ in the standard three-flavor oscillation scenario is 0.42\%, compared to the official collaboration sensitivity of 0.54\%.

JUNO plans on using the Taishan Antineutrino Observatory (TAO)~\cite{Abusleme:2020bzt} to constrain the reactor antineutrino spectrum to sub-percent energy resolution. TAO will be situated 30 meters from the core of the Taishan Nuclear Power Plant and will be able to observe roughly 2000 reactor $\overline{\nu}_e$ interactions per day. Using TAO and the 20 kt JUNO detector, the collaboration can perform a near-to-far detector ratio measurement similar to Daya Bay (see Section~\ref{subsec:DayaBay}). We also make the conservative assumption that with a comparison of TAO measurements to reactor antineutrino flux predictions, the collaboration can constrain the normalization of its near detector measurement to allow for a $10\%$ measurement of $N_e$.\footnote{Without such a near detector measurement, the near-to-far detector ratio measurement by JUNO-TAO would be subject to an overall rescaling degeneracy of the electron row elements $\absq{U_{ek}}$. Including a 10\% measurement from TAO does not impact our final joint analysis results because $N_e$ is already constrained (by KamLAND, solar neutrino measurements, and Daya Bay) at the few-percent level.}

JUNO will in principle be sensitive to each element of the electron row of the LMM, $\eone$, $\etwo$, and $\ethr$. As we will show in Section~\ref{sec:Results}, JUNO will allow for more precise measurements on $\eone$ and $\etwo$ than current experiments (specifically measuring the combination $\etwo/\eone$ very precisely), but will not improve on the precision of $\ethr$ achieved by Daya Bay.


\subsection{DUNE}

DUNE is a future beam-based neutrino experiment, scheduled to begin data collection in the late 2020s. It consists of an $\mathcal{O}(\text{GeV})$ neutrino beam, consisting primarily of $\nu_\mu$ when operating in neutrino mode and $\overline{\nu}_\mu$ when operating in anti-neutrino mode~\cite{LauraFluxes}, and a 40-kton liquid argon far detector. The baseline length is 1300 km, meaning that DUNE operates near the regime $\Delta_{21} \ll 1$. Nevertheless, it will be sensitive to the effects of $\Delta m_{21}^2$, allowing for a precise measurement of the CP-violating phase $\delta_\text{CP}$ in the PMNS matrix, among other goals.

We consider three different beam-related channels when simulating DUNE, each effectively measuring a different neutrino oscillation probability: the electron-neutrino appearance channel, sensitive to $P(\nu_\mu \to \nu_e)$ (and its CP-conjugate); muon-neutrino disappearance/survival $P(\nu_\mu \to \nu_\mu)$ (and its CP-conjugate); and tau-neutrino appearance, sensitive to $P(\nu_\mu \to \nu_\tau)$ (and its CP-conjugate). To simulate the expected event rates for these different channels, we employ simulation code developed with Refs.~\cite{Berryman:2015nua,deGouvea:2015ndi,Berryman:2016szd,deGouvea:2016pom,deGouvea:2017yvn} for $\nu_e$ appearance and $\nu_\mu$ disappearance and Ref.~\cite{deGouvea:2019ozk} for $\nu_\tau$ appearance. For all of our simulations, we assume seven years of data collection with DUNE, divided evenly between operation in neutrino mode and antineutrino mode. We include signal and background normalization uncertainties (5\% for the $\nu_e$ appearance and $\nu_\mu$ disappearance channels, and 25\% for $\nu_\tau$ appearance). As we will show in Section~\ref{sec:Results}, DUNE will allow improve on the precision of the measurements made by NOvA and T2K for both $\nu_\mu$ disappearance and $\nu_e$ appearance, as well as OPERA for $\nu_\tau$ appearance.

Finally, DUNE is capable of improving on existing measurements of the $^{8}$B solar neutrino flux using $\nu_e$ CC and elastic electron scattering~\cite{Capozzi:2018dat}. We simplify the analysis by assuming that DUNE will be able to measure $P_{ee} \equiv \absq{U_{e2}} \left( \absq{U_{e1}} + \absq{U_{e2}}\right) + \left\lvert U_{e3}\right\rvert^4$ at the $3\%$ level, consistent with the more complete analysis of Ref.~\cite{Capozzi:2018dat}.


\subsection{T2HK}

In the next decade, T2K will be upgraded with a larger water \v{C}erenkov detector and begin operating as T2HK. It will operate in a similar region of $L/E_\nu$ as DUNE, albeit at a lower length and energy. This, along with the different detection mechanism, allows for tests between the results of the two experiments, and further power in validating (or discovering new physics beyond) the three-massive-neutrinos paradigm. T2HK will also collect a very large sample of atmospheric neutrinos, which we do not include in our analysis. Its beam-based program plans to collect data in a $1{:}3$ ratio between neutrino and antineutrino modes. While T2HK intends to operate for ten years or longer, we rescale all of our expected signal and background rates to a data collection period of seven years to be consistent with our DUNE projections.

We perform simulations of the T2HK expected yields for $\nu_\mu \to \nu_e$ appearance and $\nu_\mu \to \nu_\mu$ disappearance, consistent with Refs.~\cite{Abe:2015zbg,Abe:2018uyc}, developed from Refs.~\cite{Kelly:2017kch,deGouvea:2017yvn}. As with our DUNE simulation, this has been modified to allow for a non-unitary LMM. We include expected signal and background yields in our simulations, along with energy resolutions discussed in Refs.~\cite{Kelly:2017kch,deGouvea:2017yvn}, and signal and background normalization uncertainties of 5\%.


\section{Primary Results}
\label{sec:Results}

Throughout this section, we present the results of analyzing the datasets described in Sections~\ref{sec:CurrentExps} and~\ref{sec:FutureExps}. We begin in Section~\ref{subsec:Consistency} with consistency checks where unitarity is assumed, and determine whether different datasets agree on their measurements of different parameters. These consistency checks can serve as a simple test of unitarity. Subsequently, we abandon the unitary assumption and consider the agnostic case, adopting the MP parameterization, and present results for our full analysis in Sections~\ref{app:EMuRows},~\ref{subsec:USq}, and~\ref{subsec:NormClos}. Section~\ref{app:EMuRows} shows how subsets of data contribute to the sensitivity in the electron and muon rows.  Section~\ref{subsec:USq} presents how well we can currently constrain the LMM matrix element magnitudes $\left\lvert U_{\alpha k}\right\rvert^2$ and how much we can improve with future data. Section~\ref{subsec:NormClos} demonstrates how well we can determine the normalization of the rows/columns, and closure of the row/column triangles of the LMM.

Before presenting our results, we clarify the statistical approaches taken in our analyses. In Sections~\ref{subsec:Consistency} and~\ref{app:EMuRows}, the analyses only rely on a handful of parameters. We perform frequentist analyses, scanning a given likelihood function over the parameters of interest, and determining the confidence levels (CL) in these parameter spaces. In Sections~\ref{subsec:USq} and~\ref{subsec:NormClos}, the analyses depend on 15 parameters, and we use the Bayesian inference tool {\sc pyMultiNest}~\cite{Feroz:2007kg,Feroz:2008xx,Feroz:2013hea,Buchner:2014nha} to construct credible regions (CR) based on the posterior likelihood density. For further details, see Appendix~\ref{app:Bayesian}.


\subsection{Simple Unitarity Constraints \& Consistency Checks}
\label{subsec:Consistency}

\begin{figure}[t]
\centering
\includegraphics[width=0.8\linewidth]{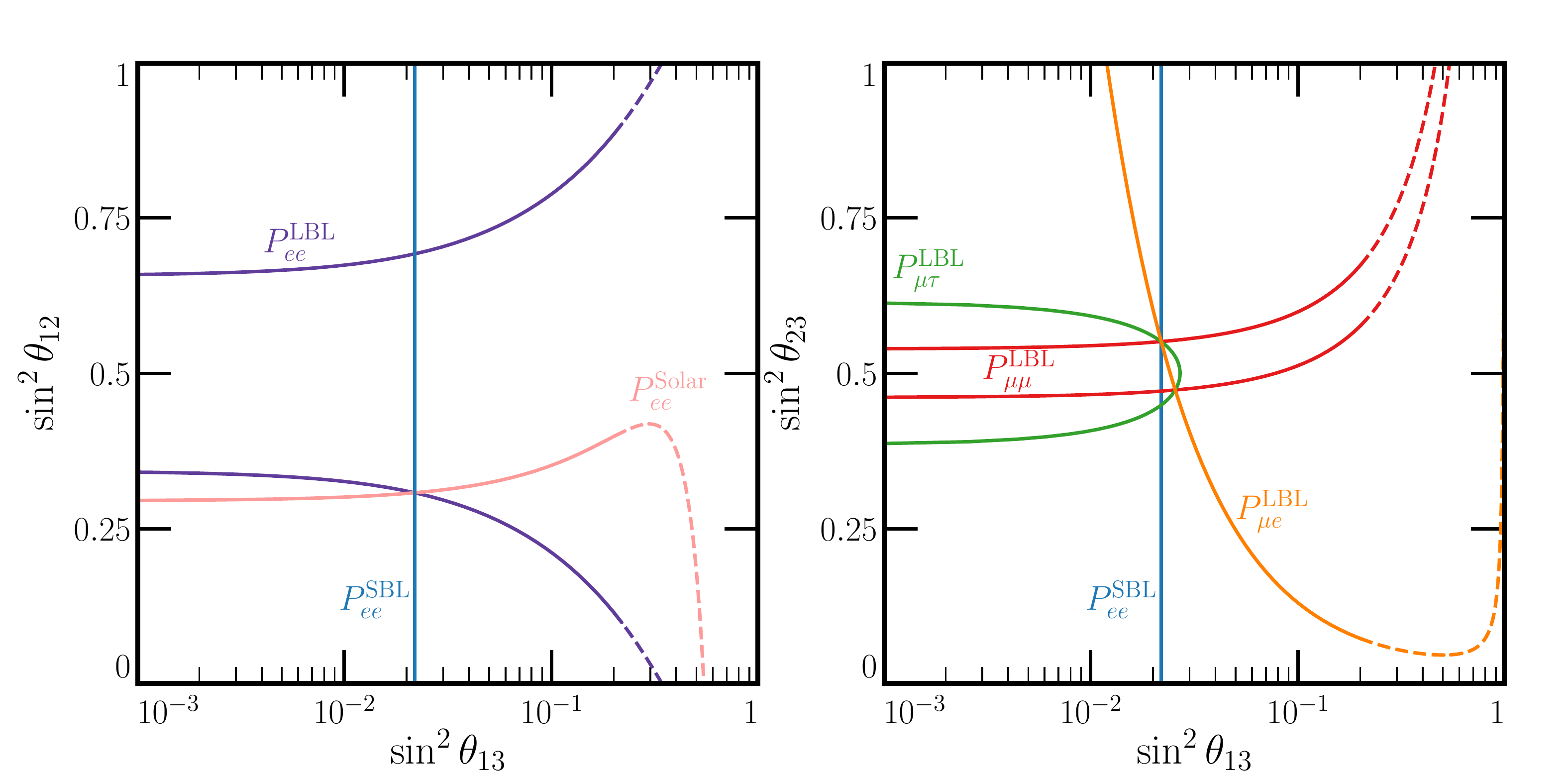}
\caption{Hypothetical perfect measurements of different oscillation coefficients as described in the test -- the true values of the mixing angles are assumed to be $\sin^2\theta_{12} = 0.316$, $\sin^2\theta_{13} = 0.022$, and $\sin^2\theta_{23} = 0.565$. We dash lines above $\sin^2\theta_{13}$ where our analytical approximations are likely no longer valid.
\label{fig:S12S13S23Analytic}} 
\end{figure}

One straightforward way to test whether the LMM is unitary is by analyzing different experimental measurements separately and checking for consistency. This is demonstrated conceptually in Fig.~\ref{fig:S12S13S23Analytic} for two pairs of mixing angles, $\sin^2\theta_{13}$ vs. $\sin^2\theta_{12}$ and $\sin^2\theta_{13}$ vs. $\sin^2\theta_{23}$.\footnote{For the third combination, $\sin^2\theta_{12}$ vs. $\sin^2\theta_{23}$, no existing or future measurement is sensitive to this combination of angles in an interesting and non-trivial way.} We assume the LMM is unitary, and interpret experimental measurements as combinations of PMNS mixing angles (see Table~\ref{tab:QuantitiesPMNSLMM}). If the LMM is indeed unitary, all of these measurements should meet at a single point in the $\sin^2\theta_{jk}$-$\sin^2\theta_{nl}$ planes, while if it not unitary, an intersection is not guaranteed.

Before continuing, we note that the analytical expressions in Table~\ref{tab:QuantitiesPMNSLMM} are good approximations of measurements near the best-fit regions of mixing parameters. If one deviates too far from best-fit regions, these approximations break down. For example, current data indicates small $\absq{U_{e3}}$, or $\sin^2\theta_{13}$ if unitarity is assumed. In what follows, when analyzing different oscillation coefficients, we will generically allow $\sin^2\theta_{13}$ to be large. This very likely produces results at large $\sin^2\theta_{13}$ that are inconsistent with real experimental observations. Nevertheless, it is instructive to see how the results of these analytic estimates change over the full, potential range of the mixing angles. We indicate the region where these expressions likely break down, $\sin^2\theta_{13} \gtrsim 0.2$, with dashed lines.

In the left panel of Fig.~\ref{fig:S12S13S23Analytic}, we compare how three types of measurements can pin down $\sin^2\theta_{13}$ vs. $\sin^2\theta_{12}$ given infinite experimental precision, under the assumption that the full oscillation probabilities are dominated by the coefficients listed in Table~\ref{tab:QuantitiesPMNSLMM}. The three measurements we incorporate are: 
\begin{itemize}
\item $P_{ee}^{\rm SBL}$: short-baseline measurement of $P(\overline{\nu}_e \to \overline{\nu}_e)$ from a reactor neutrino experiment, e.g., Daya Bay.
\item $P_{ee}^{\rm Solar}$: solar neutrinos measured via CC interactions.
\item $P_{ee}^{\rm LBL}$: long-baseline measurement of $P(\overline{\nu}_e \to \overline{\nu}_e)$ from a reactor neutrino experiment, e.g., KamLAND.
\end{itemize}
Whether all three measurements intersect is a test of the $e$-row normalization, i.e., whether $N_e = 1$.

In the right panel of Fig.~\ref{fig:S12S13S23Analytic}, we show similar hypothetical infinite-precision measurements of oscillation probabilities that are sensitive to a combination of $\sin^2\theta_{13}$ and $\sin^2\theta_{23}$. They are the following:
\begin{itemize}
\item $P_{ee}^{\rm SBL}$ as above.
\item $P_{\mu\mu}^{\rm LBL}$: long-baseline measurement of $P(\nu_\mu \to \nu_\mu)$ as performed by the current T2K and NOvA, or future DUNE and T2HK experiments.
\item $P_{\mu e}^{\rm LBL}$: long-baseline measurement of $P(\nu_\mu \to \nu_e)$ as performed by the current T2K/NOvA or future DUNE/T2HK experiments.
\item $P_{\mu \tau}^{\rm LBL}$: long-baseline measurement of $P(\nu_\mu \to \nu_\tau)$ as performed by the current OPERA experiment, upcoming IceCube results, or future DUNE experiment.
\end{itemize}
From the right panel of Fig.~\ref{fig:S12S13S23Analytic}, we see that  long-baseline experiments alone do not suffice to resolve the octant degeneracy, i.e., whether $\sin^2\theta_{23}$ is larger or smaller than $1/2$, as the green, red, and orange curves meet in two locations. Reactor data providing a precise measurement of $\sin^2\theta_{13}$ is needed to break the degeneracy. Also, we see that an infinitely precise measurement of the $\nu_\tau$ appearance probability is useful, but not necessary, to precisely determine the oscillation parameters. 

Figure~\ref{fig:S12S13S23Analytic} (right) allows for a test of the normalization of the third column of the LMM, i.e., whether $N_3 = 1$ when one relaxes the unitarity assumption. Referring to the right column of Table~\ref{tab:QuantitiesPMNSLMM}, given measurements from KamLAND, solar CC experiments, and Daya Bay, we see that $\ethr$ can be determined fairly robustly. Next, long-baseline $\nu_\mu \to \nu_e$ appearance allows us to measure $4\absq{U_{e3}} \absq{U_{\mu3}}/N_\mu^2$. Meanwhile, long-baseline $\nu_\mu \to \nu_\mu$ disappearance measures $4\absq{U_{\mu3}} (\absq{U_{\mu1}} + \absq{U_{\mu2}})/N_\mu^2$. If additional information (for instance from MINOS/MINOS+) on $N_\mu^2$ is obtained, this combination allows us to determine $\mthr$ precisely. Finally, long-baseline $\nu_\mu \to \nu_\tau$ appearance is sensitive to $4\absq{U_{\mu3}} \absq{U_{\tau3}}/N_\mu^2$, from which we can extract $\absq{U_{\tau3}}$. In tandem then, we can measure each of the elements $\absq{U_{\alpha 3}}$, which allows the placing of a constraint on $N_3$. Note that the measurement capability of $N_3$ will be mostly limited by one's measurement of long-baseline $\nu_\mu \to \nu_\tau$ appearance, as it is the least well-constrained of these measurements currently. See the discussion around Fig.~\ref{fig:S13S23_UV} for an illustration of this.
  
\begin{figure}[t]
\begin{center}
\includegraphics[width=0.75\linewidth]{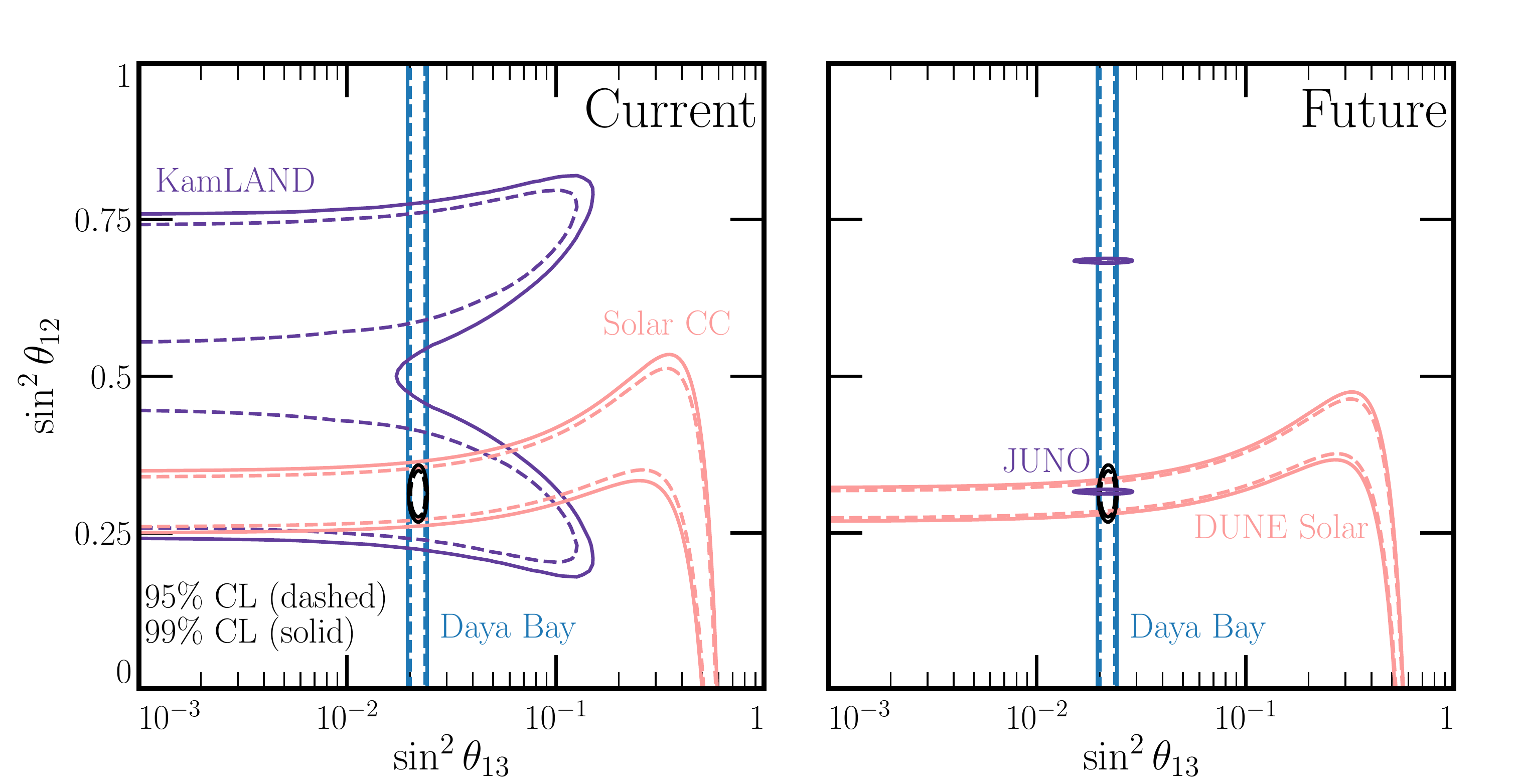}
\caption{Current (left) and projected (right) measurements of the mixing angles $\sin^2\theta_{13}$ and $\sin^2\theta_{12}$ at 95\% and 99\% CL. The black contours in both panels show the joint-fit region with current data.
\label{fig:T12T13}}
\end{center}
\end{figure}

\begin{figure}[t]
\begin{center}
\includegraphics[width=0.75\linewidth]{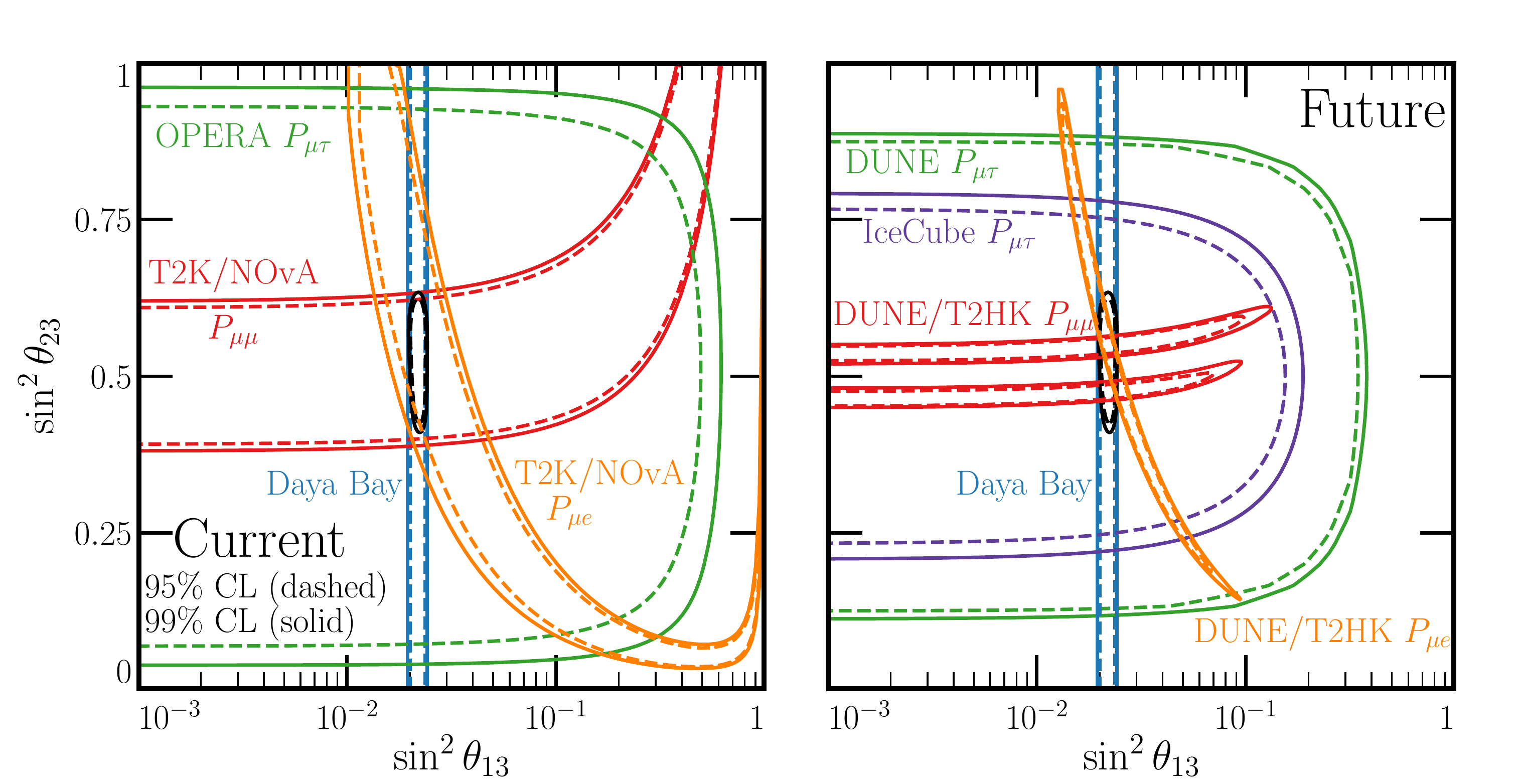}
\caption{Current (left) and projected (right) measurements of the mixing angles $\sin^2\theta_{23}$ and $\sin^2\theta_{13}$ at 95\% and 99\% CL. The black contours in both panels show the joint-fit region with current data.
\label{fig:T13T23}}
\end{center}
\end{figure}
 
We now analyze how well the combination $\sin^2\theta_{13}$ vs. $\sin^2\theta_{12}$ is measured by current experiments, as well as prospects for near-future experiments. This result is presented in Fig.~\ref{fig:T12T13}, where the left (right) panel demonstrates our current (expected future) knowledge of the two parameters. For easy comparison, the right panel also includes a joint fit of current data in black.

The shapes of the measurement regions in Fig.~\ref{fig:T12T13} match those in the left panel of Fig.~\ref{fig:S12S13S23Analytic}, as expected. We see that JUNO will significantly improve the precision on measuring $\sin^2\theta_{12}$. However, it will not measure $\sin^2\theta_{13}$ as precisely as Daya Bay. If we allow $\sin^2\theta_{12} > 1/2$, JUNO has an allowed region (analogous to KamLAND) that cannot be resolved by JUNO alone. In addition, DUNE will modestly improve on existing solar neutrino measurements.

Similarly, Fig.~\ref{fig:T13T23} demonstrates our knowledge of the combination $\sin^2\theta_{13}$ vs. $\sin^2\theta_{23}$. In the left panel, because the T2K and NOvA experiments both measure $\nu_\mu \to \nu_\mu$ disappearance and $\nu_\mu \to \nu_e$ appearance, and their measurements are qualitatively similar, we combine the two in each analysis, but separate by the two channels (all unseen parameters, including $\delta_{\rm CP}$, are marginalized over in these analyses). Like with the current set of measurements shown in the left panel of Fig.~\ref{fig:T12T13}, these measurements all agree at the $1\sigma$ level. NOvA and T2K both see modestly higher event rates than expected in $\nu_e$ appearance, driving the orange contours up and to the right relative to where the blue and red contours overlap, however, any tension is modest at best. DUNE and T2HK (like with NOvA and T2K, we combine these two experiments because of their similar sensitivity) will improve on the NOvA, T2K, and OPERA measurements in the right panel of Fig.~\ref{fig:T13T23}, leading to the tighter contours of the right panel. If non-unitarity is present, then these regions may not overlap at high CL, as they do today.

Finally, we revisit the point raised when discussing the right panel of Fig.~\ref{fig:S12S13S23Analytic}, regarding how one might use such a set of measurements to determine whether the LMM is unitary through their sensitivity to $N_3 \neq 1$. Here, we inject unitarity violation by making $\absq{U_{\tau3}}$ significantly larger than it should be, making $N_3 \approx 2$, beyond the edge of the currently allowed 3$\sigma$ range.  We then fit the simulated data using DUNE's $\nu_\tau$ appearance measurement while still assuming the mixing matrix is unitary. The resulting fit is shown in green in Fig.~\ref{fig:S13S23_UV} -- note the lack of overlap between the green (DUNE $P_{\mu\tau}$), orange (DUNE/T2HK $P_{\mu e}$), and red (DUNE/T2HK $P_{\mu\mu}$) measurements -- indicating the need for a more thorough test of unitarity violation. If such a discrepancy arises, IceCube's measurement of $4\absq{U_{\mu 3}}\absq{U_{\tau3}}/N_\mu^2$ could allow for verification of this arising from unitarity violation.

\begin{figure}
\begin{center}
\includegraphics[width=0.5\linewidth]{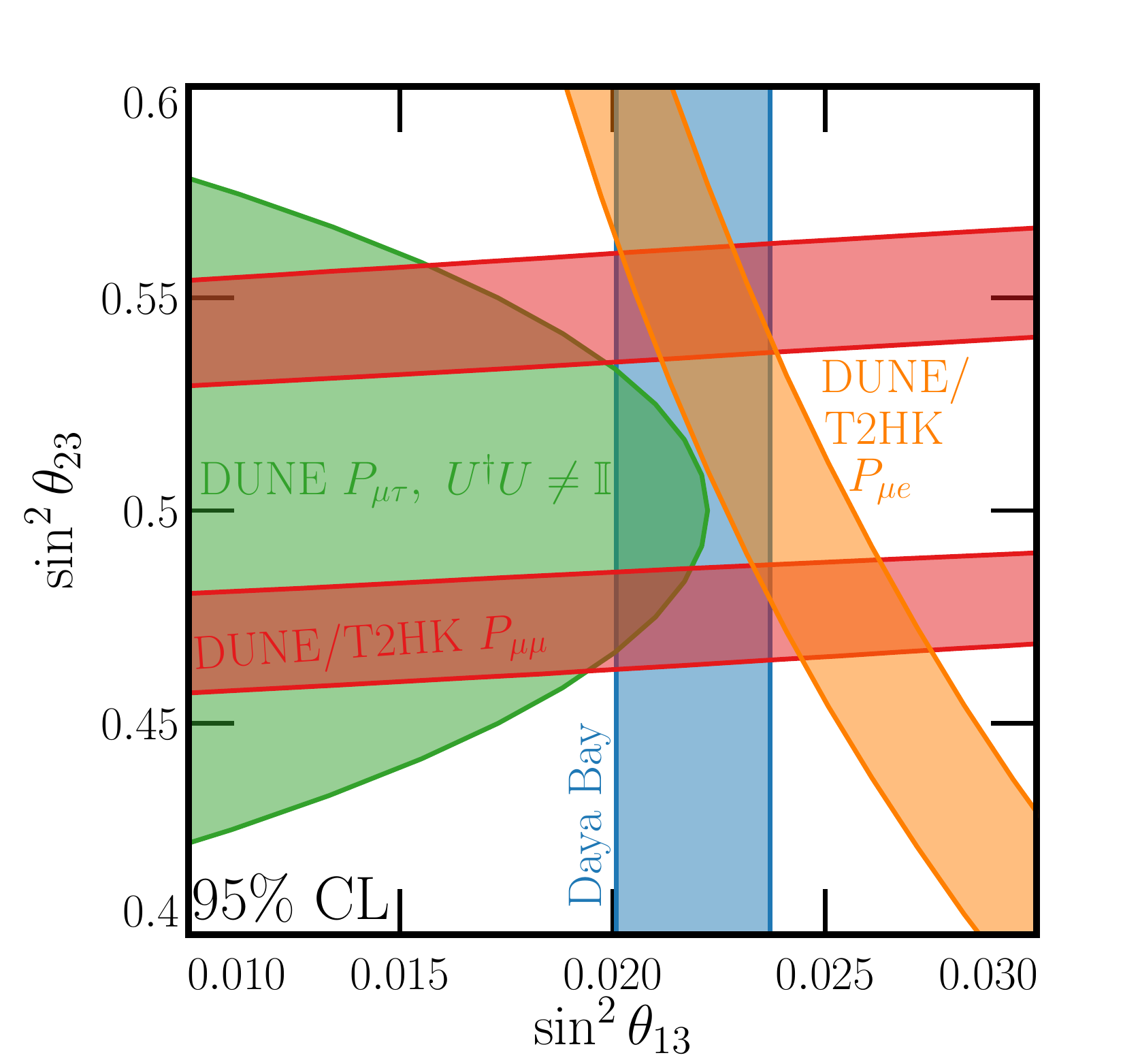}
\caption{Projected measurements of $\sin^2\theta_{13}$ vs. $\sin^2\theta_{23}$ when unitarity is violated ($N_3 \approx 2$). For DUNE's long-baseline measurement of $P_{\mu\tau}$ (green), we simulate data assuming the underlying mixing matrix is non-unitary, and extract the measurement of these parameters assuming the matrix is unitary.
\label{fig:S13S23_UV}}
\end{center}
\end{figure}

We caution the reader that the approach taken in Fig.~\ref{fig:S13S23_UV} was to illustrate how channel combinations test unitarity. However, this is not the most robust way of testing unitarity, as the sensitivities of different measurements to unitarity violation are not easily disentangled. Furthermore, this framework does not accommodate sterile neutrino searches. An alternate example of how to test unitarity when analyzing data in the PMNS paradigm can be found in Ref.~\cite{Ellis:2020ehi}, where it was demonstrated how unitarity triangles $\rho_{xy} + i\eta_{xy}$ can be used. In Ref.~\cite{Ellis:2020ehi}, we showed that, like here, separating analysis results by different oscillation channels can lead to inconsistent fits. In the following subsections, we carry out a global fit to the LMM that can directly test unitarity.


\subsection{Measurements of the Electron and Muon Rows}
\label{app:EMuRows}

Before looking at the global fit results of the LMM elements, we show the results for the absolute-value-squared of matrix elements in the electron and muon rows from a fit to certain subsets of the experiments we consider. Showing these subsets of experiments illustrates how combinations of different measurements affect our understanding of the various $\absq{U_{\alpha k}}$. This is performed for the electron and muon rows, where individual experimental measurements are only sensitive to those elements. We do not do this for the tau row, because there is not nearly as much experimental information for it, and the oscillation probability $P_{\tau\tau}$ has never been measured.

\begin{figure}[!ht]
\begin{center}
\includegraphics[width=0.49\linewidth]{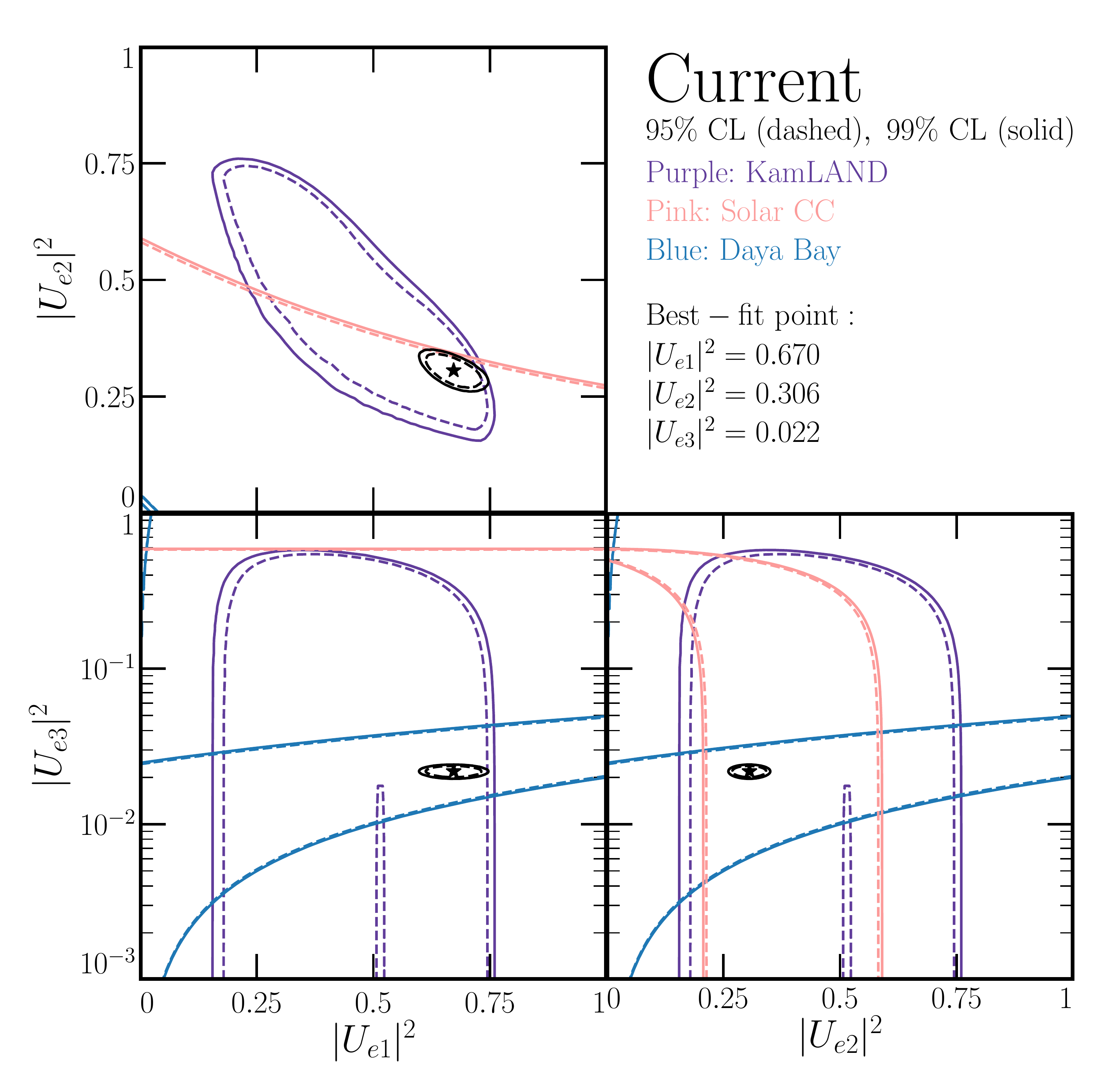}
\includegraphics[width=0.49\linewidth]{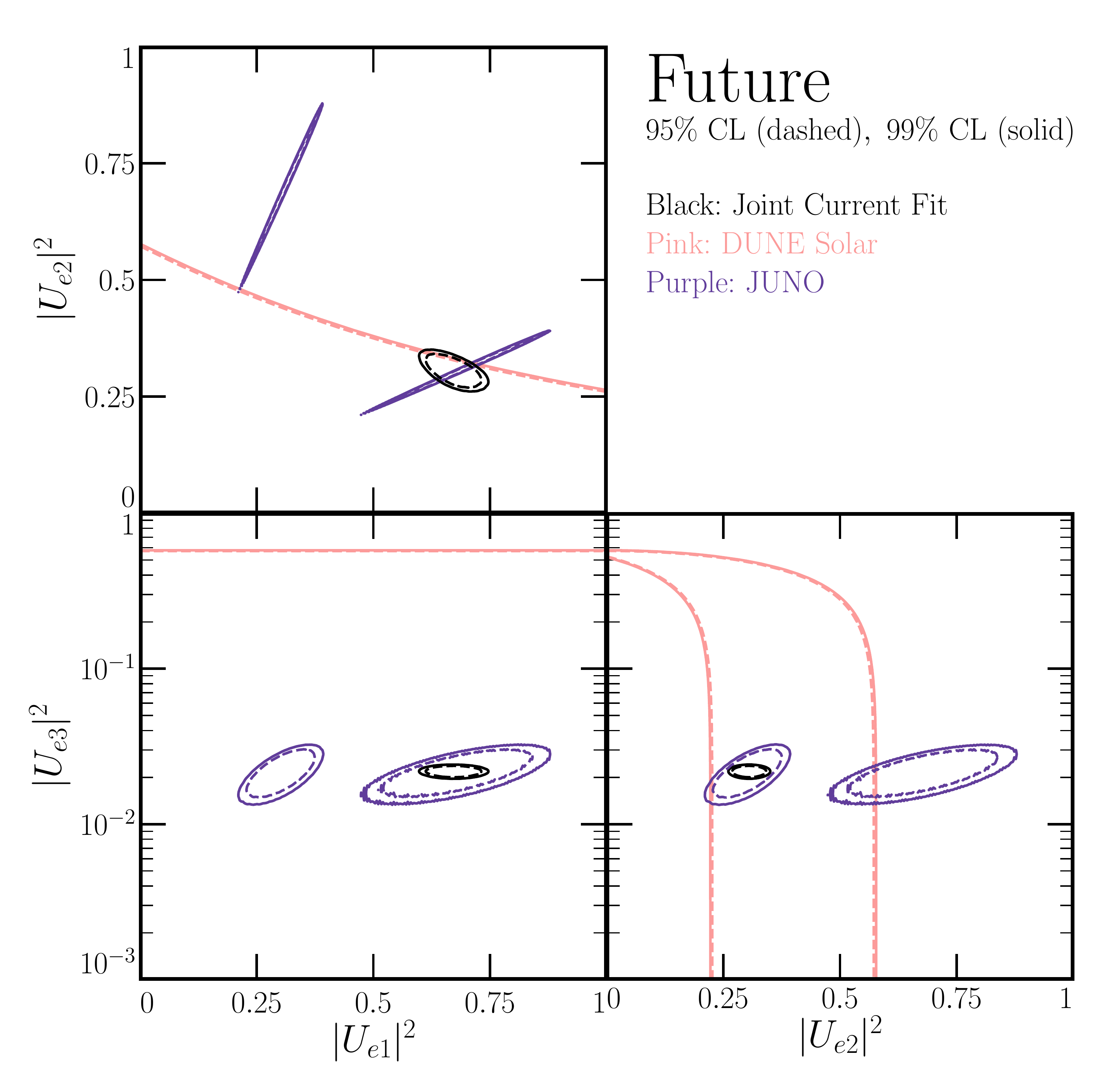}
\caption{Left: Current measurements coming from experiments that are only sensitive to electron-row parameters $\eone$, $\etwo$, and $\ethr$. Here we compare measurements from KamLAND, Solar CC, and Daya Bay, and a joint fit region (black). Note that the measurements are combined prior to marginalizing over any of the three parameters, hence the joint measurement appears stronger than expected. Right: Future projections of these parameter measurements by JUNO and DUNE's solar neutrino capabilities, compared against the joint fit of current data (black).
\label{fig:ERowCurrentFuture}}
\end{center}
\end{figure}

\textbf{Electron Row Only:} Figure~\ref{fig:ERowCurrentFuture} (left) displays the current knowledge of the electron row $|U_{ek}|^2$ for a subset of the existing experiments we discussed in Section~\ref{sec:CurrentExps}. Each panel displays two-dimensional projections of the test statistic $\Delta \chi^2$ for these CL, after marginalizing over the third, unseen parameter.

We show results individually from KamLAND, SNO and Super-K measurements of solar neutrinos from CC interactions, and Daya Bay, in addition to a joint fit to these three sets of results. Note that due to not marginalizing over any parameters before performing the joint fit, the resulting two-dimensional $\Delta \chi^2$ contours do not follow the na\"ive expectation as a result of degeneracies in the parameter space. For example, Daya Bay measures the combination $4\ethr(\eone + \etwo)/N_e^2$, which is degenerate under $\eone \leftrightarrow \etwo$, such that it appears that Daya Bay places no constraint in the upper-left panel of Fig.~\ref{fig:ERowCurrentFuture} (left), showing constraints in the $\eone-\etwo$ plane. However, Daya Bay constraints on $\ethr$ as a function of $\eone,~\etwo$ combine with solar measurements in the $\etwo-\ethr$ plane to bound $\etwo$ and $\eone$, such that the best-fit regions are the small black elliptical contours shown in the figure. Note that in this procedure the only constraint we impose is that we require each of the parameters satisfy $0 \leqslant |U_{ek}|^2 \leqslant 1$. If one were to impose the constraint that the sum not exceed one, as would apply in the sub-matrix case we discussed in Section~\ref{sec:LMM}, the upper-right triangular half of the $\absq{U_{e1}}$ vs. $\absq{U_{e2}}$ panel would be forbidden, somewhat limiting the KamLAND, solar CC, and joint fit contours. Compared to Fig.~1 in Ref.~\cite{Antusch:2006vwa}, the $U_{e1}$-$U_{e2}$ panel is similar, but with some differences due to how data sets are combined. The $U_{e3}$ panel is qualitatively different due to the addition of the Daya Bay measurement. The stars in Fig.~\ref{fig:ERowCurrentFuture} represent the best-fit point in this parameter space given by this combination of datasets: $\eone = 0.670$, $\etwo = 0.306$, and $\ethr = 0.022$. 

The upcoming JUNO experiment will measure a combination of these parameters as well, with the oscillation probability given by Eq.~\eqref{eq:Pdis}. JUNO will operate in the regime of $L/E_\nu$ in which the effects of both mass-squared splittings are relevant, and therefore, with enough statistical power, can be independently sensitive to each mixing element $|U_{e k}|^2$. JUNO's capacity to measure each of these three parameters is depicted in Fig.~\ref{fig:ERowCurrentFuture}, alongside the combined current measurement from Fig.~\ref{fig:ERowCurrentFuture} (left) in black. JUNO will precisely measure the combination $\etwo/\eone$ leading to the very sharp regions in the $\eone$ vs. $\etwo$ panel. When combined with the current fit region, this will lead to impressive measurements of both $\eone$ and $\etwo$.

On its own, JUNO suffers the same degeneracy discussed above for Daya Bay under the interchange $\absq{U_{e1}} \leftrightarrow \absq{U_{e2}}$, and requires solar neutrino experiments to break the degeneracy. However, this interchange $\absq{U_{e1}} \leftrightarrow \absq{U_{e2}}$ also requires changing the neutrino mass ordering (or the sign of $\Delta m_{31}^2$). If the mass ordering can be determined independently of JUNO at high enough significance (for instance, by DUNE, T2HK) then the solution where $\absq{U_{e2}} > \absq{U_{e1}}$ may be eliminated. The availability of redundant but independent data to select the right $(\eone, \etwo)$ is a powerful tool to test new physics scenarios such as possible non-standard interactions of neutrinos.

In Fig.~\ref{fig:ERowCurrentFuture} (right), we do not perform a joint analysis of current and future data, and simply note that once future data from JUNO are included, the measurements of $\absq{U_{e1}}$ and $\absq{U_{e2}}$ will be dominated by JUNO, whereas the measurements of $\absq{U_{e3}}$ will be dominated by current experiments, specifically Daya Bay. JUNO will measure $\absq{U_{e3}}$ at slightly worse precision than Daya Bay. This measurement, since it is performed at a significantly different baseline from Daya Bay, serves as an important cross-check.

\begin{figure}
\begin{center}
\includegraphics[width=0.49\linewidth]{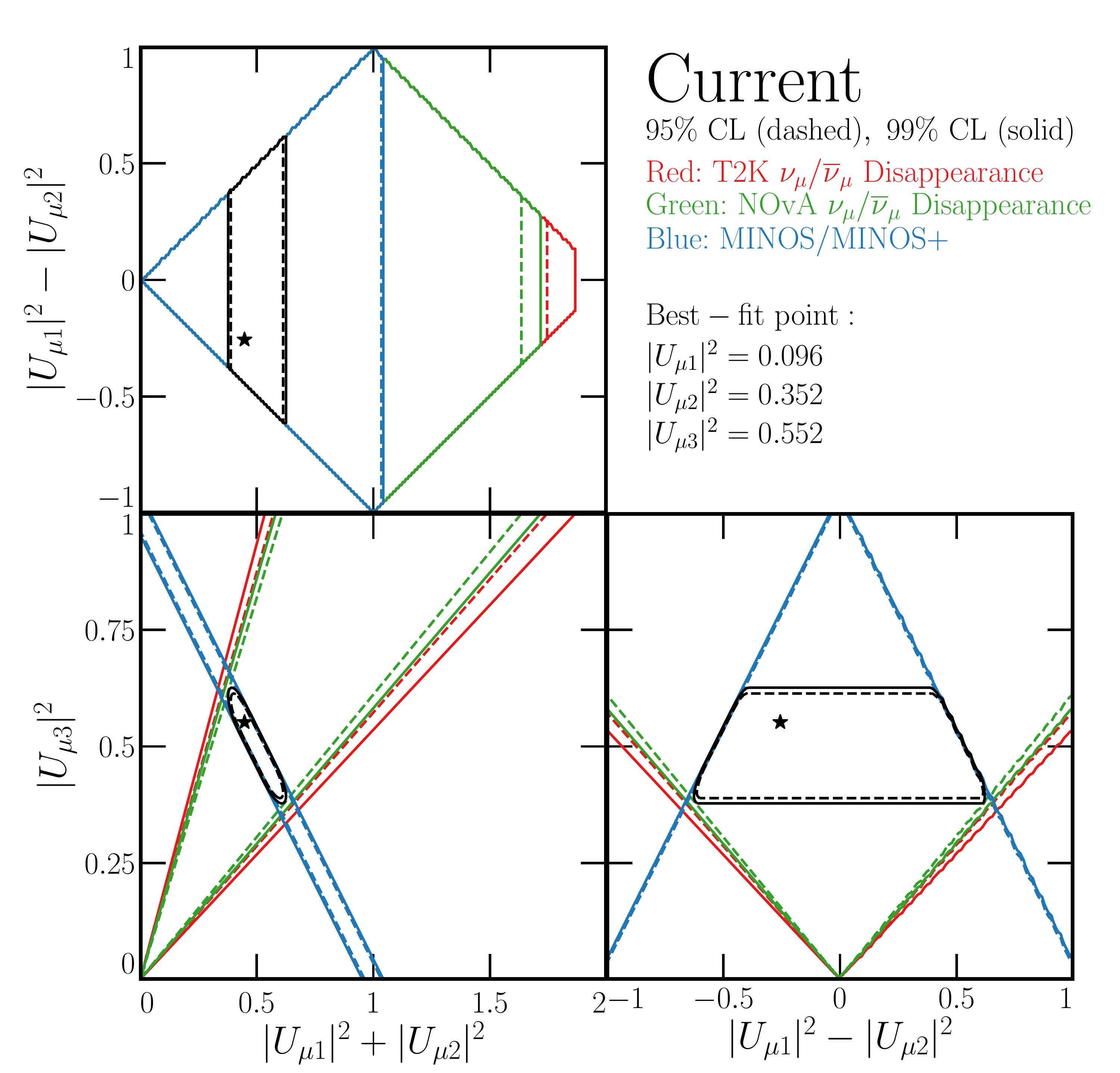}
\includegraphics[width=0.49\linewidth]{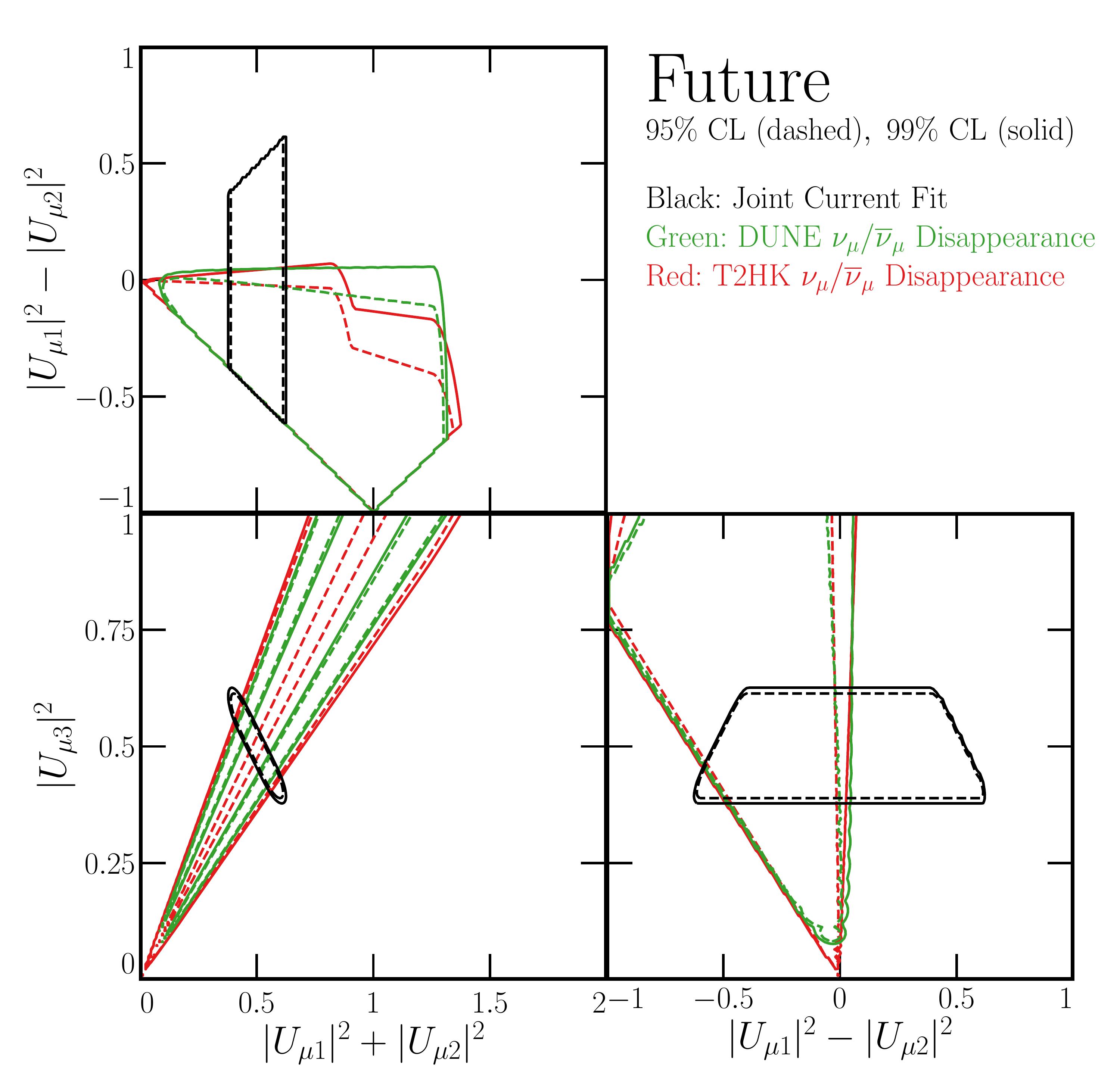}
\caption{Left: Current measurements coming from experiments only sensitive to the muon-row parameters $\mthr$ and the combinations $\mone \pm \mtwo$. Here we compare measurements from T2K $\nu_\mu$ and $\overline{\nu}_\mu$ disappearance, NOvA $\nu_\mu$ and $\overline{\nu}_\mu$ disappearance, and the MINOS/MINOS+ sterile neutrino search, as well as a combined fit of these (black). Right: Comparison between the current, joint fit (black) and the future DUNE and T2HK $\nu_\mu$ and $\overline{\nu}_\mu$ measurements of these parameters.
\label{fig:MuRowCurrentFuture}}
\end{center}
\end{figure}

\textbf{Muon Row Only:} We perform a similar procedure focusing on the elements $|U_{\mu k}|^2$ in Fig.~\ref{fig:MuRowCurrentFuture}. In the left panel of Fig.~\ref{fig:MuRowCurrentFuture}, we include the MINOS/MINOS+ sterile neutrino search, NOvA $\nu_\mu$ and $\overline{\nu}_\mu$ disappearance, and T2K $\nu_\mu$ and $\overline{\nu}_\mu$ disappearance. The combined fit of these data sets is shown in black. Instead of presenting these measurements in terms of the elements $\absq{U_{\mu 1}}$, $\absq{U_{\mu 2}}$, and $\absq{U_{\mu 3}}$, we present the measurements in terms of the combinations $\absq{U_{\mu1}} \pm \absq{U_{\mu2}}$ and $\absq{U_{\mu3}}$. This is because the combination $\absq{U_{\mu1}} + \absq{U_{\mu2}}$ is more precisely measured by MINOS/MINOS+, NOvA, and T2K, where their difference is not well-constrained by current data. Again, if we compare against the analogous parameter space in Ref.~\cite{Antusch:2006vwa}, we find overall consistency, with the stronger constraints from T2K, NOvA, and MINOS+ contributing to stronger measurements in the $\absq{U_{\mu 1}} + \absq{U_{\mu 2}}$ vs. $\absq{U_{\mu 3}}$ space than those in Ref.~\cite{Antusch:2006vwa}.

Again, parameter degeneracies cause the combined fit to appear stronger than na{\"i}ve expectations. We see that $\mthr$ is measured the most precisely of these three parameters, and the combination $\absq{U_{\mu1}} + \absq{U_{\mu2}}$ is well measured by the combination of datasets in the bottom-left panel. The difference $\absq{U_{\mu1}} - \absq{U_{\mu2}}$ is not well-measured, and is predominantly constrained by the requirement that both $\absq{U_{\mu1}}$ and $\absq{U_{\mu2}}$ are both between $0$ and $1$. This forces the difference $\absq{U_{\mu1}} - \absq{U_{\mu2}}$ to be less in magnitude than the sum $\absq{U_{\mu1}} + \absq{U_{\mu2}}$. The stars in Fig.~\ref{fig:MuRowCurrentFuture} (left) correspond to the best-fit points of these elements from the full analysis discussed in the main text: $\mone = 0.096$, $\mtwo = 0.352$, $\mthr = 0.552$.

Figure~\ref{fig:MuRowCurrentFuture} (right) displays our future projections on the muon row element measurements, where we compare the current joint fit (black) to projections of DUNE (green) and T2HK (red) $\nu_\mu$ and $\overline{\nu}_\mu$ disappearance.\footnote{In order to study how DUNE and T2HK muon-neutrino disappearance are sensitive to the muon row elements, and only the muon row elements, we perform this simulation assuming oscillations of $\nu_\mu$ occur in vacuum. This allows us to use the expression in Eq.~\eqref{eq:Pdis} for our calculations. We find this to be a good approximation for the muon neutrino/antineutrino channels.} DUNE and T2HK measure oscillations over a wider range of $L/E_\nu$ than their predecessors, thus they are sensitive to more than just the ``dominant'' term in the disappearance channel oscillation probability, namely the prefactor of $\sin^2\left(\Delta_{31}/2\right)$ in Eq.~\eqref{eq:Pdis}. Indeed, these experiments have some sensitivity to the interference term, namely the final term of Eq.~\eqref{eq:Pdis}. This allows DUNE and T2HK to be sensitive to $\absq{U_{\mu1}} - \absq{U_{\mu2}}$ unlike the current data considered in Fig.~\ref{fig:MuRowCurrentFuture} (left). We see that these experiments can both demonstrate $\absq{U_{\mu2}} > \absq{U_{\mu1}}$ at high significance.


\subsection{Joint Measurement of All Matrix-Elements-Squared}
\label{subsec:USq}

In this subsection, we present the current constraints on the parameterization-independent $\left \lvert U_{\alpha k}\right\rvert^2$, and project how well these will be constrained once future data from DUNE/T2HK, JUNO, and IceCube Upgrade are included. In Appendix~\ref{app:Phases}, we show how well the parameterization-dependent phases $\left\lbrace \phi_{e2}, \phi_{e3}, \phi_{\tau 2}, \phi_{\tau 3}\right\rbrace$ are and will be constrained. 

\begin{figure}
\centering
\includegraphics[width=\linewidth]{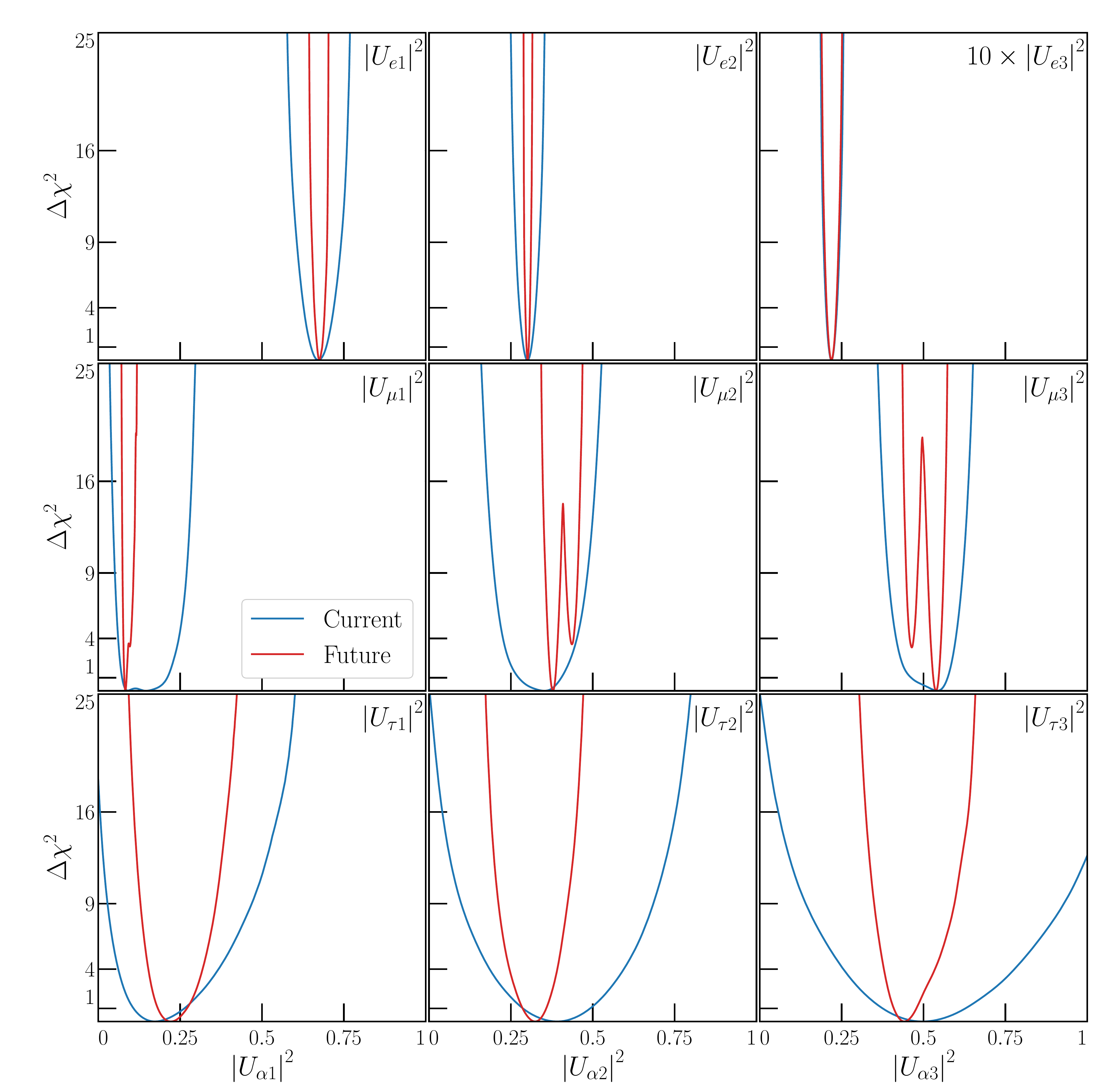}
\caption{Measurements (and projections) of the matrix-elements-squared $\left \lvert U_{\alpha k}\right\rvert^2$ of the LMM. These are one-dimensional $\Delta \chi^2$ measurements of each parameter after marginalizing a 15 parameter fit down to each individual one. Results are generated using {\sc pyMultiNest}~\cite{Feroz:2007kg,Feroz:2008xx,Feroz:2013hea,Buchner:2014nha}. All results here are obtained under the \textit{\textbf{agnostic}} assumption.
\label{fig:CurrentFutureUs}}
\end{figure}

Figure~\ref{fig:CurrentFutureUs} displays the individual measurements of $\left\lvert U_{\alpha k}\right\rvert^2$ including current data (blue) and current data with future data (red).\footnote{Note that we present the top-right axis in terms of $10\times \absq{U_{e3}}$ rather than $\absq{U_{e3}}$ for presentation purposes so that it can share axes with the $\absq{U_{\mu3}}$ and $\absq{U_{\tau3}}$ panels.} Each panel displays the one-dimensional $\Delta \chi^2$ measurement of the element after marginalizing the 15-parameter fit down to the individual element. Here, we define $\Delta \chi^2 = -2\Delta \mathcal{L}$, where $\mathcal{L}$ is the posterior likelihood obtained in our analysis. The improvement in $\eone$ and $\etwo$ is driven predominantly by JUNO, while the improvement in the muon row is driven by DUNE and T2HK $\nu_\mu \to \nu_\mu$ measurements.The tau row improvements are driven by $\nu_\mu \to \nu_\tau$ measurements in IceCube and DUNE. The IceCube measurement will reach a higher precision than DUNE because of the large systematic uncertainties on neutrino-nucleus cross sections at DUNE's beam energies.

Using these results, we can determine the allowed $3\sigma$ CR for each of the nine mixing-matrix-elements squared according to the current data, as well as projections to including future data. The allowed ranges for these are
\begin{align}
\left\lvert U_{\rm LMM}^{\rm Current} \right\rvert^2_{3\sigma} &= \left( \begin{array}{c c c} \left[0.606, 0.742\right] & \left[0.265, 0.337\right] & \left[0.020, 0.024\right] \\ \left[0.051, 0.270\right] & \left[0.198, 0.484\right] & \left[0.392, 0.620\right] \\ \left[0.028, 0.469\right] & \left[0.098, 0.685\right] & \left[0.140, 0.929\right] \end{array}\right), \\
\left\lvert U_{\rm LMM}^{\rm Future} \right\rvert^2_{3\sigma} &= \left( \begin{array}{c c c} \left[0.653, 0.699\right] & \left[0.291, 0.311\right] & \left[0.020, 0.024\right] \\ \left[0.074, 0.108\right] & \left[0.355, 0.454\right] & \left[0.447, 0.561\right] \\ \left[0.129, 0.359\right] & \left[0.212, 0.423\right] & \left[0.349, 0.595\right] \end{array}\right).
\end{align}

We can interpret the amount by which each of these measurements will improve, at the $3\sigma$ level, by computing the reduction in size of the allowed $3\sigma$ range of each $\absq{U_{\alpha k}}$, $\Delta \absq{U_{\alpha k}^{\rm Current}}/\Delta \absq{U_{\alpha k}^{\rm Future}}$:
\begin{align}
\mathrm{Improvement\ Factor:}\ \left(\begin{array}{c c c} 3.0 & 3.6 & 1.0 \\ 6.4 & 2.9 & 2.0 \\ 1.9 & 2.8 & 3.2 \end{array}\right).
\end{align}
As is evident in Fig.~\ref{fig:CurrentFutureUs}, the improvement is noticeable especially for the elements $\absq{U_{e1}}$, $\absq{U_{e2}}$, $\absq{U_{\mu1}}$, and the $\tau$ row elements. This analysis is performed under the agnostic case -- we compare these results with those obtained under the sub-matrix case in Section~\ref{sec:AgnosticSterile}.


\subsection{Constraining the Normalization and Closure Conditions with Current and Future Data}
\label{subsec:NormClos}
In this subsection, we check the consistency of data with the requisite conditions to determine whether the LMM is unitary. Specifically, we measure the row/column normalizations $N_\alpha$ and $N_k$ and triangle closures $t_{\alpha\beta}$ (between two rows) and $t_{kl}$ (between two columns), using the same analyses as in the previous subsection.

\begin{figure*}
\begin{center}
\includegraphics[width=0.49\linewidth]{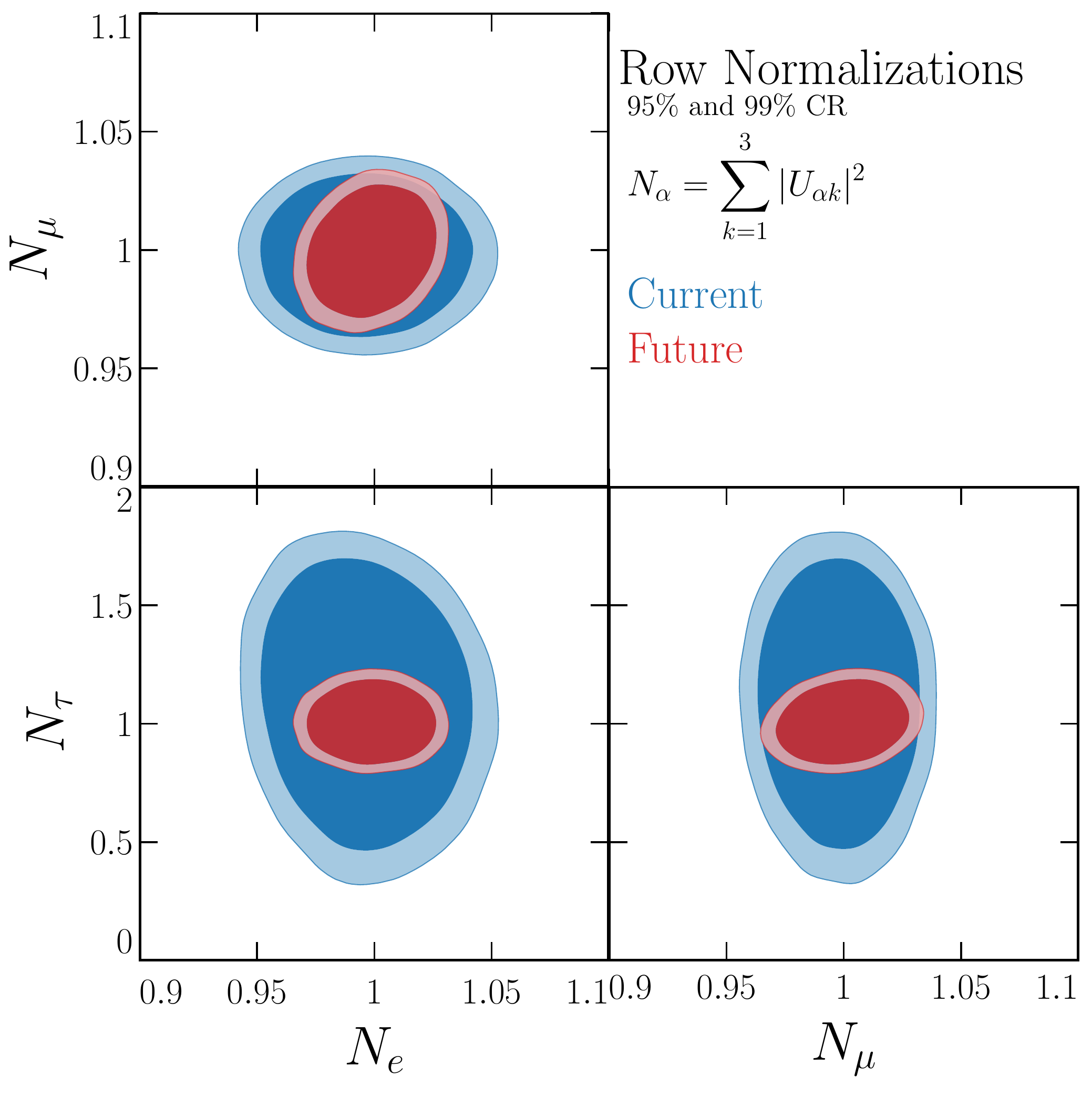}
\includegraphics[width=0.49\linewidth]{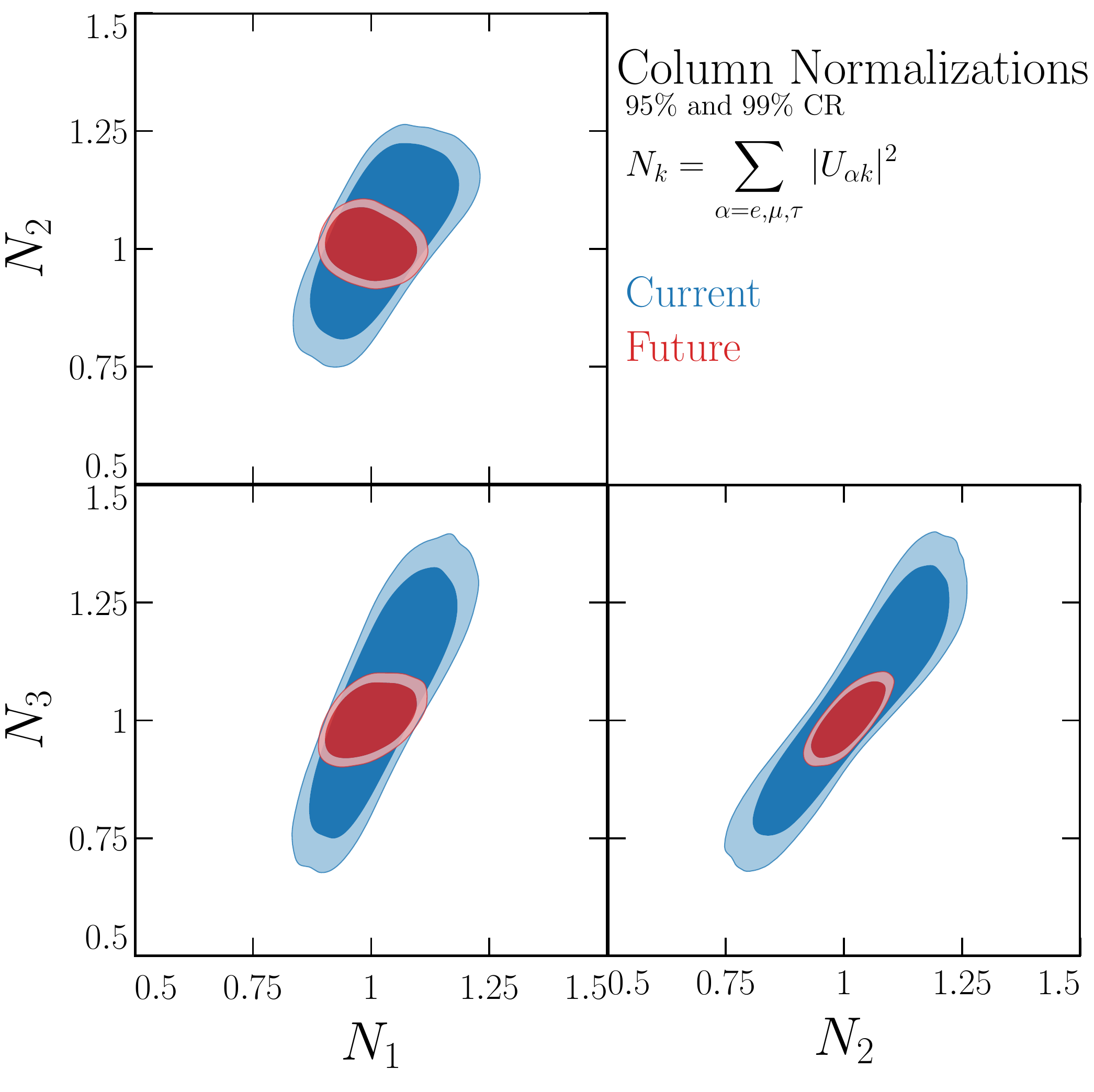}
\caption{Left: constraints (and projected constraints) on the row normalizations $N_e$, $N_\mu$, and $N_\tau$ at 95\% (dark) and 99\% (faint colors) credibility. Right: constraints on the column normalizations $N_1$, $N_2$, and $N_3$ at 95\% (dark) and 99\% (faint colors) credibility. All results here are obtained under the \textit{\textbf{agnostic}} assumption.
\label{fig:RowColNorms}}
\end{center}
\end{figure*}

The left panel of Fig.~\ref{fig:RowColNorms} displays the results of this analysis, projecting down to two-dimensional CR measuring the row normalizations $N_e$, $N_\mu$, and $N_\tau$ at 95\% and 99\% credibility. We see that the analysis of all current data is consistent with unitarity for these values. Future data will lead to a modest improvement in the constraint on $N_\mu$, some improvement in $N_e$, and significant improvement in $N_\tau$. 

Similarly, the right panel of Fig.~\ref{fig:RowColNorms} presents the current constraints, as well as projected future ones, on the column normalizations $N_1$, $N_2$, and $N_2$, at 95\% and 99\% credibility. The correlation between measurements of each pair of column normalizations is due to the fact that these constraints are limited by the measurement of the tau-row elements, $\left\lvert U_{\tau k}\right\rvert^2$. Future data will improve the constraint on each column normalization by a factor of roughly 3.

Table~\ref{tab:NormSummary} summarizes the current and expected future measurements of the row and column normalizations of the LMM. Here, we give the current best-fit (maximum likelihood point) value of each normalization, as well as the extents of its current $3\sigma$ CR. We also show the projected future $3\sigma$ CR, assuming a true value of $N_{X} = 1$, demonstrating the improvement attributable to future data. Our projected constraint on $N_e$ is 1.1\%, consistent with the official JUNO analysis, which reports a 1.2\% constraint on $N_e$.

\begin{table}
\begin{center}
\caption{Summary of current and expected future constraints on the row ($N_\alpha$) and column ($N_k$) normalizations, under the \textit{\textbf{agnostic}} assumption.
\label{tab:NormSummary}}
\begin{tabular}{|c||c|c|c|}\hline
 & Best-fit (current) & $3\sigma$ (current) & $3\sigma$ (future) \\ \hline \hline
 $N_e$ & $1.00$ & $[0.94,1.05]$ & $[0.97, 1.03]$ \\ \hline
 $N_\mu$ & $0.99$ & $[0.96,1.04]$ & $[0.96, 1.03]$ \\ \hline
 $N_\tau$ & $1.12$. & $[0.32,1.82]$ & $[0.79, 1.23]$ \\ \hline\hline
 $N_1$ & $1.01$ & $[0.84,1.22]$ & $[0.89, 1.12]$ \\ \hline
 $N_2$ & $1.05$ & $[0.75,1.27]$ & $[0.92, 1.10]$ \\ \hline
 $N_3$ & $1.05$ & $[0.67,1.40]$ & $[0.90, 1.10]$ \\ \hline
\end{tabular}
\end{center}
\end{table}

\begin{figure}[!htbp]
\begin{center}
\includegraphics[width=\linewidth]{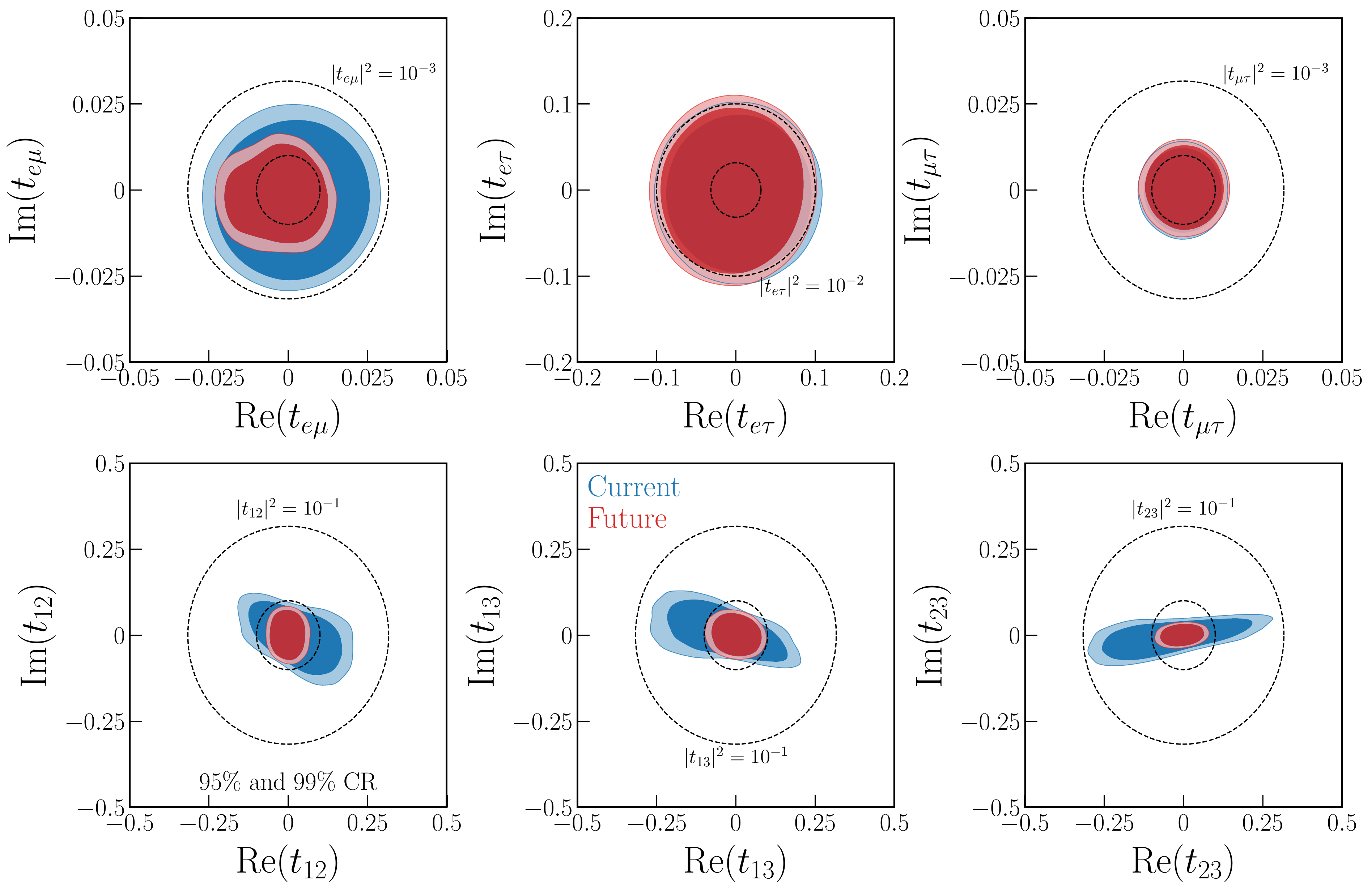}
\caption{Constraints (and projected constraints)  on the real ($x$-axes) and imaginary ($y$-axes) parts of the closures of the six unitarity triangles. Dashed circles indicate contours corresponding to fixed $\absq{t_{xy}}$, with the outer one in each panel as labeled. The inner dashed circles are an order of magnitude smaller $\absq{t_{xy}}$ than the outer ones. Here we analyze data under the \textit{\textbf{agnostic}} assumption.
\label{fig:Closures:CurrentFuture}}
\end{center}
\end{figure}

Figure~\ref{fig:Closures:CurrentFuture} presents the results on the closures of different triangles $t_{\alpha\beta}$ and $t_{kl}$. Each panel in this figure presents constraints on the real and imaginary part of $t_{\alpha\beta}$ (top row) or $t_{kl}$ (bottom row) at $95\%$ credibility (dark) and $99\%$ credibility (faint). We draw circles corresponding to constant values of the magnitude of $\left\lvert t_{\alpha\beta}\right\rvert^2$ and $\left\lvert t_{kl}\right\rvert^2$ as labeled, where each successive inward circle is an order of magnitude smaller. When constraining $t_{e\tau}$ and $t_{\mu\tau}$, the expected future constraints are nearly degenerate with the current ones -- constraints here are dominated by the sterile neutrino searches discussed in Section~\ref{sec:CurrentExps}, specifically the NOMAD and CHORUS results discussed in Table~\ref{tab:Steriles}. Constraints on $t_{\mu e}$ will improve modestly once information from DUNE and JUNO are incorporated. In contrast, measurements of the closures of the different pairs of columns will improve significantly with future data. Currently, each of these can be constrained $\left\lvert t_{kl}\right\rvert^2 \lesssim 10^{-1}$ at $95\%$ credibility. With future data, this will improve to roughly $\left\lvert t_{kl}\right\rvert \lesssim 10^{-2}$ for each of the three triangles. We summarize the current and future $3\sigma$ credibility upper limits on the triangle closures in Table~\ref{tab:ClosureSummary}.

\begin{table}
\begin{center}
\caption{Summary of current and expected future constraints on the row closures $\left\lvert t_{\alpha\beta}\right\rvert$ and column closures $\left\lvert t_{kl}\right\rvert$, under the \textbf{\textit{agnostic}} assumption.
\label{tab:ClosureSummary}}
\begin{tabular}{|c||c|c|}\hline
& Current $3\sigma$ Upper Limit & Future $3\sigma$ Upper Limit \\ \hline \hline
$\left\lvert t_{e\mu}\right\rvert$ & $3.2 \times 10^{-2}$ & $2.5 \times 10^{-2}$ \\ \hline
$\left\lvert t_{e\tau}\right\rvert$ & $1.3 \times 10^{-1}$ & No Improvement \\ \hline
$\left\lvert t_{\mu\tau}\right\rvert$ & $1.6 \times 10^{-2}$ & No Improvement \\ \hline \hline
$\left\lvert t_{12}\right\rvert$ & $2.5\times 10^{-1}$ & $1.0 \times 10^{-1}$ \\ \hline
$\left\lvert t_{13}\right\rvert$ & $3.2 \times 10^{-1}$ & $1.2 \times 10^{-1}$ \\ \hline
$\left\lvert t_{23}\right\rvert$ & $3.3\times 10^{-1}$ & $1.1 \times 10^{-1}$ \\ \hline\end{tabular}
\end{center}
\end{table}

The analysis yielding Figs.~\ref{fig:RowColNorms} and~\ref{fig:Closures:CurrentFuture} was conducted assuming the agnostic case of Section~\ref{sec:Parameterizations}, whereby the matrix of which the LMM is a sub-matrix need not be unitary. The sub-matrix approach was taken in Ref.~\cite{Parke:2015goa}, where it was pointed out that assuming unitarity of the larger matrix leads to strong constraints from Cauchy-Schwarz inequalities. By remaining agnostic about the larger matrix, the improved measurement capability of future data is more apparent. An analysis assuming the larger matrix is unitary is contained in Section~\ref{sec:AgnosticSterile}.


\section{Secondary Results with Alternate Assumptions}
\label{sec:AltResults}

As discussed throughout this work, different assumptions regarding the origin of unitarity violation, as well as which datasets are included in the analysis, can have significant impact on the resulting constraints on the unitarity of the LMM. The primary results of our work, where we analyzed all possible data under the agnostic case, were shown in Section~\ref{sec:Results}. In this section, we explore two alternate assumptions. In Section~\ref{appendix:SterileSearches}, we repeat our analysis without including any short-baseline sterile neutrino searches (discussed in Section~\ref{subsec:SterileSearch} and Table~\ref{tab:Steriles}). In Section~\ref{sec:AgnosticSterile}, we conduct an analysis in the sub-matrix case of Section~\ref{sec:Parameterizations}, comparing the results with those obtained in the agnostic case presented above.

\subsection{Impact of Short-Baseline Sterile Neutrino Searches}
\label{appendix:SterileSearches}

In the bulk of the analyses performed in our work, we have included results of short-baseline sterile neutrino searches, with results adapted from these sterile neutrino searches reinterpreted as limits on unitarity violation (see Table~\ref{tab:Steriles} for a summary of these results). To better understand how unitarity constraints rely on sterile neutrino searches, we repeat the analyses of the main text surrounding Fig.~\ref{fig:Closures:CurrentFuture} without short-baseline results.

\begin{figure}
\begin{center}
\includegraphics[width=0.75\linewidth]{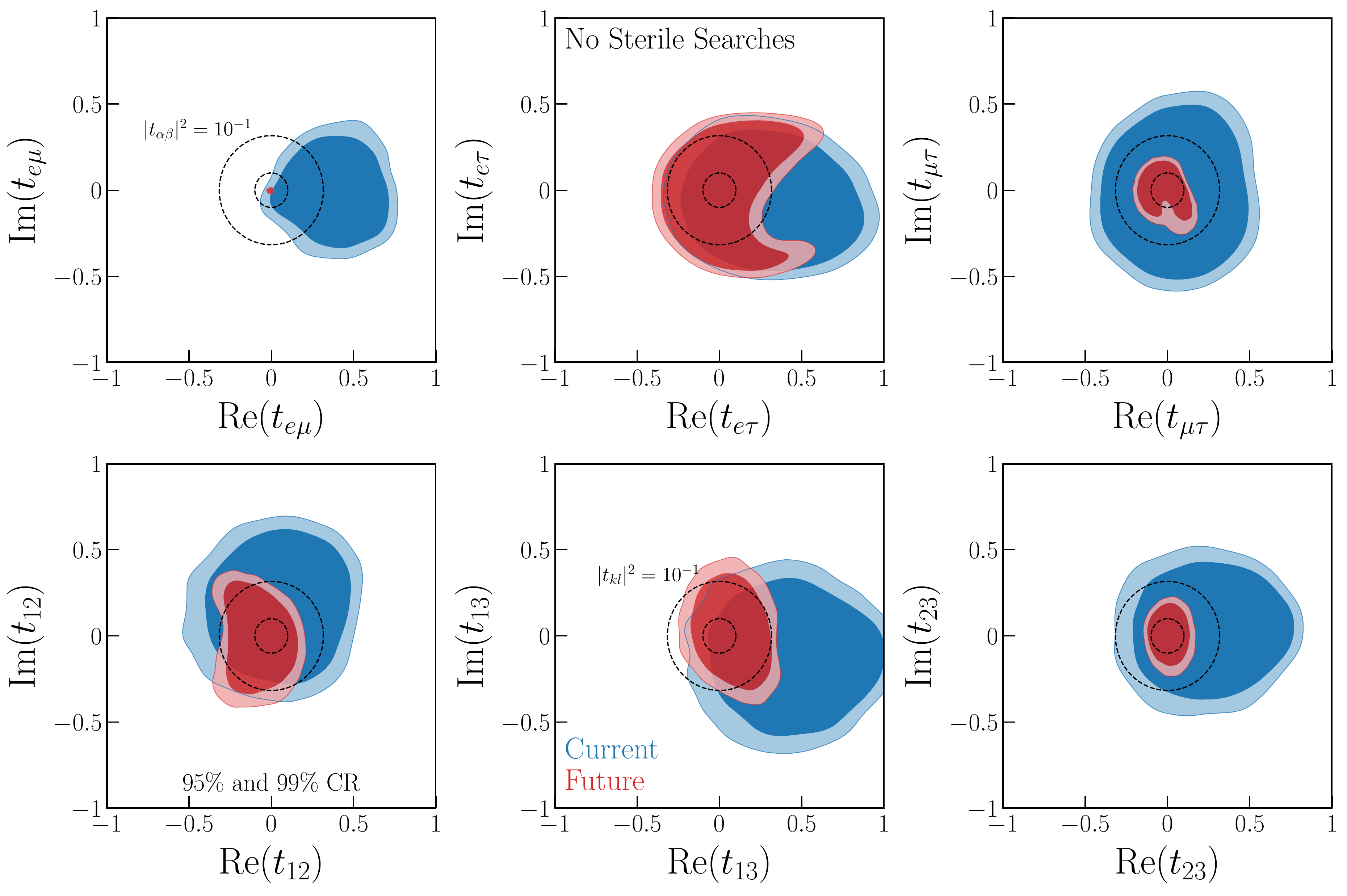}
\caption{Constraints (and projected constraints) on the closures of the six unitarity triangles excluding data from short-baseline sterile neutrino searches. In each panel, the outer (inner) dashed circle corresponds to constant $\absq{t_{\alpha\beta}} = 10^{-1}$ ($10^{-2}$) for the row/column closures.
\label{fig:Closures_NoSterileSearches}}
\end{center}
\end{figure}

Figure~\ref{fig:Closures_NoSterileSearches} shows the results. Here we note that the ranges on each of the panels in Fig.~\ref{fig:Closures_NoSterileSearches} measuring $t_{\alpha\beta}$ and $t_{kl}$ are much larger than the corresponding ranges in Fig.~\ref{fig:Closures:CurrentFuture}. However, it is apparent that in the absence of sterile searches, future data from IceCube, DUNE, JUNO, and T2HK would nevertheless allow us to understand the closure of all triangles of the LMM considerably better than current data allow. As in Fig.~\ref{fig:Closures:CurrentFuture}, we draw lines of constant $\absq{t_{\alpha\beta}}$ and $\absq{t_{kl}} = 10^{-1}$ and $10^{-2}$ in each panel, where the outer (inner) dashed line corresponds to a constant $10^{-1}$ ($10^{-2}$) in these planes.

Table~\ref{tab:ClosureSummaryNS} summarizes the numerical results. Comparing Tables~\ref{tab:ClosureSummary} and~\ref{tab:ClosureSummaryNS}, the improvement in the absence of sterile searches is much more dramatic, highlighting the importance of such experiments. We have not included any additional short-baseline searches in our future projections, as we do not expect any upcoming experiments to provide stronger sensitivity in the ``averaged-out'' regime~\cite{Berryman:2015nua,Kelly:2017kch,Machado:2019oxb,Ghosh:2019zvl} (as discussed in Section~\ref{subsec:SterileSearch}) than those summarized in Table~\ref{tab:Steriles}.

Comparing the measurements of the individual matrix-elements-squared $\absq{U_{\alpha k}}$, as well as the row and column normalization conditions $N_\alpha$ and $N_k$, is difficult in this scenario. Short-baseline sterile neutrino searches, particularly the information from $\nu_\tau$ appearance that $\absq{t_{e\tau}}$ and $\absq{t_{\mu\tau}}$ are small, provide significant information on the elements $\absq{U_{\tau k}}$. Additionally, the constraint from MINOS/MINOS+ that $N_\mu \approx 1$ is very important for determining the muon elements $\absq{U_{\mu k}}$. If this information is discarded, every other probe of $\absq{U_{\mu k}}$ we consider is subject to a rescaling degeneracy. This is a direct result of the discussion of normalization effects throughout Section~\ref{sec:LMM}. Again, this highlights the importance of short-baseline sterile neutrino searches, such as MINOS/MINOS+ $\nu_\mu$ disappearance, for precise tests of leptonic unitarity. This analysis without short-baseline measurements results in a lower limit on $N_\mu$ comparable to the one given in Table~\ref{tab:NormSummary}, however, $N_\mu$ can be as large as $\approx 2$.

\begin{table}
\begin{center}
\caption{Summary of current and expected future constraints on the row closures $\left\lvert t_{\alpha\beta}\right\rvert$ and column closures $\left\lvert t_{kl}\right\rvert$, under the \textbf{\textit{agnostic}} case regarding the LMM, when short-baseline sterile neutrino search results are \textbf{\textit{not}} included.
\label{tab:ClosureSummaryNS}}
\begin{tabular}{|c||c|c|}\hline
& Current $3\sigma$ Upper Limit & Future $3\sigma$ Upper Limit \\ \hline \hline
$\left\lvert t_{e\mu}\right\rvert$ & $8.2 \times 10^{-1}$ & $3.0 \times 10^{-2}$ \\ \hline
$\left\lvert t_{e\tau}\right\rvert$ & $1.0 \times 10^{0}$ & $8.1 \times 10^{-1}$ \\ \hline
$\left\lvert t_{\mu\tau}\right\rvert$ & $6.8 \times 10^{-1}$ & $3.2 \times 10^{-1}$ \\ \hline \hline
$\left\lvert t_{12}\right\rvert$ & $8.2\times 10^{-1}$ & $5.5 \times 10^{-1}$ \\ \hline
$\left\lvert t_{13}\right\rvert$ & $1.1 \times 10^{0}$ & $5.2 \times 10^{-1}$ \\ \hline
$\left\lvert t_{23}\right\rvert$ & $8.7\times 10^{-1}$ & $2.7 \times 10^{-1}$ \\ \hline\end{tabular}
\end{center}
\end{table}


\subsection{Impact of the Sub-matrix Case Assumption}
\label{sec:AgnosticSterile}

As discussed in Section~\ref{sec:Parameterizations}, the sub-matrix case places certain Cauchy-Schwarz restrictions on the elements of $U_{\rm LMM}$, specifically requiring
\begin{align}
N_\alpha &\leqslant 1,\ \quad N_k \leqslant 1, \\
\absq{t_{\alpha\beta}} &\leqslant \left( 1 - N_\alpha\right) \left(1 - N_\beta\right), \\
\absq{t_{kl}} &\leqslant \left(1 - N_k\right) \left(1 - N_l\right).
\end{align}

\begin{figure}
\centering
\includegraphics[width=0.75\linewidth]{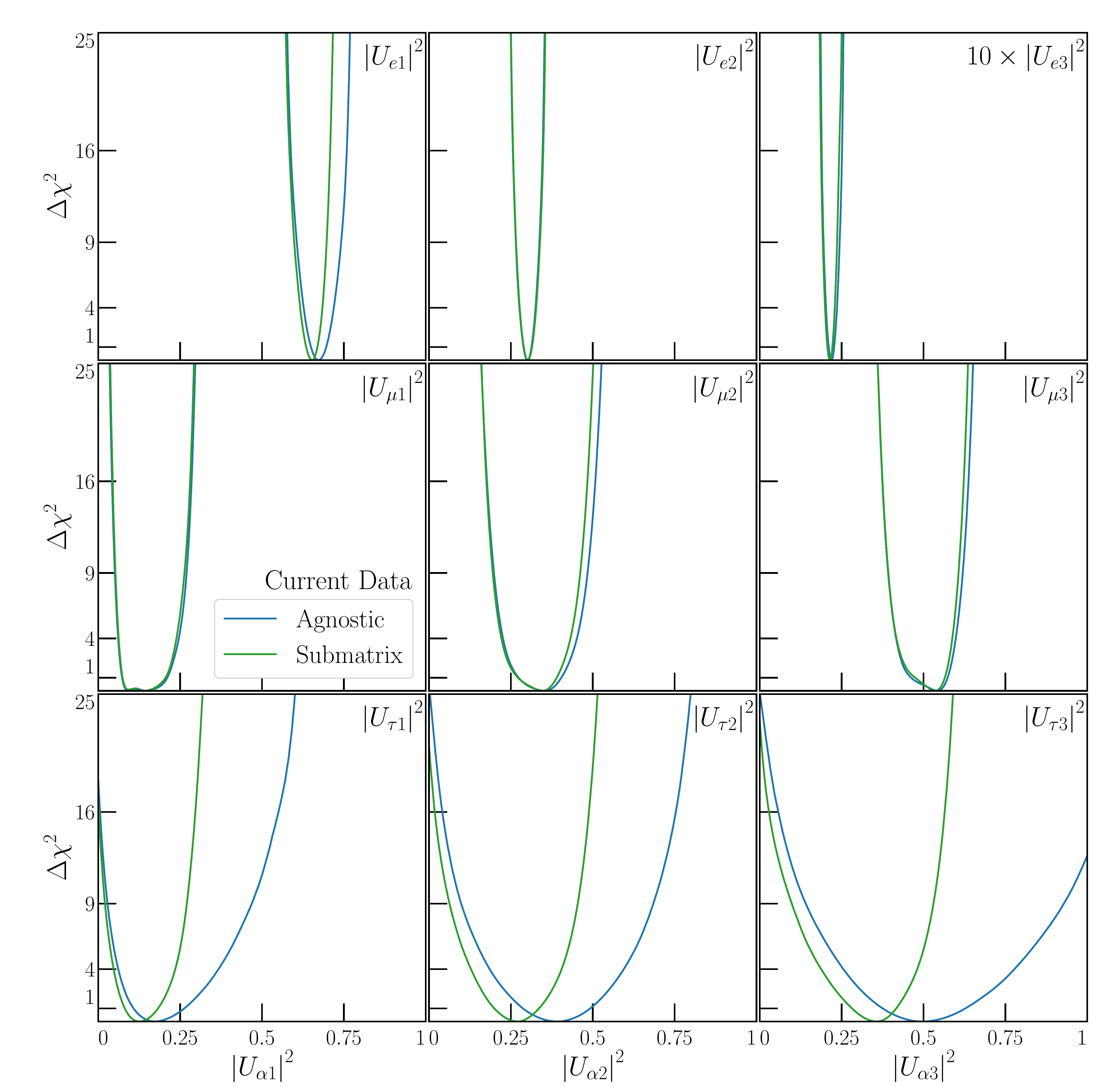}
\caption{Measurements (and projections) of $\left \lvert U_{\alpha k}\right\rvert^2$ with current experimental data under the \textbf{\textit{agnostic}} assumption (blue) and the \textbf{\textit{sub-matrix}} case (green). Similar to Fig.~\ref{fig:CurrentFutureUs}, these are one-dimensional $\Delta \chi^2$ measurements of each parameter after marginalizing a 15 parameter fit down to each individual one.
\label{fig:UsAgnosticSterile}}
\end{figure}

For our main analyses in Sections~\ref{subsec:USq} and~\ref{subsec:NormClos}, we assumed the agnostic case. In this subsection, we compare the results analyzed under the agnostic and sub-matrix hypotheses. First, we repeat the process that generates Fig.~\ref{fig:CurrentFutureUs} for the current data analyzed, under the two different case assumptions. The results of this procedure are shown in Fig.~\ref{fig:UsAgnosticSterile}.

We note several effects of the sub-matrix case in Fig.~\ref{fig:UsAgnosticSterile}. First, the measurement of the electron row elements is largely unchanged -- this is because the combination of KamLAND, solar neutrino measurements, and Daya Bay measure these elements very precisely regardless of the sub-matrix or agnostic assumptions, as discussed in Sec.~\ref{app:EMuRows}. We note that the measurement $\absq{U_{e1}}$ results in a slightly lower preferred value under the sub-matrix hypothesis than the agnostic one -- this arises from the Bayesian approach where we place a prior forbidding $N_e > 1$ in the sub-matrix analysis. In looking at the muon row elements, we see marginal improvement in the measurement capability in the sub-matrix case than the agnostic one. This is again unsurprising, as restricting $N_\mu \leqslant 1$ allows for improved measurements of these parameters. Finally, the largest difference between the two cases is in the tau row elements. Under the agnostic assumption, due to the mild excess over expectation of OPERA $\nu_\tau$ appearance events, the current data prefer $N_\tau > 1$ (at very low significance). When we analyze the data under the sub-matrix assumption, this solution is forbidden, and the $\absq{U_{\tau k}}$ elements are both preferred to be lower in magnitude, and we end up with smaller measurement regions. 

In Section~\ref{subsec:NormClos}, we analyzed how current and future data can constrain the normalizations of the LMM rows and columns to be close to 1, asking how well we measure $N_\alpha$ and $N_k$. If we were to repeat this analysis under the sub-matrix assumption, the resulting figure analogous to Fig.~\ref{fig:RowColNorms} would look very similar, modulo the regions $N_\alpha > 1$ and $N_k > 1$ being forbidden.

\begin{figure}[!ht]
\centering
\includegraphics[width=0.75\linewidth]{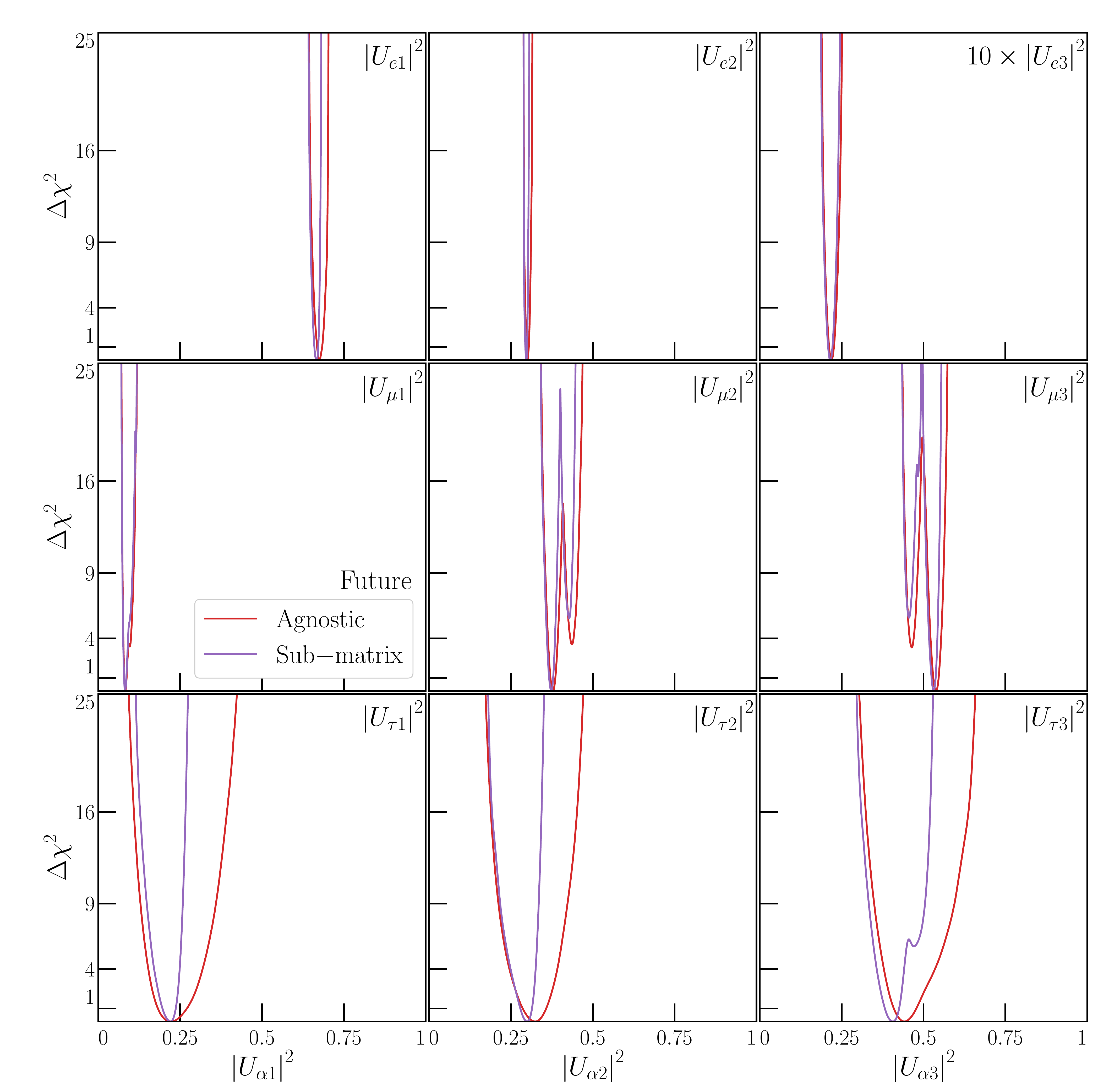}
\caption{Projections of future measurements of $\left \lvert U_{\alpha k}\right\rvert^2$ under the \textbf{\textit{agnostic}} case assumption (red) and the \textbf{\textit{sub-matrix}} case (purple). Similar to Figs.~\ref{fig:CurrentFutureUs} and~\ref{fig:UsAgnosticSterile}, these are one-dimensional $\Delta \chi^2$ measurements of each parameter after marginalizing a 15 parameter fit down to each individual one.
\label{fig:UsAgnosticSterileFuture}}
\end{figure}

For completeness, Fig.~\ref{fig:UsAgnosticSterileFuture} presents the projected future measurements of $\absq{U_{\alpha k}}$ under the agnostic (red) and sub-matrix (purple) assumptions. The red lines are identical to those in Fig.~\ref{fig:CurrentFutureUs}. We see that the sub-matrix hypothesis improves constraints on $\absq{U_{\mu k}}$ somewhat, and $\absq{U_{\tau k}}$ significantly.

\begin{figure}[!htbp]
\begin{center}
\includegraphics[width=0.75\linewidth]{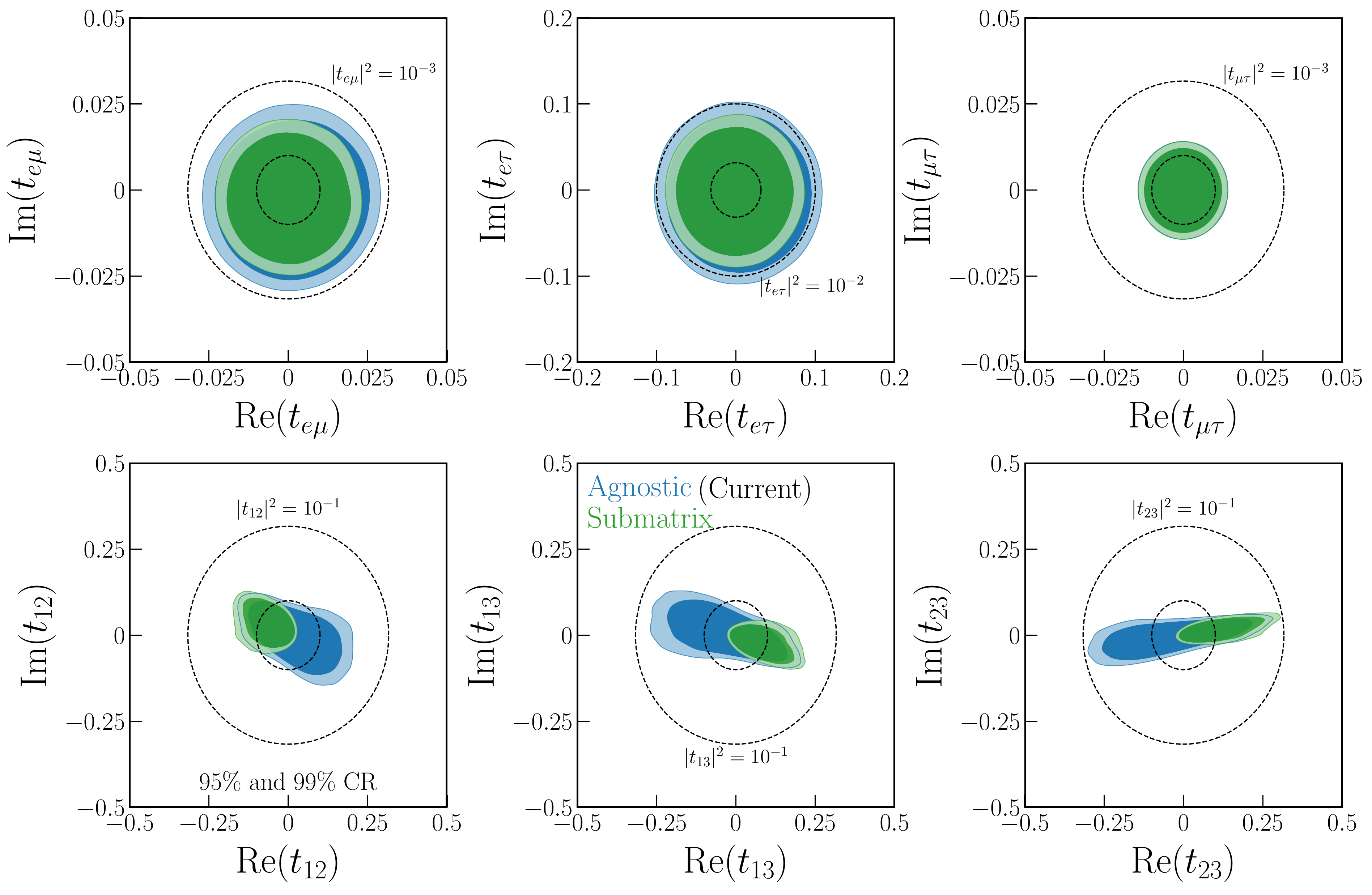}
\caption{Constraints (and projected constraints) on the closures of the six unitarity triangles with data analyzed under the \textbf{\textit{agnostic}} (blue) and \textbf{\textit{sub-matrix}} (green) assumptions. Dashed circles indicate contours corresponding to fixed $\absq{t_{xy}}$, with the outer one in each panel as labeled. The inner dashed circles are an order of magnitude smaller $\absq{t_{xy}}$ than the outer ones.
\label{fig:ClosuresAgnosticSterile}}
\end{center}
\end{figure}

Finally, we repeat the procedure that generated Fig.~\ref{fig:Closures:CurrentFuture}, which determined how well we can constrain the closure of the six different pairs of rows/columns focusing on current data only, we compare the results of this process when data are analyzed under the agnostic or sub-matrix case, in Fig.~\ref{fig:ClosuresAgnosticSterile}.

\begin{table}
\begin{center}
\caption{Summary of current and expected future constraints on the row closures $\left\lvert t_{\alpha\beta}\right\rvert$ and column closures $\left\lvert t_{kl}\right\rvert$, under the \textbf{\textit{sub-matrix}} assumption.
\label{tab:ClosureSummarySterile}}
\begin{tabular}{|c||c|c|}\hline
& Current $3\sigma$ Upper Limit & Future $3\sigma$ Upper Limit \\ \hline \hline
$\left\lvert t_{e\mu}\right\rvert$ & $2.7 \times 10^{-2}$ & $2.1 \times 10^{-2}$ \\ \hline
$\left\lvert t_{e\tau}\right\rvert$ & $1.0 \times 10^{-1}$ & $6.5 \times 10^{-2}$ \\ \hline
$\left\lvert t_{\mu\tau}\right\rvert$ & $1.6 \times 10^{-2}$ & No Improvement \\ \hline \hline
$\left\lvert t_{12}\right\rvert$ & $2.2\times 10^{-1}$ & $8.0 \times 10^{-2}$ \\ \hline
$\left\lvert t_{13}\right\rvert$ & $2.5 \times 10^{-1}$ & $1.0 \times 10^{-1}$ \\ \hline
$\left\lvert t_{23}\right\rvert$ & $3.3\times 10^{-1}$ & $1.0 \times 10^{-1}$ \\ \hline\end{tabular}
\end{center}
\end{table}

Again, we note several features of this result. First, when looking at the closure between two rows $\absq{t_{\alpha\beta}}$ (the top three panels of Fig.~\ref{fig:ClosuresAgnosticSterile}), we see that the resulting constraint on $\absq{t_{\alpha\beta}}$ is largely unchanged. This is because both analyses include the searches for short-baseline neutrino appearance discussed in Section~\ref{subsec:SterileSearch} which directly constrain the closures $\absq{t_{e\mu}}$ (from $\nu_\mu \to \nu_e$ appearance), $\absq{t_{e\tau}}$ (from $\nu_e \to \nu_\tau$ appearance), and $\absq{t_{\mu\tau}}$ (from $\nu_\mu \to \nu_\tau$ appearance). The mild improvement seen in each of these panels comes from the Cauchy-Schwarz constraint $\absq{t_{\alpha\beta}} \leqslant (1-N_\alpha) (1-N_\beta)$, where the normalization constraints assist in these planes. For the bottom panels, the closure of triangles formed between two columns $\absq{t_{kl}}$, the difference between the agnostic and sub-matrix analyses is more drastic. Here, the Cauchy-Schwarz constraints are of the form $\absq{t_{kl}} \leqslant (1-N_k) (1-N_l)$, and because there are no direct experimental constraints on the closure, these Cauchy-Schwarz inequalities play a much more significant role. This feature was observed in Ref.~\cite{Parke:2015goa}, where they analyzed data under the sub-matrix case and noted that these inequalities place the strongest constraints on the closures between two columns.

\begin{figure}[!htbp]
\begin{center}
\includegraphics[width=0.75\linewidth]{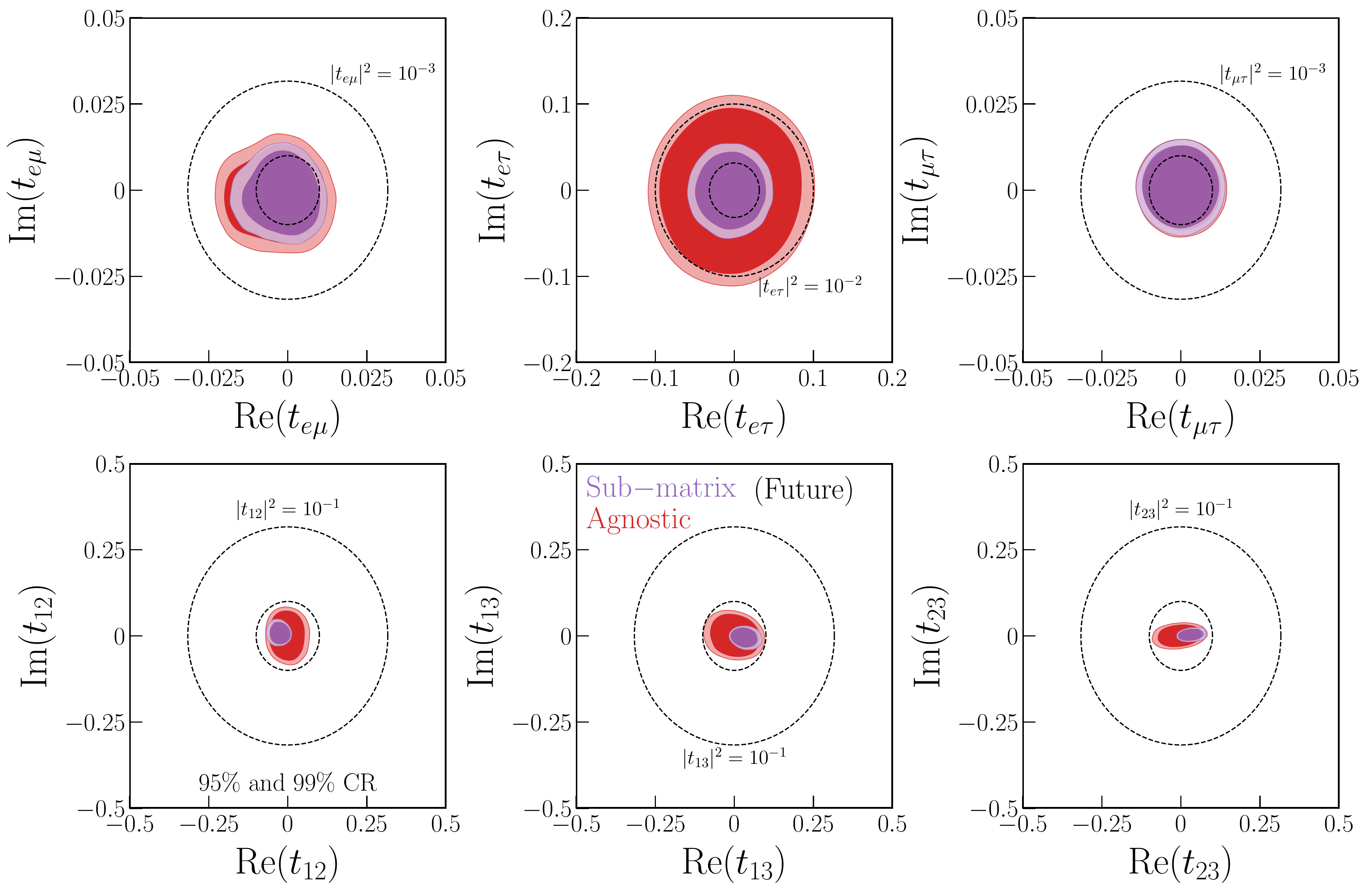}
\caption{Projected future constraints on the closures of the six different unitarity triangles with data analyzed under the \textbf{\textit{agnostic}} (red) and \textbf{\textit{sub-matrix}} (purple) assumptions. Dashed circles indicate contours corresponding to fixed $\absq{t_{xy}}$, with the outer one in each panel as labeled. The inner dashed circles are an order of magnitude smaller $\absq{t_{xy}}$ than the outer ones.
\label{fig:ClosuresAgnosticSterileFuture}}
\end{center}
\end{figure}

Likewise, Fig.~\ref{fig:ClosuresAgnosticSterileFuture} presents projected future constraints on $\absq{t_{\alpha\beta}}$ and $\absq{t_{kl}}$ under these two hypotheses, similar to Fig.~\ref{fig:ClosuresAgnosticSterile}. Moderate improvement on each parameter going from the agnostic to the sub-matrix cases is present in each panel here.

Table~\ref{tab:ClosureSummarySterile} summarizes the results of these analyses -- current and projected constraints on $\left\lvert t_{\alpha\beta}\right\rvert$ and $\left\lvert t_{kl}\right\rvert$ when operating under the sub-matrix hypothesis. Comparing Tables~\ref{tab:ClosureSummary} and~\ref{tab:ClosureSummarySterile}, we see the improved constraints that are obtained under the sub-matrix hypothesis compared to the agnostic case, coming from the Cauchy-Schwarz constraints and forbidding $N_\alpha, N_k > 1$ in the fits. We do not reproduce an analogous version of Table~\ref{tab:NormSummary} here -- the same $3\sigma$ lower limits on $N_\alpha$ and $N_k$ apply here, where the upper limits are now simply $1$ from the theoretical constraints of the sub-matrix case.

\section{Discussion \& Conclusions}
\label{sec:Conclusions}

We have comprehensively analyzed how current and future neutrino oscillation data can be used to constrain non-unitarity of the LMM. Neutrino oscillation probabilities are functions of the LMM elements, the mass-squared differences of the neutrino mass eigenstates, and the interaction potential of the neutrinos in matter. Measurements of the oscillation probabilities are therefore particularly useful for probing the structure of the LMM, as they are generally less sensitive to other new physics than non-oscillation probes (e.g. lepton universality, lepton-flavor violation, etc.).

In our analysis we have characterized our ignorance of the LMM structure by separating our analysis according to three hypotheses: 
\begin{enumerate}
\item The ``standard'' case, with only 3 neutrino flavors, and the LMM is the PMNS matrix, and therefore unitary. 
\item The ``sub-matrix" case, where there are $n> 3$ neutrino flavors, and the $3\times3$ LMM is non-unitary, but the greater $n\times n$ matrix is unitary. 
\item The ``agnostic" case where we assume nothing regarding the unitarity of a possibly $m\times n$ neutrino mixing matrix.
\end{enumerate}
In the bulk of our analysis, we choose not to impose any a priori theoretical biases, and therefore compute constraints on non-unitarity in the agnostic case. However, we provide comparisons and sanity checks with the other two cases as necessary as validation of this approach.

When performing fits to current data, we include a representative sample of experiments that, when combined, provide the strongest set of constraints on the unitarity of the LMM possible. This set of data is as follows: solar neutrino experiments (SNO and Super-K), reactor antineutrino experiments (KamLAND and Daya Bay), long-baseline muon-neutrino disappearance (T2K and NOvA), long-baseline electron-neutrino appearance (T2K and NOvA), and long-baseline tau-neutrino appearance (OPERA). Where possible, up to being able to adapt experimental results to account for non-unitary mixing, we include the most recent results from each experiment. We also include results from short-baseline searches for anomalous neutrino disappearance/appearance from KARMEN, NOMAD, CHORUS, and MINOS/MINOS+.

Projecting to the next decade or so, we include the upcoming experiments that will qualitatively improve our understanding of leptonic unitarity. First is JUNO, where we simulate its reactor antineutrino capabilities. Second, we include T2HK's long-baseline beam-based searches for muon-neutrino disappearance and electron-neutrino appearance. Next, we include near-future projections for IceCube's DeepCore and/or Upgrade measurements of tau-neutrino appearance. Finally, we consider all DUNE searches currently under study -- its long-baseline beam-based searches for muon-neutrino disappearance, electron-neutrino appearance, and tau-neutrino appearance; as well as its solar neutrino capabilities.

Our main results are visually represented in Figs.~\ref{fig:CurrentFutureUs},~\ref{fig:RowColNorms} and \ref{fig:Closures:CurrentFuture} for matrix elements-squared, column/row normalizations (Eqs.~\eqref{eq:N_k} and~\eqref{eq:N_a}) and column/row closures (Eqs.~\eqref{eq:t_kl} and~\eqref{eq:t_ab}) respectively, all in the agnostic case. The $3\sigma$ constraints on $|U_{\mu1}|^2$ will improve by almost an order of magnitude, and further significant improvements are expected for $|U_{e1}|^2$, $|U_{e2}|^2$ and $|U_{\mu2}|^2$. These are driven primarily by the improved precision of JUNO and DUNE over current experiments.  The $\tau$-row elements will be measured better by a factor of 2, owing to the expected sensitivity of DUNE and IceCube's $\tau$-appearance search. Of note is that while the constraints on most substructures of the LMM will improve with future experiments, some do not. In particular, $|U_{e3}|^2$ has been extremely well-measured by the Daya Bay experiment, and the NOMAD constraints on $|t_{e\tau}|^2$ and $|t_{\mu\tau}|^2$ will not be improved on in the near future.

Our results highlight the improvements achievable by long-baseline oscillation experiments. However, they also emphasize that many constraints on the LMM are dominated by sterile neutrino searches capable of measuring triangle closure and row/column normalizations, in some cases extremely precisely. Indeed, the results of Fig.~\ref{fig:RowColNorms} demonstrate that much of the power in constraining $N_\mu$ arises from the MINOS/MINOS+ search for $\nu_\mu$ disappearance. This is suggestive that dedicated sterile searches measuring $\nu_e$ and $\nu_\tau$ disappearance precisely could be important for future improvements in our understanding of the LMM. A dedicated $\tau$ beam in particular would be extremely useful in understanding many of the poorly-constrained components of the LMM. An electron disappearance sterile search (specifically one free of large systematics that allows for stronger constraints in the ``averaged-out'' regime), meanwhile, would provide an alternative handle on the improvements expected from long-baseline experiments, and would therefore provide greater understanding of possible systematic effects in the two types of measurements.

Obtaining a detailed understanding of the LMM is critical to developing a theory of neutrino masses, as LMM unitarity or non-unitarity can be the result of non-intersecting subsets of neutrino mass mechanisms. Our results highlight the power of neutrino oscillation measurements to provide theoretically clean constraints on non-unitarity of the LMM, and therefore act as probes of these mechanisms. The neutrino puzzle is rapidly being untangled, as evidenced by the significant improvements expected between the current and future generation of neutrino oscillation experiments. Continuing improvement of both our theoretical and experimental understanding of neutrino oscillations, and therefore of the LMM, is critical to solving fundamental questions left unanswered by the Standard Model.


\section*{Acknowledgements}

We thank Carlos Arg{\"u}elles, Doug Cowen, Peter Denton, Teppei Katori, Pedro Machado, Stephen Parke, Michael Peskin, Xin Qian, and Natalia Toro for helpful discussions, and especially Francesco Capozzi for sharing DUNE solar simulation codes.  We acknowledge the 7th LCTP Spring Symposium: Neutrino Physics. SARE and SWL are supported by the U.S. Department of Energy under Contract No. DE-AC02-76SF00515. SARE is also supported in part by the Swiss National Science Foundation, SNF project number P400P2$\_$186678. KJK is supported by Fermi Research Alliance, LLC under contract DE-AC02-07CH11359 with the U.S. Department of Energy.


\appendix


\section{Rare Charged Lepton Decays as Probes of Lepton Unitarity}
\label{app:RareDecays}

Elements of the LMM are relevant in physical processes beyond neutrino oscillations, especially in decay processes involving leptons. Take, for example, the decay $\mu \to e \nu \overline{\nu}$. The final state neutrino and antineutrino are typically considered to be flavor eigenstates $\ket{\nu_\mu}$ and $\ket{\overline{\nu}_e}$, respectively. If the LMM is not unitary, however, the partial width for the decay must be expressed in terms of mass eigenstates: $\mu \to e \nu_i \overline{\nu}_j$, and the partial width is summed over these unmeasurable indices:
\begin{equation}
\Gamma_{\mu\to e\nu\overline{\nu}} \propto G_F^2 \to G_F^2 N_\mu N_e .
\end{equation}
When $N_\mu$ and $N_e$ are not $1$, the value of $G_F$ inferred from muon decay would be different from other processes in which it is measured. Measurements such as leptonic universality, the weak mixing angle, $Z$ decays, and rare charged lepton decays place constraints on this type of scenario. We direct the reader to Refs.~\cite{Antusch:2006vwa,Antusch:2014woa,deGouvea:2015euy,Fernandez-Martinez:2016lgt} for a thorough discussion of these types of constraints, as well as the model-dependent cases in which they apply.

We will discuss the model-dependence of such probes through the example of rare charged lepton decays. These processes would be forbidden in the SM if neutrinos were massless. The most constrained of these rare decays are loop-mediated processes such as $\mu \to e \gamma$, $\mu \to e$ conversion and $\mu \to 3e$.\footnote{Radiative $\tau$ decays violating lepton flavor are also searched for, but these constraints are significantly weaker than those involving $\mu$ decays.} Unlike in cases where the neutrinos are in the final state, in these loop-mediated processes, neutrinos only appear within the loop. These processes are therefore in principle able to probe contributions from all contributing neutrino states, including weakly-coupled sterile states, as well as the structure of the full lepton mixing matrix $\mathcal{U}_{\alpha k}$. However, as we explain below, the interpretation of such processes as constraints on unitarity are somewhat subtle and model-dependent.

The measurement of loop-mediated lepton flavor-violating decays is often expressed as a branching ratio between $\ell_\alpha \to \ell_\beta \gamma$ and $\ell_\alpha \to \ell_\alpha \nu\nu$. The branching ratio for this decay is~\cite{Petcov:1976ff,Bilenky:1977du,Lee:1977qz,Marciano:1977wx}
\begin{align}
\mathcal{B}(\ell_\alpha \to \ell_\beta \gamma) =\frac{\Gamma(\ell_\alpha \to \ell_\beta \gamma)}{\Gamma(\ell_\alpha \to \ell_\beta \nu \bar{\nu})}  = \frac{3 \alpha}{32\pi}\Big\vert \sum_k \mathcal{U}_{\alpha k} \mathcal{U}_{\beta k}^* F(x_k) \Big\vert^2 \ ,
\label{eq:meg}
\end{align}
where $F(x_k)$ is a loop function of the ratio $x_k \equiv (m_{\nu_k} / m_W)^2$, whose relevant limits are
\begin{align}
F(x_k) \sim \begin{dcases}
\frac{10}{3} - x_k + 2 x_k^2 + \mathcal{O}(x_k^3), & x_k \ll 1 \ ,\\
\frac{17}{6} - \frac{3(x_k - 1)}{10} + \frac{(x-1)^2}{10} + \mathcal{O}((x_k-1)^3), & x_k \simeq 1 \ , \\
\frac{4}{3} + \frac{6 \log x_k - 11}{x_k} + \frac{2(12 \log x_k -13)}{x_k^2} + \mathcal{O}(x_k^{-3}), & x_k \gg 1 \ .
\end{dcases}\ 
\end{align} 
Note that the matrix in Eq.~\eqref{eq:meg} is not the $3\times 3$ sub-matrix $U_{\alpha k}$, but the full matrix including all sterile states $\mathcal{U}_{\alpha k}$, where $k$ runs through 1, 2, ... $n$. From here we may proceed to interpret constraints from rare leptonic decays according to the three cases we discussed in Section~\ref{sec:LMM}. 

In the standard case with three neutrinos and $\mathcal{U} \equiv U_{\rm PMNS}$, by construction, any observation of charged lepton flavor violation (CLFV) should be interpreted as arising due to new physics that does not affect the unitarity of the LMM.

In the sub-matrix case, let us first examine the expected level of unitarity violation in a canonical type-I seesaw scenario, where there are three sterile neutrinos with masses much larger than the weak scale. Equation~\eqref{eq:meg} can consquently be written as~\cite{Cheng:1980tp}:
\begin{align}
\mathcal{B}(\ell_\alpha \to \ell_\beta \gamma) \simeq \frac{3\alpha}{32\pi}\Bigg\vert U_{\alpha k} U^*_{\beta k} \Big( F(x_k) \cos \theta_k^2 + F(X_k) \sin\theta_k^2 \Big) \Bigg\vert \ ,
\end{align}
where $X_k = (m_{\nu_{H,k}}/m_W)^2$ and $U_{\alpha k}$ is the unitary PMNS matrix. The newly introduced mixing angle is defined as $\tan2\theta_k \sim m^D_k / m^M$ in terms of the Dirac masses $m^D_k$ and Majorana mass scale $m^M$ of the canonical type-I seesaw mechanism. This manner of writing the branching ratio is dependent on the assumption that the Majorana mass matrix is diagonal with identical entries in the diagonal. A more complex ultraviolet structure will undoubtedly lead to variations on the estimates we present here, but should not affect the conclusion as long as the assumptions we make hold to a good approximation (e.g. small off-diagonal terms in the Majorana mass matrix).

Since by construction, $U_{\alpha k}$ is unitary, the leading non-zero terms arising due to the inclusion of heavy neutrinos are $\mathcal{O}(\theta_k^4)$ and $\mathcal{O}(x_k \theta_k^2)$.  Because $\theta_k^2 \sim m_{\nu_L} / m_{\nu_H}$, and we assumed $m_{\nu_H} \gg m_W$, both these terms lead to unobservably small CLFV.

One could interpret CLFV in the sub-matrix case without model-dependence, and therefore allow CLFV to impose a constraint on the LMM unitarity. This is often referred to as the Minimal Unitarity Violating approach, see e.g., Ref~\cite{Antusch:2014woa}. Remaining agnostic about the scale of neutrino masses, but imposing that all the sterile states are in the same limit compared to the weak scale, the leading contribution to Eq.~\eqref{eq:meg} can be written as
\begin{align}
\nonumber \mathcal{B}(\ell_\alpha \to \ell_\beta \gamma) &\sim \frac{3 \alpha}{32\pi} \Big\vert \frac{10}{3} \sum_{k=1}^3 \mathcal{U}_{\alpha k} \mathcal{U}_{\beta k}^* + c_1 \sum_{k=4}^n\mathcal{U}_{\alpha k} \mathcal{U}_{\beta k}^* \Big\vert^2 \ \\
\nonumber  &= \frac{3 \alpha}{32\pi}\Big\vert \frac{10}{3}\sum_{k=1}^n \mathcal{U}_{\alpha k} \mathcal{U}_{\beta k}^* + c_2 \sum_{k=4}^n\mathcal{U}_{\alpha k} \mathcal{U}_{\beta k}^*\Big\vert^2 \\
&= \frac{3 \alpha}{32\pi}\Bigg(\left(\frac{10}{3}\right)^2\Big\vert\sum_{k=1}^n \mathcal{U}_{\alpha k} \mathcal{U}_{\beta k}^* \Big\vert^2+ c_2^2 \Big\vert \sum_{k=4}^n\mathcal{U}_{\alpha k} \mathcal{U}_{\beta k}^*\Big\vert^2 + \frac{10 c_2}{3}\mathrm{Re}\left(\sum_{k=1}^n \mathcal{U}_{\alpha k} \mathcal{U}_{\beta k}^*\right)\left(\sum_{k=4}^n\mathcal{U}_{\alpha k}^* \mathcal{U}_{\beta k}\right)\Bigg) \ ,
\label{eq:meg_branch}
\end{align}
where $c_1 = 4/3,~17/6,~10/3$ depending on the mass scale of the sterile neutrinos, and $c_2 = c_1-10/3$.\footnote{Note that if the steriles are light, it would suggest $c_2 = 0$. One must then include the next term in the expansion in $x_k$ to find a non-zero contribution to CLFV from leptonic non-unitarity.} In going from the first line to the second above, we make use of the fact that by virtue of studying the sterile case, the overall mixing matrix $\mathcal{U}$ must be unitary. This also results in the first and third terms in Eq.~\eqref{eq:meg_branch} being zero due to row closure of the full $\mathcal{U}$ matrix. Then,
\begin{align}
\nonumber \mathcal{B}(\ell_\alpha \to \ell_\beta \gamma) &\sim \frac{3 \alpha c_2^2}{32\pi} \Big\vert \sum_{k=4}^n\mathcal{U}_{\alpha k} \mathcal{U}_{\beta k}^*\Big\vert^2 = \frac{3 \alpha c_2^2}{32\pi} \Big\vert\sum_{k=1}^n\mathcal{U}_{\alpha k} \mathcal{U}_{\beta k}^* - \sum_{k=1}^3\mathcal{U}_{\alpha k} \mathcal{U}_{\beta k}^*\Big\vert^2 \\
&=\frac{3 \alpha c_2^2}{32\pi} \Big\vert -\sum_{k=1}^3\mathcal{U}_{\alpha k} \mathcal{U}_{\beta k}^*\Big\vert^2 = \frac{3 \alpha c_2^2}{32\pi} |t_{\alpha\beta}|^2 \ .
\end{align}
In this case, we can see that the non-observation of CLFV can indeed be viewed as a constraint on the closure of the $\alpha$-$\beta$ triangle, and therefore a test of unitarity. Note that our analysis of this scenario assumes that all steriles have masses in the same limit with respect to the weak scale. This would therefore not apply to e.g., models with one light and two heavy steriles.

Finally, in the agnostic case, we can see that owing to the fact that the leading terms of $F(x_k)$ are all constants, CLFV is a clear test of unitarity of the full $\mathcal{U}_{\alpha k}$ matrix. Below we quote the constraints on closure of the $\alpha-\beta$ rows in terms of the full matrix in the agnostic case. It is then straightforward to map this onto a constraint on $t_{\alpha\beta}$ when assuming Minimal Unitarity Violation for the sub-matrix case as above.

For the $\mu-e$ row, the strongest such constraint comes from the MEG collaboration, and leads to the requirement that $\lvert \sum_k \mathcal{U}_{\mu k} \mathcal{U}_{e k}^* \rvert \lesssim 10^{-5}$~\cite{TheMEG:2016wtm}. Future measurements of $\mu \to e \gamma$ will improve this limit by roughly one order of magnitude~\cite{Baldini:2013ke}. Planned searches for the related process $\mu \to 3e$, which are expected to have a similar sensitivity~\cite{Blondel:2013ia}, while searches for $\mu \to e$ conversion in nuclei can improve on this future constraint by a further order of magnitude~\cite{Abrams:2012er}. 

For the $\tau-e$ and $\tau-\mu$ rows, the strongest current constraints come from the BaBar collaboration and require that $\lvert \sum_k \mathcal{U}_{\tau k} \mathcal{U}_{(e,\mu) k}^* \rvert \lesssim 10^{-3}$~\cite{Aubert:2009ag}. These measurements will be improved at B-factories, leading to a factor of $\sim3$ improvement in the $\tau-e$ row, and an order of magnitude improvement in the $\tau-\mu$ row~\cite{Aushev:2010bq}.


\section{Derivation of Vacuum Oscillation Probabilities Without Assuming Unitarity}
\label{app:Derivations}

In this appendix, we provide derivations for three-neutrino oscillations in vacuum where unitarity is not assumed. We analyze these oscillation probabilities in different distance and neutrino energy regimes that are appropriate for a variety of experiments. 
Experiments operating in particular length/energy limits are most sensitive to only certain matrix elements, as we will explain below.

In Section~\ref{subsec:OscProbs} we introduced the formalism for time-evolving a neutrino flavor eigenstate $\ket{\nu_\alpha}$ when its mixing is not unitary, beginning with Eq.~\eqref{eq:state_evolve}. We can then project this onto a flavor eigenstate $\ket{\nu_\beta}$ to determine the transition amplitude for $\nu_\alpha \to \nu_\beta$, which we express as
\begin{equation}
\label{eq:Amp}
\mathcal{A}_{\alpha\beta} \equiv \langle\nu_\beta|\nu_\alpha(L)\rangle = \frac{e^{-im_1^2 L/2E_\nu}}{\sqrt{N_\alpha N_\beta}} \Big(U_{\alpha 1}^* U_{\beta 1} + U_{\alpha 2}^* U_{\beta 2} e^{-i\Delta_{21}} + U_{\alpha 3}^* U_{\beta 3} e^{-i\Delta_{31}}\Big),
\end{equation}
where $\Delta_{kl} \equiv \Delta m_{kl}^2 L/2E_\nu$, $\Delta m_{kl}^2 \equiv m_k^2 - m_l^2$, $E_\nu$ is the neutrino energy, and $N_\alpha$ is defined in Eq.~\eqref{eq:N_a}. The overall phase, $\exp(-im_1^2 L/2E_\nu)$, does not enter oscillation probabilities so we will drop it henceforth. We separate our discussion into oscillation probabilities for disappearance/survival channels ($\alpha=\beta$) and appearance channels ($\alpha\neq\beta$).

For antineutrino oscillations, $U_{\alpha k}^* \to U_{\alpha k}$ and $U_{\beta l} \to U_{\beta l}^*$ everywhere, and the matter potential $V_{\alpha k} \to -V_{\alpha k}$.


\subsection{Disappearance/Survival Probabilities}

If $\alpha=\beta$, Eq.~\eqref{eq:Amp} becomes
\begin{equation}
\mathcal{A}_{\alpha\alpha} = \frac{1}{N_\alpha}\left(\left\lvert U_{\alpha 1}\right\rvert^2 + \left\lvert U_{\alpha 2}\right\rvert^2e^{-i\Delta_{21}} + \left\lvert U_{\alpha 3}\right\rvert^2 e^{-i\Delta_{31}}\right).
\end{equation}
This can be rewritten as
\begin{align}
\mathcal{A}_{\alpha\alpha} &= \frac{1}{N_\alpha}\left(N_\alpha + \left\lvert U_{\alpha 2}\right\rvert^2 \left( e^{-i\Delta_{21}} - 1\right) + \left\lvert U_{\alpha 3}\right\rvert^2 \left( e^{-i\Delta_{31}} - 1\right)\right) \label{eq:amp1}\\
&= 1 - 2i \frac{\left\vert U_{\alpha 2}\right\rvert^2}{N_\alpha} e^{-\frac{i\Delta_{21}}{2}}\sin\left(\frac{\Delta_{21}}{2}\right) - 2i \frac{\left\vert U_{\alpha 3}\right\rvert^2}{N_\alpha} e^{-\frac{i\Delta_{31}}{2}} \sin\left(\frac{\Delta_{31}}{2}\right) .
\label{eq:amp2}
\end{align}
To obtain the oscillation probability $P_{\alpha\alpha} \equiv P(\nu_\alpha \to \nu_\alpha) \equiv \left\lvert \mathcal{A}_{\alpha\alpha}\right\rvert^2$, we square the transition amplitude. After rearranging terms, we obtain
\begin{eqnarray}
P_{\alpha\alpha} &=& 1 - \frac{4\left\lvert U_{\alpha 2}\right\rvert^2 \left( N_\alpha - \absq{U_{\alpha 2}} \right)}{N_\alpha^2} \sin^2\left(\frac{\Delta_{21}}{2}\right) - \frac{4\left\lvert U_{\alpha 3}\right\rvert^2 \left( N_\alpha - \absq{U_{\alpha 3}} \right)}{N_\alpha^2}\sin^2\left(\frac{\Delta_{31}}{2}\right) \nonumber \\
&+& \frac{8 \left\lvert U_{\alpha 2}\right\rvert^2 \left\lvert U_{\alpha 3}\right\rvert^2}{N_\alpha^2} \sin\left(\frac{\Delta_{21}}{2}\right) \sin\left(\frac{\Delta_{31}}{2}\right) \cos\left(\frac{\Delta_{32}}{2}\right ). 
\end{eqnarray}
After minor rearrangements, this becomes
\begin{eqnarray}
P_{\alpha\alpha} &=& 1 - \frac{4\left\lvert U_{\alpha 2}\right\rvert^2 \left( \left\lvert U_{\alpha 1}\right\rvert^2 + \left\lvert U_{\alpha 3}\right\rvert^2 \right)}{N_\alpha^2} \sin^2\left(\frac{\Delta_{21}}{2}\right) - \frac{4\left\lvert U_{\alpha 3}\right\rvert^2 \left( \left\lvert U_{\alpha 1}\right\rvert^2 + \left\lvert U_{\alpha 2}\right\rvert^2 \right)}{N_\alpha^2}\sin^2\left(\frac{\Delta_{31}}{2}\right) \nonumber \\
&+& \frac{8 \left\lvert U_{\alpha 2}\right\rvert^2 \left\lvert U_{\alpha 3}\right\rvert^2}{N_\alpha^2} \sin\left(\frac{\Delta_{21}}{2}\right) \sin\left(\frac{\Delta_{31}}{2}\right) \cos\left(\frac{\Delta_{32}}{2}\right ). 
\label{eq:Pdis}
\end{eqnarray}

The majority of oscillation experiments operate in one of two regimes, where $\Delta_{21} \ll 1$ or $\Delta_{31} \gg 1$. We explore these two regimes separately.

\textbf{$\Delta_{21} \ll 1$ Regime:} For experiments where
\begin{equation}
\frac{L}{E_\nu} \ll 5\times 10^{3}\ \frac{\mathrm{km}}{\mathrm{GeV}},
\end{equation}
oscillations associated with $\Delta m_{21}^2 \approx 7.5 \times 10^{-5}$ eV$^2$ have yet to develop. Taking the limit $\Delta_{21} \ll 1$ in Eq.~\eqref{eq:Pdis}, we obtain the oscillation probability
\begin{equation}
P_{\alpha\alpha} = 1 - \frac{4\absq{U_{\alpha 3}} \left(\absq{U_{\alpha 1}} + \absq{U_{\alpha 2}}\right)}{N_\alpha^2} \sin^2\left(\frac{\Delta_{31}}{2}\right).
\label{eq:Dis:SBL}
\end{equation}
In the limit of unitarity, this has a familiar form $1 - 4\absq{U_{\alpha 3}}(1-\absq{U_{\alpha 3}})\sin^2(\Delta_{31}/2)$.\footnote{Even more familiar, if $\alpha = e$ and this is the electron-(anti)neutrino disappearance probability $P_{ee}$, we obtain the limit $P_{ee} \to 1 - \sin^2\left(2\theta_{13}\right) \sin^2\left(\Delta_{31}/2\right)$ as measured by Daya Bay.} This regime is appropriate for reactor neutrino experiments with relatively short baselines, such as Daya Bay, and long-baseline accelerator neutrino experiments (MINOS, NOvA, T2K). This expression of disappearance oscillation probabilities in vacuum approximates all of the appropriate experiments well, except for JUNO, where matter effects can shift the central values of $\sin^2\theta_{12}$ by $\mathcal{O}(1\%)$~\cite{An:2015jdp}.

\textbf{$\Delta_{31} \gg 1$ Regime:} Meanwhile, when
\begin{equation}
\frac{L}{E_\nu} \gg 2\times 10^{2}\ \frac{\mathrm{km}}{\mathrm{GeV}},
\end{equation}
oscillations associated with $\Delta m_{31}^2 \approx 2.5 \times 10^{-3}$ eV$^2$ will have averaged out over the energy resolution of an experiment. In this regime, the terms proportional to $\sin^2(\Delta_{31}/2)$ will average out to $1/2$. The term with $\sin(\Delta_{31}/2)\cos(\Delta_{32}/2)$ will not, as one might naively expect, average out to $0$, but
\begin{align}
&8\absq{U_{\alpha 2}}\absq{U_{\alpha 3}} \sin\left(\frac{\Delta_{21}}{2}\right) \sin\left(\frac{\Delta_{31}}{2}\right) \cos\left(\frac{\Delta_{32}}{2}\right)\\
 &= 8\absq{U_{\alpha 2}}\absq{U_{\alpha 3}} \left[\frac{1}{4} \sin(\Delta_{21})\sin(\Delta_{31}) + \sin^2\left(\frac{\Delta_{21}}{2}\right) \sin^2\left(\frac{\Delta_{31}}{2}\right)\right] \\
&\to 4\absq{U_{\alpha 2}}\absq{U_{\alpha 3}} \sin^2\left(\frac{\Delta_{21}}{2}\right).
\end{align}
Putting this into the complete oscillation probability in Eq.(\ref{eq:Pdis}), we arrive at
\begin{equation}
P_{\alpha\alpha} = 1 - \frac{2\absq{U_{\alpha 3}} \left( \absq{U_{\alpha 1}} + \absq{U_{\alpha 2}} \right)}{N_\alpha^2} - \frac{4\absq{U_{\alpha 1}} \absq{U_{\alpha 2}}}{N_\alpha^2} \sin^2\left(\frac{\Delta_{21}}{2}\right).
\label{eq:Dis:LBL}
\end{equation}

The $\Delta_{31} \gg 1$ approximation is valid for the KamLAND experiment's measurement of $P_{ee}$. The proposed JUNO experiment, which aims to measure the mass-ordering of neutrinos, will not operate in this regime. Instead, JUNO will operate in a regime between the two limiting cases discussed here, and as such we must adopt the full oscillation probability.


\subsection{Appearance Probabilities}

Now we discuss the appearance oscillation probabilities, where $\alpha\neq \beta$. We define the phases associated with each quantity as $\varphi^{\alpha\beta}$, $\varphi_2^{\alpha\beta}$, and $\varphi_3^{\alpha\beta}$ with $t_{\alpha\beta} \equiv |t_{\alpha\beta}| e^{i\varphi^{\alpha\beta}}$ and $U_{\alpha k}^*U_{\beta k} \equiv |U_{\alpha k}||U_{\beta k}| e^{i\varphi_k^{\alpha\beta}}$. In analogy to Eqs.~(\ref{eq:amp1}) and~(\ref{eq:amp2}), we write the transition amplitude as
\begin{align}
\mathcal{A}_{\alpha\beta} &= \frac{1}{\sqrt{N_\alpha N_\beta}}\left(t_{\alpha\beta} + U_{\alpha 2}^* U_{\beta 2} \left(e^{-i\Delta_{21}} - 1\right) + U_{\alpha 3}^* U_{\beta 3} \left(e^{-i\Delta_{31}} - 1\right)\right) \\
&= \frac{1}{\sqrt{N_\alpha N_\beta}}\left(|t_{\alpha\beta}| e^{i\varphi^{\alpha\beta}} - 2i |U_{\alpha 2}| |U_{\beta 2}| e^{i\left(\varphi_2^{\alpha\beta} - \frac{\Delta_{21}}{2}\right)} \sin\left(\frac{\Delta_{21}}{2}\right) - 2i |U_{\alpha 3}| |U_{\beta 3}| e^{i\left(\varphi_3^{\alpha\beta} - \frac{\Delta_{31}}{2}\right)} \sin\left(\frac{\Delta_{31}}{2}\right)\right) . 
\end{align}
Squaring this, we arrive at the oscillation probability,
\begin{eqnarray}
P(\nu_\alpha \rightarrow \nu_\beta) &=& \frac{\left\vert t_{\alpha\beta}\right\rvert^2}{N_\alpha N_\beta} + \frac{4 \left\lvert U_{\alpha 2}\right\rvert^2 \left\lvert U_{\beta 2}\right\rvert^2}{N_\alpha N_\beta} \sin^2\left(\frac{\Delta_{21}}{2}\right) + \frac{4 \left\lvert U_{\alpha 3}\right\rvert^2 \left\lvert U_{\beta 3}\right\rvert^2}{N_\alpha N_\beta} \sin^2\left(\frac{\Delta_{31}}{2}\right) \nonumber \\
&+& \frac{8\left\lvert U_{\alpha 2}\right\rvert \left\lvert U_{\beta 2}\right\rvert \left \lvert U_{\alpha 3}\right\rvert \left \lvert U_{\beta 3}\right \rvert}{N_\alpha N_\beta} \sin\left(\frac{\Delta_{21}}{2}\right) \sin\left(\frac{\Delta_{31}}{2}\right) \cos\left(\frac{\Delta_{32}}{2} + \varphi_{2}^{\alpha\beta} - \varphi_{3}^{\alpha\beta} \right) \nonumber \\
&+& \frac{4\left\lvert t_{\alpha \beta}\right\rvert}{N_\alpha N_\beta} \left[\left\lvert U_{\alpha 2}\right\rvert \left\lvert U_{\beta2}\right\rvert \sin\left( \frac{\Delta_{21}}{2}\right) \sin\left(\frac{\Delta_{21}}{2} + \varphi^{\alpha\beta} - \varphi^{\alpha\beta}_{2}\right) \right. \nonumber \\
&+& \left. \left\lvert U_{\alpha 3}\right\rvert \left\lvert U_{\beta3}\right\rvert \sin\left( \frac{\Delta_{31}}{2}\right) \sin\left(\frac{\Delta_{31}}{2} + \varphi^{\alpha\beta} - \varphi^{\alpha\beta}_{3}\right) \right] 
\label{eq:App}
\end{eqnarray}
In the limit of unitarity, $\left\lvert t_{\alpha\beta}\right\rvert^2 \to 0$ and only the first two lines of Eq.~\eqref{eq:App} (sans the first term) remain.

\textbf{$\Delta_{21} \ll 1$ Regime:} Here the second term in Eq.~\eqref{eq:App} goes to zero, as do the second and third lines:
\begin{equation} 
P_{\alpha\beta} = \frac{\left\vert t_{\alpha\beta}\right\rvert^2}{N_\alpha N_\beta} + \frac{4\absq{U_{\alpha 3}}\absq{U_{\beta 3}}}{N_\alpha N_\beta} \sin^2\left(\frac{\Delta_{31}}{2}\right) - \frac{4|t_{\alpha\beta}||U_{\alpha 3}| |U_{\beta 3}| } {N_\alpha N_\beta} \sin\left(\frac{\Delta_{31}}{2}\right) \sin\left(\frac{\Delta_{31}}{2} + \varphi^{\alpha\beta}-\varphi_3^{\alpha\beta} \right).
\end{equation}
This regime serves as a decent approximation for long-baseline electron-neutrino and tau-neutrino appearance oscillation probabilities (T2K, NOvA, DUNE, and OPERA). Note that unlike for the case of disappearance/survival probabilities, matter effects make a significant impact on these expressions, and we account for that in our simulations.

\textbf{$\Delta_{31} \gg 1$ Regime:} If the $\Delta m_{31}^2$-driven oscillations are averaged out, the oscillation probability becomes
\begin{align}
P_{\alpha\beta} &= \frac{4|U_{\alpha 2}|^2|U_{\beta 2}|^2}{N_\alpha N_\beta} \sin^2\left(\frac{\Delta_{21}}{2}\right) - \frac{4|U_{\alpha 2}||U_{\beta 2}||U_{\alpha 3}||U_{\beta 3}|}{N_\alpha N_\beta} \sin\left(\frac{\Delta_{21}}{2}\right)  \sin\left(\frac{\Delta_{21}}{2}  - \varphi_2^{\alpha\beta} + \varphi_3^{\alpha\beta}\right) \nonumber \\ 
&-\frac{4|t_{\alpha\beta}||U_{\alpha 2}||U_{\beta 2}|}{N_\alpha N_\beta} \sin\left(\frac{\Delta_{21}}{2}\right) \sin\left(\frac{\Delta_{21}}{2} + \varphi^{\alpha\beta} - \varphi_2^{\alpha\beta}\right) \nonumber \\
&+\frac{2\absq{U_{\alpha 3}}\absq{U_{\beta 3}}}{N_\alpha N_\beta} + \frac{2|t_{\alpha\beta}||U_{\alpha 3}||U_{\beta 3}| }{N_\alpha N_\beta} \cos\left(\varphi^{\alpha\beta} - \varphi_3^{\alpha\beta}\right)+ \frac{\absq{t_{\alpha\beta}}}{N_\alpha N_\beta}.
\end{align}
The appearance channel oscillation probability in the $\Delta_{31} \gg 1$ regime has various terms capable of probing the second column of the LMM, and interference between terms from the third and second columns, as well as row triangle closure. Unfortunately, no experiments probing neutrino appearance in this distance/neutrino energy regime exist, or are planned.


\section{Bayesian Priors for Fifteen Parameter Fits}
\label{app:Bayesian}

In many of our results, we use the Bayesian inference tool {\sc pyMultiNest}~\cite{Feroz:2007kg,Feroz:2008xx,Feroz:2013hea,Buchner:2014nha} in order to analyze current and future measurements of the parameters of interest. We include all fifteen parameters in the following -- nine matrix-elements-squared, four phases, and two mass-squared splittings. In this appendix, we explain the assumptions made in our Bayesian analysis and how they impact the applicable results.

We allow the mass-squared splittings to vary within the ranges
\begin{align}
\Delta m_{21}^2 &\in \left[2,10\right] \times 10^{-5}\ \mathrm{eV}^2, \\
\Delta m_{31}^2 &\in \left( \left[-3, -2\right] \cup \left[2,3\right]\right) \times 10^{-3}\ \mathrm{eV}^2,
\end{align}
with flat priors in those ranges on the two splittings. The current measurements pull the posterior likelihood of the fit to the current best-fit values of $\Delta m_{21}^2 \approx 7.5 \times 10^{-5}$ eV$^2$ and $\left\lvert \Delta m_{32}^2\right\rvert \approx 2.45 \times 10^{-3}$ eV$^2$.

For the matrix-elements-squared $\left\lvert U_{\alpha k}\right\rvert^2$, we require that they lie in the range $\left[ 0, 1\right]$. We include flat priors on each of these elements-squared between $0$ and $1$. For the electron- and muon-row elements $\left\lvert U_{ek}\right\rvert^2$ and $\left\lvert U_{\mu k}\right\rvert^2$, we find that the current data are powerful enough that the posterior likelihood is largely independent of the choice of prior (unless anything atypical is adopted), whereas the tau-row elements $\left\lvert U_{\tau k}\right\rvert^2$ still do not have strong information from the OPERA experiment and are mildly sensitive to the choice of prior. Once future data are included, this point becomes unimportant.

Lastly, the phases $\left\lbrace \phi_{e2}, \phi_{e3}, \phi_{\tau 2}, \phi_{\tau 3}\right\rbrace$ are defined to span $[0, 2\pi]$ and the prior included is flat in terms of the phase. 
The relative sizes of some of the matrix element magnitude combinations require the phases to take on certain values. For instance, the phase $\phi_{e2}$ is quite constrained by the closure of the $e$-$\mu$ triangle, so it is insensitive to the choice of the prior. On the other hand, since any current constraint on $\phi_{e3}$ is relatively weak, its posterior likelihood can be sensitive to the choice of prior. As with the case of the matrix-elements-squared above, this issue is unimportant once future data are included in the analysis.


\section{Measurement of LMM Phases}
\label{app:Phases}

In the main text (Section~\ref{subsec:USq}), we presented the results of our analysis for the current and future measurements of the elements-squared $\absq{U_{\alpha k}}$, which are parameterization-independent quantities. Here, we discuss the current and future measurements of the phases $\left\lbrace \phi_{e2},\ \phi_{e3},\ \phi_{\tau2},\ \phi_{\tau3}\right\rbrace$, the four phases we consider in our analyses. These measurements are parameterization-dependent and only apply in the MP parameterization we employ.

\begin{figure}[!htbp]
\begin{center}
\includegraphics[width=0.7\linewidth]{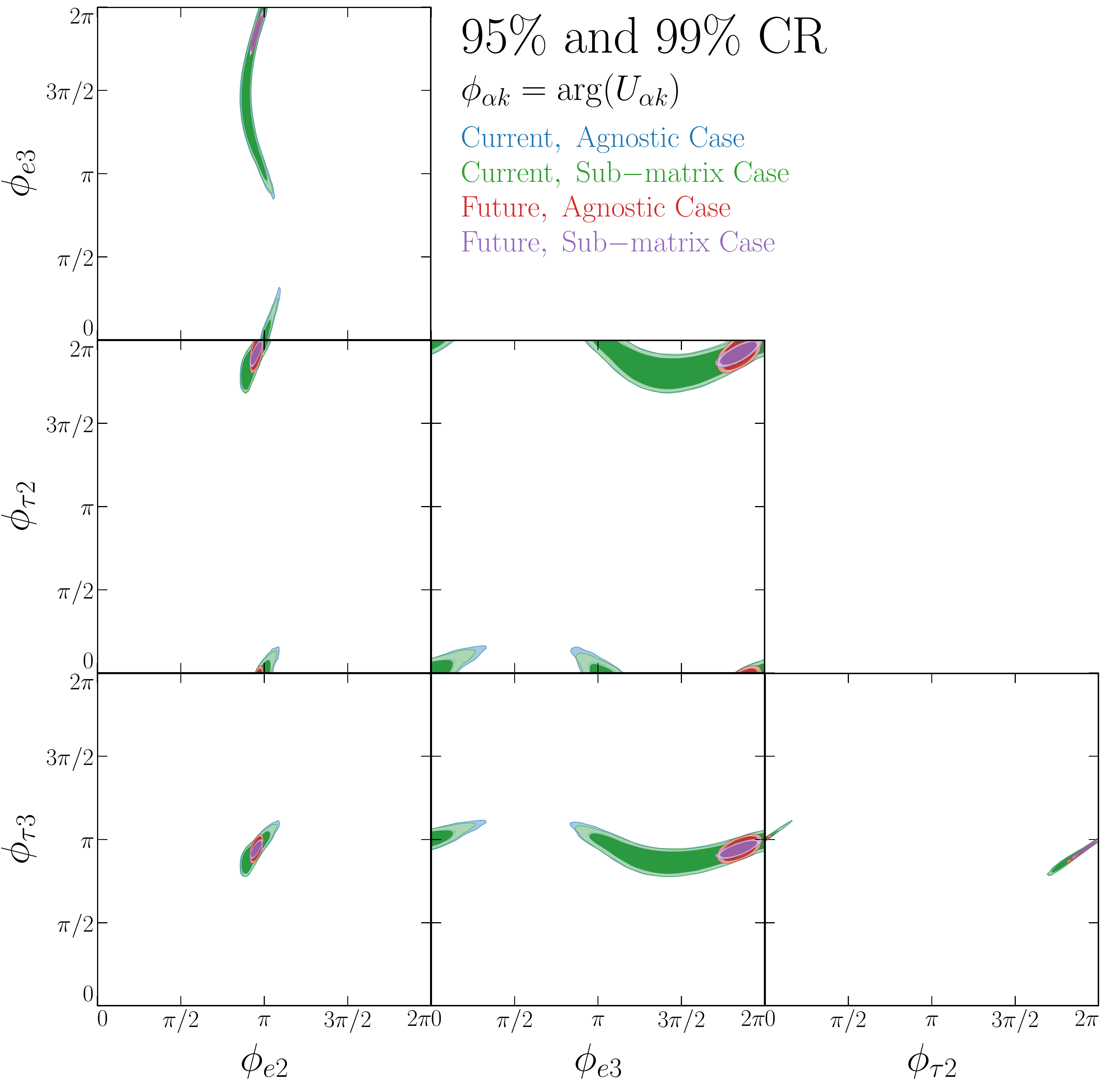}
\caption{Current and future measurements of the four phases $\phi_{e2}$, $\phi_{e3}$, $\phi_{\tau2}$, and $\phi_{\tau3}$ in the MP parameterization. We show the measurements given current data under the \textbf{\textit{agnostic}} case assumption (blue) and the \textbf{\textit{sub-matrix}} case assumption (green), compared with future data analyzed under the \textbf{\textit{agnostic}} case (red) and the \textbf{\textit{sub-matrix}} case (purple). All contours shown are 95\% (dark colors) and 99\% (fainter colors) credibility.
\label{fig:Phases}}
\end{center}
\end{figure}

Figure~\ref{fig:Phases} presents each set of two-dimensional measurements at 95\% and 99\% credibility of these four phases. We show this for four sets of data/assumptions -- all current data are analyzed in blue and green, where the blue (green) contours correspond to data analyzed under the agnostic (sub-matrix) case. Future data are analyzed under the agnostic (sub-matrix) case in red (purple).

We see in Fig.~\ref{fig:Phases} that, even with current data, $\phi_{e2}$ and $\phi_{\tau2}$ are constrained to be close to $\pi$, and $\phi_{\tau3}$ must be close to $\pi$. These requirements come from the current constraints on the closures between different rows $\absq{t_{\alpha\beta}}$ and the relative sizes of different products of magnitudes of elements. For instance, the relative size of the legs of the $e$-$\mu$ triangle are $|U_{e1}| |U_{\mu 1}| \approx 0.25$, $|U_{e2}| |U_{\mu2}| \approx 0.33$, and $|U_{e3}| |U_{\mu 3}| \approx 0.11$. This requires $\phi_{e2}$ to be near $\pi$ if the triangle is to close. Because the product $|U_{e3}| |U_{\mu3}|$ is relatively small, the constraints on $\phi_{e3}$ from the closure of this triangle are relatively weak. As precision experiments that are sensitive to CP-violation begin collecting data, all of the phases will be measured more precisely -- with DUNE and JUNO data included, $\phi_{e3}$ will be measured to much higher precision. We also see that the assumptions of sub-matrix or agnostic do not impact the measurement precisions of the phases, with current or future data.


\section{The LSND/MiniBooNE Anomalies}
\label{app:LSNDMB}

In Section~\ref{subsec:SterileSearch} we discussed how various short-baseline searches for anomalous neutrino disappearance/appearance are used to constrain the unitarity of the $3\times3$ LMM. We briefly commented on the LSND and MiniBooNE anomalies before ultimately \textit{not} including them in our more complete analysis, due to tensions between their anomalous $\nu_\mu \to \nu_e$ appearance results with searches for $\nu_\mu$ and $\nu_e$ disappearance.

If we assume that the anomalous appearance of $\nu_\mu \to \nu_e$ in LSND and MiniBooNE is instead due to non-unitarity, we must use the ``averaged-out'' region of their sterile neutrino parameter space to draw connections. Their combined result points to a preference for nonzero $\absq{t_{e\mu}} \approx 2.6 \times 10^{-3}$, excluding $\absq{t_{e\mu}} = 0$ at roughly $6\sigma$ confidence. Under the sub-matrix assumption, this closure is constrained using
\begin{equation}
\label{eq:tCompare}
\absq{t_{e\mu}} \leqslant \left(1-N_\mu\right) \left(1 - N_e\right),
\end{equation}
This is in tension with the current constraints of $|1-N_e|\leqslant0.022$ and $|1-N_\mu|\leqslant0.016$ from other oscillation measurements. However, if we abandon the sub-matrix case and instead focus on the agnostic case, the constraint in Eq.~\eqref{eq:tCompare} no longer holds. This implies that, despite tight constraints on $N_e$ and $N_\mu$, we can still have large $\absq{t_{e\mu}}$, allowing for short-baseline $\nu_\mu \to \nu_e$ appearance at LSND and MiniBooNE.

In this appendix, we explore the ramifications, under the agnostic assumption, of including the $6\sigma$ preference for $\absq{t_{e\mu}} \neq 0$ in our analysis. When repeating the analysis of all current data, we find no meaningful change to the measurements of $\absq{U_{\alpha k}}$ of Fig.~\ref{fig:CurrentFutureUs}, the row/column normalizations of Fig.~\ref{fig:RowColNorms}, and, with the exception of the triangle $t_{e\mu}$, no significant change to the results shown in Fig.~\ref{fig:Closures:CurrentFuture}. Unsurprisingly, including the $6\sigma$ preference for $\absq{t_{e\mu}}$ results in a ring-shaped region forming in the $\mathrm{Re}(t_{e\mu})$ vs. $\mathrm{Im}(t_{e\mu})$ plane.

In order to determine what future data, i.e., DUNE, T2HK, and JUNO, can say regarding LSND and MiniBooNE, we perform an alternative analysis of our future data compared to the one in the main text. So as not to bias ourselves regarding the short-baseline appearance and disappearance results, we perform this fit without any short-baseline information whatsoever, in contrast to Section~\ref{subsec:NormClos} in the main text. This allows us to determine how well $t_{e\mu}$ can be constrained using measurements of ``oscillated'' neutrinos only, driven primarily by measurements in DUNE/T2HK ($\nu_\mu$ disappearance and $\nu_e$ appearance, as well as the corresponding antineutrino channels) and JUNO ($\overline{\nu}_e$ disappearance).

\begin{figure}[!htbp]
\begin{center}
\includegraphics[width=0.4\linewidth]{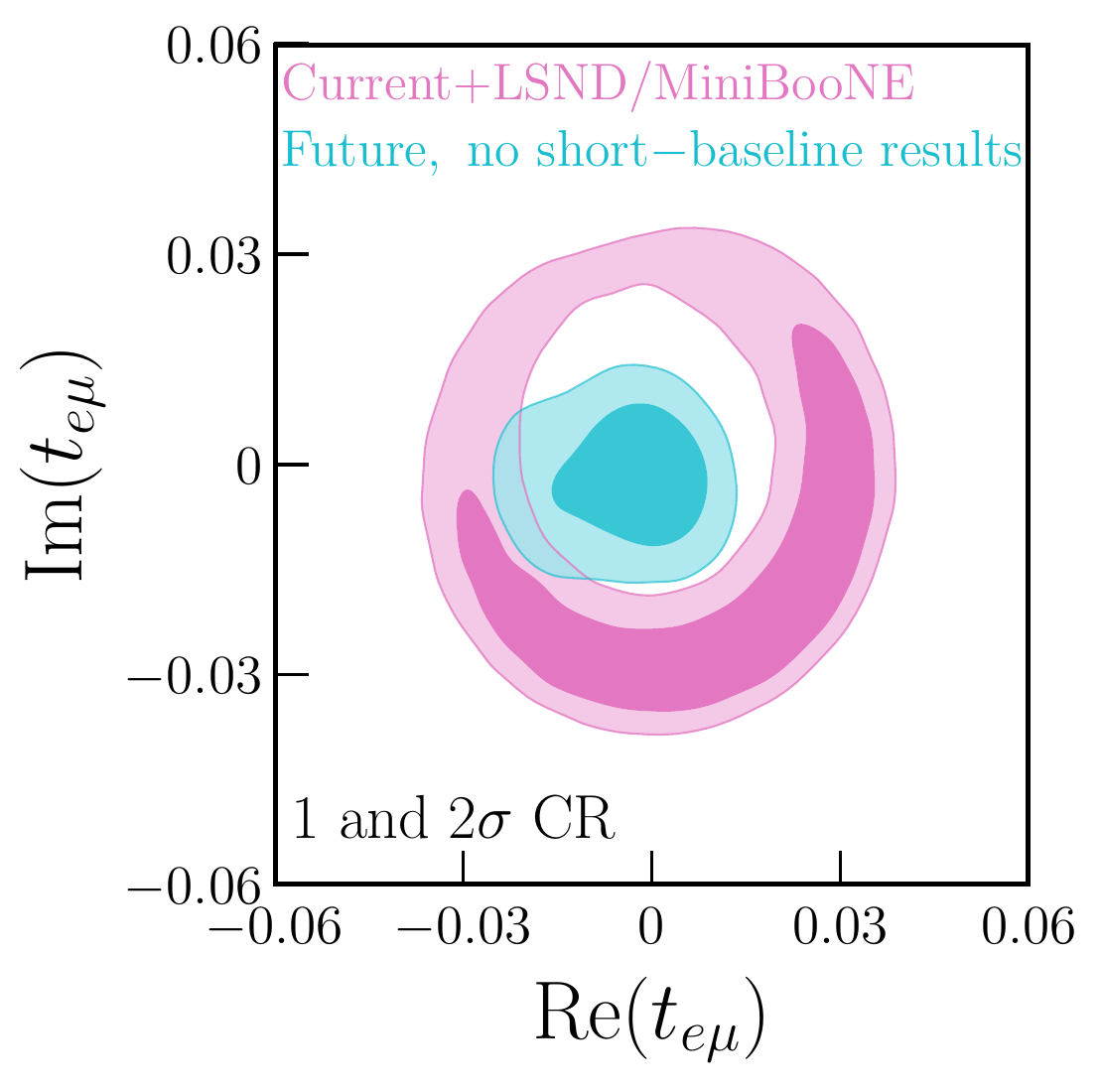}
\caption{Measurement capability of future results (without any short-baseline information at all) to the closure of the $e$-$\mu$ triangle $t_{e\mu}$ at $1$ and $2\sigma$ credibility (light blue), compared to the preferred region by the current experimental data with the LSND and MiniBooNE preference for $\absq{t_{e\mu}} \neq 0$ included (pink). Here, data are analyzed under the \textbf{\textit{agnostic}} assumption.
\label{fig:AnomalyComp}}
\end{center}
\end{figure}

We compare the results of these two analyses -- current data plus the LSND/MiniBooNE preference for $\absq{t_{e\mu}} \neq 0$ in pink vs. current and future data with no short-baseline information in light blue -- in Fig.~\ref{fig:AnomalyComp}.

We see here that, even without the direct constraints on $\absq{t_{e\mu}}$ from short-baseline searches, JUNO, DUNE, and T2HK will be able to constrain $\absq{t_{e\mu}}$ fairly well. Because of the relative strengths of these two measurements, we choose to display $1$ and $2\sigma$ CR. Assuming $\absq{t_{e\mu}} = 0$ (the future data are simulated under this assumption), the combination of DUNE, T2HK, and JUNO will be able to exclude the LSND/MiniBooNE preference for $\absq{t_{e\mu}}$ at somewhere between $1$ and $2\sigma$ credibility. While this is by no means a definitive test of the LSND and MiniBooNE anomalous appearance results, we find this worth highlighting. Note that, in order for this result to be realized, we had to work under the agnostic case, and whether one can construct a model in which we have large $\absq{t_{e\mu}}$ with $N_e$ and $N_\mu$ close to 1 (i.e., $U_{\rm LMM}$ is \textit{not} a subset of a larger, unitary matrix) is unclear.


\bibliographystyle{JHEP}
\bibliography{refs}

\end{document}